\documentclass[12pt]{iopart}
\usepackage{natbib}
\usepackage{iopams}   
\usepackage{epsfig}  
\begin{document}
\newcommand{\anu}{$\rm\bar{\nu}$}         
\newcommand{\neutrino}{$\rm \nu$} 
\newcommand{\Ge}{$\rm^{71}Ge\;$}
\newcommand{\Ga}{$\rm^{71}Ga\;$}  
\newcommand{\Ger}{$\rm^{69}Ge\;$}         
\newcommand{\Gal}{$\rm^{69}Ga\;$}   
\newcommand{\Germ}{$\rm^{68}Ge\;$}
\newcommand{\Gall}{$\rm^{68}Ga\;$}
\newcommand{\Gacl}{$\rm GaCl_3\;$} 
\newcommand{\Gecl}{$\rm GeCl_4\;$} 
\newcommand{\Germane}{$\rm GeH_4\;$}
\newcommand{\Cr}{$\rm ^{51}Cr\;$}
\newcommand{\Crc}{$\rm^{50}Cr\;$}
\newcommand{\Zn}{$\rm ^{65}Zn\;$}   
\newcommand{\Cl}{$\rm ^{37}Cl\;$}
\newcommand{\Ar}{$\rm ^{37}Ar$}   
\newcommand{\In}{$\rm ^{115}In\;$}
\newcommand{\Sn}{$\rm ^{115}Sn\;$}
\newcommand{\iode}{$\rm ^{127}I\;$}
\newcommand{\xenon}{$\rm ^{127}Xe\;$}
\newcommand{\Mo}{$\rm ^{98}Mo\;$}
\newcommand{\Te}{$\rm ^{98}Te\;$}
\newcommand{\Pb}{$\rm ^{205}Pb\;$}
\newcommand{\Tl}{$\rm ^{205}Tl\;$}
\newcommand{\bonze}{$\rm ^{11}B\;$}
\newcommand{\bonzex}{$\rm ^{11}B^*\;$}
\newcommand{\conze}{$\rm ^{11}C\;$}
\newcommand{\conzex}{$\rm ^{11}C^*\;$}
\newcommand{\arq}{$\rm ^{40}Ar\;$}
\newcommand{\arqx}{$\rm ^{40}Ar^*\;$}
\newcommand{\Kq}{$\rm ^{40}K\;$}
\newcommand{\Kqx}{$\rm ^{40}K^*\;$}
\newcommand{\fer}{$\rm ^{55}Fe\;$}
\newcommand{\ttt}{$\rm T_{1/2}\;$}
\newcommand{\tttt}{$\rm t_{1/2}\;$}
\newcommand{\ft}{$\rm ft_{1/2}\;$}
\newcommand{\bd}{$\rm \beta$-decay }
\newcommand{\bds}{$\rm \beta$-decays }
\newcommand{\mum}{$\rm \mu ^{-}\;$}
\newcommand{\mup}{$\rm \mu ^{+}\;$}      
\newcommand{\about}{$\rm \sim\,$} 
\newcommand{\numero}{$\rm n^o$}
\newcommand{\nB}{$\rm \nu_B\;$}
\newcommand{\npp}{$\rm \nu_{pp}\;$}
\newcommand{\npep}{$\rm \nu_{pep}\;$}
\newcommand{\nBe}{$\rm \nu_{Be}\;$}
\newcommand{\Be}{$\rm ^7Be\;$}
\newcommand{\B}{$\rm ^8B\;$}
\newcommand{\nue}{$\rm \nu_e\;$}
\newcommand{\numu}{$\rm \nu_{\mu}\;$}
\newcommand{\nutau}{$\rm \nu_{\tau}\;$}
\newcommand{\nua}{$\rm \nu_1\;$}
\newcommand{\nub}{$\rm \nu_2\;$}
\newcommand{\nuc}{$\rm \nu_3\;$}
\newcommand{\nux}{$\rm \nu_x\;$}
\newcommand{\CCl}{$\rm C_2 Cl_4$}
\newcommand{\He}{$\rm ^4He\;$}
\newcommand{\dmd}{$\rm \Delta m^2\;$}
\newcommand{\absdm}{$\rm |\Delta m^2 |\,(eV^2)\;$}       
\newcommand{\sindt}{$\rm sin^22\theta\;$}
\newcommand{\sint}{$\rm sin^2\theta\;$}
\newcommand{\osplane}{$\rm (\Delta m^2\, ,\,sin^22\theta)\;plane\;$}
\newcommand{\osonly}{$\rm (\Delta m^2\, ,\,sin^22\theta)\;$}
\newcommand{\emin}{$\rm E_{min}\,$}
\newcommand{\emax}{$\rm E_{max}\,$}
\newcommand{\emoins}{$\rm e^-\;$}
\newcommand{\eau}{$\rm H_2O\;$}
\newcommand{\eaul}{$\rm D_2O\;$}
\newcommand{\rar}{$\rightarrow$}
\newcommand{\calcium}{$\rm ^{40}Ca$}
\newcommand{\cano}{$\rm Ca(NO_3)_2$}
\newcommand{\micron}{$\rm \mu m\;$}
\newcommand{\cmt}{$\rm cm^3$}   
\newcommand{\cmd}{$\rm cm^2$}
\newcommand{\evi}{$\rm 10^{21}$} 
\newcommand{\onesig}{$\rm 1\,\sigma\;$}
\newcommand{\twosig}{$\rm 2\,\sigma\;$}
\newcommand{\threesig}{$\rm 3\,\sigma\;$}
\newcommand{\enu}{$\rm E_{\nu}\,$}
\newcommand{\sinw}{$\rm sin^2\theta_w$}
\newcommand{\Bep}{$\rm ^7Be(p,\gamma)^8B$}
\noindent
{\it Report on Progress in Physics,  final version 30 May 2011}

\def\ie{{\it i.e.}}
\def\eg{{\it e.g.}}
\def\etal{{\it et al.}}
\def\cf{{\it cf.}}
\def\TC{{Turck-Chi\`eze et al.}}
\def\TConly{{Turck-Chi\`eze}}
 \vspace{-1.2cm}
\title[Solar neutrinos, helioseismology and the solar internal dynamics
]{Solar neutrinos, helioseismology and the solar internal dynamics}
 \vspace{-0.3cm}
\author{Sylvaine Turck-Chi\`eze$^1$ \& S\'ebastien Couvidat$^2$} 
\footnote{\small sylvaine.turck-chieze@cea.fr, couvidat@stanford.edu}
 \vspace{-0.6cm}
\address{$^1$Service d'Astrophysique/IRFU/DSM/CEA, 91191 Gif sur Yvette Cedex, France\\
$^2$HEPL, Stanford University, Stanford, CA 94305, USA  }
\begin{abstract}
Neutrinos are fundamental particles ubiquitous in the Universe and whose properties remain elusive despite more than 50 years of 
intense research activity.  This review illustrates the importance of solar neutrinos in 
Astrophysics, Nuclear Physics, and  Particle Physics. After a description of the historical context, we remind the reader of the noticeable properties of 
these particles and of the stakes of the solar neutrino puzzle.  The Standard Solar Model triggered persistent efforts in fundamental Physics to predict the solar neutrino fluxes, and its constantly evolving predictions have been regularly compared to the detected neutrino signals. Anticipating that this standard model could not reproduce the internal solar dynamics, a Seismic Solar Model was developed which enriched theoretical neutrino flux predictions with in situ observation of acoustic and gravity waves propagating in the Sun. This seismic model contributed  to the stabilization of the neutrino flux predictions.
This review reminds the main historical steps, from the pioneering Homestake mine experiment and the GALLEX-SAGE experiments capturing the first pp neutrinos. It  emphasizes
the importance of the Superkamiokande and SNO detectors. Both experiments demonstrated that the solar-emitted electronic neutrinos are 
partially transformed into other neutrino flavors before reaching the Earth. 
This sustained experimental effort opens the door to Neutrino Astronomy, with long-base lines and underground detectors. The success of BOREXINO in detecting the $^7Be$ neutrino signal alone instills confidence in the physicists ability to detect each neutrino source separately. It justifies the building of a new generation of detectors to measure the entire solar neutrino spectrum with greater detail, as well as supernova neutrinos.  
A coherent picture emerged from neutrino physics and helioseismology. Today, new paradigms take shape in these two fields: the neutrinos are  massive particles, but their masses are still unknown, and the research on the solar interior is focusing on the dynamical aspects and on signature of dark matter. The magnetic moment of the neutrino begins to be an actor of stellar evolution. The third part of the review is dedicated to this prospect. The understanding of the crucial role of both rotation and magnetism in solar physics benefit from SoHO, SDO, and PICARD space observations, and from new prototype like GOLF-NG. The magneto-hydrodynamical view of the solar interior is a new way of understanding the impact of the Sun on the Earth environment and climate. For now, the particle and stellar challenges seem decoupled, but this is only a superficial appearance. The development of asteroseismology ---with the COROT and KEPLER spacecrafts--- and of neutrino physics will both contribute to improvements in our understanding of, for instance, supernova explosions. This shows the far-reaching impact of Neutrino and Stellar Astronomy.
 \end{abstract}
\vspace{-0.5cm}
\pacs{\small solar neutrinos, neutrino properties, global helioseismology, local helioseismology, internal solar rotation, internal solar magnetic fields,  Sun-Earth relationship.}
\maketitle
\newpage
Contents

1. Introduction
   
   \indent 
   
2. The historical context: the genesis of this field

\indent 

3. The weak interaction and the general neutrino properties

 \indent
{\it  \indent 3.1 Vacuum neutrino oscillations

\indent 
\indent 3.2 Neutrino oscillations in matter

\indent 
\indent
3.3 Characteristics of the different sources of neutrinos}

\indent 
  							
4. The classical view of the Sun through the Standard Solar Model

\indent 
{\it  \indent  4.1 The fundamental equations


\indent 
\indent 4.2 The energy transport

\indent 
\indent 4.3 Improvements in the physics of the SSM. Evolution of the $^8B$ neutrino flux

}
\indent 
    
5. The seismic view of the solar interior and comparison with solar models

\indent 
{\it \indent 5.1 The formalism

\indent 
 \indent  5.2 The sound speed: a very useful but demanding quantity
 
\indent 
\indent 5.3 The sound speed and density profiles: observations and predictions

\indent
 \indent 5.4 The gravity modes }

\indent 

6. The detection of solar neutrinos and comparison with predictions	

\indent 
{\it \indent 6.1 The radiochemical experiments

\indent 
 \indent  6.2 The real time experiments
 
\indent 
\indent 6.3 Toward a neutrino spectrum

\indent 
\indent 6.4  Neutrino predictions from the standard and seismic models

\indent 
\indent 6.5 Comparison with predictions including Neutrino Mixing Parameters}

\indent

7. Beyond the standard solar model: a dynamical view of the Sun  
   
\indent 
{\it \indent 7.1 The internal rotation and its consequences

\indent 
 \indent  7.2 The internal magnetic field
 
 \indent
 \indent 7.3 Toward a dynamical model of the Sun}
 

\indent 
							
8. Conclusion, open questions \& perspectives			 

\indent
{\it \indent 8.1 Secondary effects of the neutrino properties

\indent
\indent 8.2 The neutrino masses

\indent
\indent8.3 Future experiments: Neutrino Astronomy from the Sun to Supernovae

\indent
\indent 8.4 Solar core and Dark Matter 

\indent
\indent 8.5 Last remarks }

\newpage							
\section{Introduction}
The Sun is an outstanding and permanent source of neutrinos of different energy. Understanding this source is crucial 
to different scientific disciplines, and a parallel progress on neutrino properties is necessary to allow important 
developments in Neutrino Astronomy. The Sun is the nearest neutrino source in the Universe (see table 1 and Figure 1) and the first source in term of detectable neutrinos (it produces about  $\rm 6.7 \times 10^{10}\nu/cm^2/s$),
hence its importance. The strongest neutrino source, the  cosmic neutrino background radiation, has a flux of $\rm 10^{22} \nu/cm^2/s$, but the energy of 
these neutrinos is too low (about $\rm 10^{-4}- 10^{-3}$ eV) to currently allow a direct detection.

In the late sixties, it became clear that we did not understand the solar-neutrino properties. From there, it took about half a century to provide 
solutions to the  ``solar neutrino puzzle'', or deficit of neutrinos detected on Earth 
compared with the ``theoretical estimate'' of the emitted fluxes. We now reached a 
coherent picture of this problem. We 
know unambiguously that neutrinos have a mass, unlike what was thought for a long time. However a lot of questions remain unanswered regarding the real nature of the neutrinos, their magnetic interaction with plasma,  
with the internal solar dynamics, and their connection to the solar activity. 

During the last twenty years, helioseismology provided Astrophysics with an opportunity to 
describe with an unprecedented accuracy the different solar neutrino sources and to look for coherence between the two existing 
probes 
of the deep solar core. In parallel, this discipline stimulated the development of research on the internal Stellar Dynamics. Today an impressive agreement between ``helioseismic'' neutrino flux predictions and 
neutrino detections by the existing detectors has been reached.  Thanks to this agreement, it becomes possible to describe the solar  
neutrino sources with an accuracy equivalent to the one reached from particle accelerators or nuclear reactors producing neutrinos or antineutrinos. 
Combined with the improvements in neutrino detectors, these facts favor the development of Neutrino Astronomy. Considering the vitality of the two disciplines (including asteroseismology, the  seismology of stars), important discoveries are bound to be made in the upcoming decades. 

With hindsight, it is clear that using the second best source of neutrinos in the Universe, the Sun, was especially judicious 
because it allowed great advances in numerous fields of physics which contribute to a better description of
both stars and neutrino properties. This review details the main aspects of this research (sections 3 to 7). Section 8 details currently unanswered and interesting questions justifying 
upcoming developments,  including the question of dark matter. First, we start with an historical review of the genesis of this research field (section 2).

\begin{table*}
\caption{Nuclear processes of relevance in the central region of the Sun.}\label{tab:exp}
  \begin{center}
 \begin{tabular}{||c|l||}\hline\hline 
 {\small \bf pp chain: 98.8\% of the total energy produced by the present Sun}\\
 {\small $\;\; \rm p + p \rightarrow D + e^+ + \nu_e $  { \it called  \bf pp neutrinos}} \\
{\small $(0.25 \%)\; : \rm p + p + e^- \rightarrow D + \nu_e $  { \it called  \bf pep neutrinos}} \\
{\small $\rm pp \,I \,chain \,(86\%)\,:\,  ^3He +^3He \rightarrow ^4 He + 2p  \; Q_{eff}= 26.2  \,MeV$} \\
 {\small $ \rm pp  \,II  \,chain:(14\%)\; : \,^3He + ^4He \rightarrow ^7 Be + \gamma$ }\\  
{\small  $\rm ^7Be + e^- \rightarrow ^7Li + \nu_e$ \;\; 
{ \it called  \bf $^7Be$ neutrinos}} \\
{\small  $\rm ^7Li + p \rightarrow 2 ^4He \;\; Q_{eff}= 25.66  \,MeV $}\\
 {\small  pp III chain: (0.2\%):$\;$  $\rm ^7Be +p \rightarrow ^8B \;\; Q_{eff}= 19.17  \,MeV $ }\\
{\small   $ \rm ^8B  \rightarrow ^8Be^* + e^+  + \nu_e \;  \,^8Be^* 
\rightarrow 2 ^4$He \;  { \it called  \bf $^8$B  \,neutrinos}} \\
{\small $ \rm   pp  \,IV  \,chain:     (0.002\%) \;\;       p + ^3He \rightarrow \nu_e + e^+ + ^4 He $ (hep neutrinos)}  \\
\\
{\small \bf CNO cycle: 1.2\% of the total energy produced by the present Sun}\\
{\small  CNO $\;$ I cycle:  $\;$     $\rm ^{13}N \rightarrow ^{13}C + e^+ + \nu_e $ 
{ \it called  \bf $^{13}N$ neutrinos}} \\
{\small $\rm              ^{15}O \rightarrow ^{15}N + e^+ + \nu_e$  \, { \it called  \bf $^{15}O$ neutrinos}}\\
{\small   $ \rm CNO \; II \,cycle:  \;\;   ^{17}F \rightarrow ^{17}O + e^+ +\nu_e$  \,{ \it called  \bf$^{17}F$ neutrinos}} 
\\
{\small  $\rm Q_{eff} \; CNO = 26.73 \,MeV $ }
\\
 
{}\\
\hline\hline
  \end{tabular} 
  \end{center}
  \vspace{-5mm}
    \end{table*} 
\begin{figure}
\vspace{1.5cm}
\begin{center}
\rotatebox{90}{\includegraphics[width=9cm]{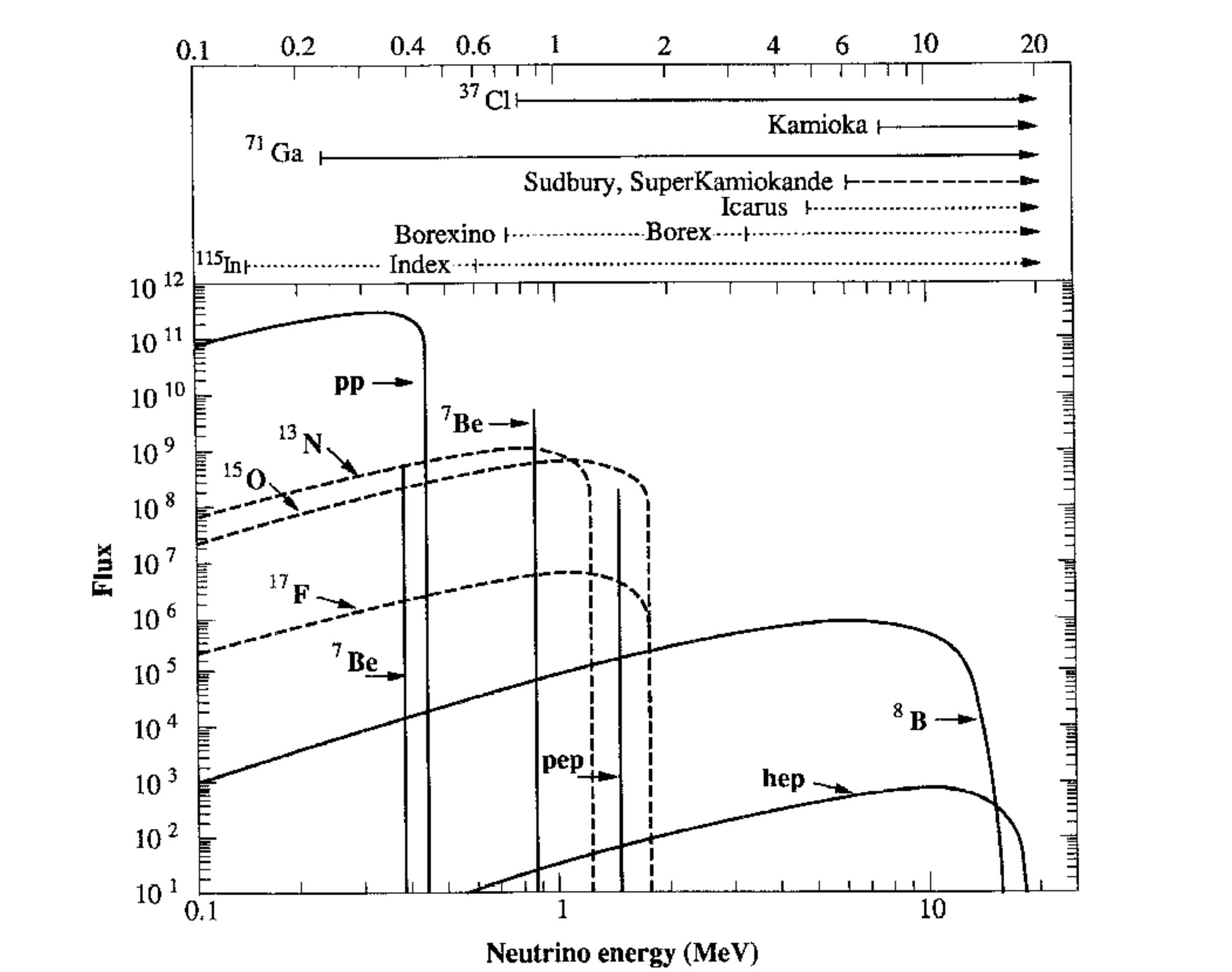}}
\caption{Energy dependence of the neutrino sources and the range of energy covered by the neutrino detectors.}
\end{center}
\end{figure}	
\section{Historical context: the genesis of this field}
Pauli was still young, on December 4$^{th}$, 1930, when he suggested the existence of a ``small neutral particle'' to interpret the continuum energy spectrum 
of the electrons in the $\beta$  radioactivity, through a letter publicly read at the
Tubingen conference. It is difficult to believe that the existence of this ``neutrino''  particle 
was proposed before the discovery of the neutron (Chadwick 1932). This brilliant idea was immediately supported by 
Fermi in 1931 as an important piece in the weak interaction theory, and he coined the name of ``the 
little neutral''. Nevertheless, the first demonstration of its existence by F. 
Reines and C. Cowan,  near the reactor of Savannah River, came only 25 years later, in 1956.  

Pontecorvo, in 1946, suggested to use the Sun as an important source of neutrinos, and Crane (1948) 
showed the potentiality of neutrino physics. In parallel, the knowledge on stellar evolution was steadfastly progressing. \cite{eddington26}
had already understood that the continuous 
brightness of the stars was of nuclear origin. Then, Gamow, Weizsacker, and Bethe solved the conundrum of these nuclear interactions: the competition between the 
Maxwellian tail distribution and the presence of the Coulomb barrier, and the existence of different chains of 
reactions (see table 1 and section 4). The number of neutrinos produced in the Sun can be  
directly deduced from the fact that  4 protons convert hydrogen to helium while releasing 2 neutrinos and some energy.
The neutrino flux is obtained from the ratio of the solar luminosity to the quantity of energy released. $$\rm 4p \rightarrow ^4He + 2 e^+ + 2 \nu_e + 26.20 MeV - E_{\nu}$$ where 26.20 MeV is the most common effective energy and $E_{\nu}$ has a mean value of 0.26 MeV (in 85\% of the cases, see tables 1 and 2). Based on this reaction, about 7 $\rm \times 10^{10} neutrinos /cm^2$/s escape from the Sun. Undoubtedly these developments fascinated the young R. Davis, who received a formation in physical chemistry and was looking for a good research topic, when he joined the Brookhaven National Laboratory in the aftermath of the second world war.   

After several years spent on looking for the best detection technique, 
Davis decided  in 1951 to follow the idea of Pontecorvo and to try capturing neutrinos through the reaction: $\rm ^{37}Cl + \nu_e \rightarrow 
^{37}Ar +e^-$, with a decay back to $^{37}$Cl by the inverse capture process with a half-life of 35 
days.  There were at that time two possible sources of detectable neutrinos: the nuclear fission reactors and the Sun. Therefore Davis first tried to validate this method with a 3900-liter tank of carbon-tetrachloride (CCl$_4$) as 
the target material, located near the  Brookhaven nuclear reactor. Back then, the fact that only antineutrinos, and  no  neutrinos, were produced was 
not deemed a good enough reason to pass this opportunity. The reactor was considered not powerful enough to produce any neutrino signal.   In parallel, the background noise resulting from cosmic rays was clearly identified, and a first experimental upper limit for the solar neutrino flux was set at 40000 SNU \citep{davis55}(where 1 SNU, the Solar 
Neutrino Unit, is defined as $10^{-36}$ captures per target atom per second). Even at the time, this result appeared uninformative. This first attempt also highlighted the necessity to recourse to underground experiments. 
Davis decided to repeat this experiment twice at 
Savannah River, the most powerful site of  nuclear reactors in the world, first with the same quantity of carbon-tetrachloride, and then with a 11400-liter tank. The absence of any reactor neutrino signal lowered the 
upper limit for neutrino capture to a factor 20 below the antineutrino capture rate. The first 
detection of a free antineutrino by \cite{cowan56,reines60}, using a different technique, coupled with the results 
of Davis, showed that the neutrino was not its own antiparticle (see  \cite{davis02}). They will win the Nobel prize in 1995 for this discovery. See section 3 for a more general consideration on neutrino properties).

The Sun produces a great many neutrinos. Unfortunately these neutrinos come mainly from the fundamental transformation of a proton into a neutron: $\rm p+ p \rightarrow 2D + \nu_e+ e^+$ (where D is a deuteron), with a resulting neutrino energy lower than $0.42$ 
MeV (see table 1 and figure 1). With such low energy these neutrinos were not accessible to the detector of R. Davis, since the energy threshold of his experiment was $0.814$ 
MeV. At that time, the ppI chain converting hydrogen into helium through ($\rm ^3He, ^3He$) interaction was considered 
as the main actor of the proton-neutron transformation. But when Holmgren and Johnston reported in 1958 that the 
reaction rate of ($\rm ^3He + ^4He$) was 1000 times greater than previously thought, the production of $\rm ^7Be$ was 
then increased by a factor 7000, immediately raising the odds of a  $\rm ^7Be$-proton interaction producing $\rm ^8B$ with emission of an energetic neutrino, and also the odds of an electronic capture producing $^7$Li and another neutrino. Both reactions are energetic sources of neutrinos and the resulting prospect of several 
neutrino captures per day was a strong motivation for building a 3800-liter tank of perchloroethylene $\rm(C_2Cl_4)$ in the 
Barberton Limestone mine in Ohio. This solution of perchloroethylene was widely used in dry cleaner shops and less dangerous for the health. Unfortunately, once more,  Davis did not detect any neutrino signal. 

The different reaction cross sections were measured. \cite{kavanagh60}
measured the 
$\rm ^7Be(p,\gamma)$ reaction rate and it appeared that this interaction cross-section was much smaller than the electron 
capture cross-section. Due to the energy range of the neutrinos produced by the two reactions (0 to $14.1$ MeV for neutrinos associated with 
$\rm ^8B$ and two narrow lines of respectively $0.38$ and $0.86$ MeV for those associated with $\rm ^7$Li), the likelihood of detecting solar neutrinos dropped, as written by Reines: ``The probability of a negative 
result even with detectors of thousands or possibly hundreds of thousands of gallons of 
perchloroethylene tends to dissuade experimentalists from making the attempts''.

However,  \cite{bahcall62} calculated the electron capture rate and the capture rate of $^8$B  and showed that the latter rate was 20 times larger than previously thought. Therefore there was a real hope of detecting solar neutrinos: the theory became a guidance for determining the first neutrino fluxes \citep{bahcall63} and
the size of the detector \citep{bahcall64}
: for instance, 378000 liters of perchloroethylene could detect 4 
to 11 $\rm ^{37}Ar$ atoms per day (about 28 SNU). In parallel, the work on muon cosmic-ray at Barberton was useful to determine
the depth of the set up needed to kill the main source of noise \citep{davis64}
Davis, in collaboration with Bahcall, put a tank in the Homestake 
Gold Mine in Dakota at the -1478 m level of the mine. When the measurements started in 1968, it was clear 
that the main contributor to the detected neutrinos was the proton capture rate on $\rm ^7Be$. The 
neutrino flux predictions \citep{bahcall68}
included an error bar, due to the uncertainties on the ingredients 
of the solar model, of about 60\%. In the following decades, as shown in this review, John Bahcall turned out to be the most active theoretician sustaining and stimulating work on new neutrino detection techniques, thanks to his extremely well documented neutrino flux predictions. He is at the root of the solar neutrino puzzle.

In 1968, R. Davis and collaborators derived an upper limit of 3 SNU for the solar neutrino emission \citep{davis68}
To obtain this result, he had first to use some $\rm ^{36}$Ar in order to demonstrate his ability to extract argon atom at the 95\% level. 
In 1971, he announced a detected flux of $1.5$ $\pm$ 1 SNU. This value has not significantly changed over time, and except a reduction in the error bar partly due to the integration over a longer timespan, the 
statistical accuracy of each run remained very low (about 7 events per month on average). 
In the meantime, J. Bahcall, N. Bahcall, \& G. Shaviv, in a letter to Physical Review Letters in 1968, produced different models of the Sun and favored a model 
leading to a chlorine neutrino flux prediction of $7.7$ SNU, much more than the detected flux. Immediately, Pontecorvo mentioned the possibility of oscillation 
between neutrino flavors \citep{pontecorvo68}. 1968 was undeniably a very inspiring year for neutrino physics.  For his lifetime efforts, R. Davis received the Nobel prize in 2002, 34 years later. This Nobel prize was shared with M. Koshiba for the outstanding success of the Kamiokande and Superkamiokande collaborations (discussed later in this review).  The detection of atmospheric and solar neutrinos contributed to the evidence of neutrino oscillation (see below).

    The prediction of solar neutrino fluxes from a solar model results from complex calculations,  and J. Bahcall played a major role in this field. Such predictions require a good estimate 
of the nuclear reaction rates listed in tables 1 or 2, a good estimate of the neutrino cross sections  \citep{bahcall64b,bahcall89}, a good knowledge of the Sun, and 
a strong understanding of the neutrino properties. Nevertheless, convergence of the various estimates of the emitted neutrino flux was rather quickly achieved, and the predictions have not changed by more than a factor of 2 (for $^8B$ neutrinos) once the first period of ``instability'' settled (see all the publications dedicated to this topic between 1985 and 1998, and mentioned in the present article). However, 
the deficit of neutrino detection compared to the prediction, coined the neutrino puzzle as early as the seventies, clearly meant that a great deal of interesting physics was yet to be discovered: new properties of the neutrinos and/or new insight in the solar (stellar) 
physics beyond the standard framework. Since this standard framework (ignoring, among others, the role of the magnetic field) could not reproduce the observations, this raised the question of whether or not the Sun should be treated  as a magnetic star. That means including a lot of processes that are potentially present in the Sun, such as mixing in the solar core, rotation, magnetic field \citep{shaviv71,schatzman69,schatzman81,roxburgh85,gough90} but that are not easy to quantify without constraints in the radiative zone.
This period was very stimulating for all the research fields linked to the neutrino puzzle. 
Variations by up to 30\% in the prediction of the $^8$B neutrino flux were partly due to the strong temperature 
dependence of this flux (see section 4) and to the separate influence of numerous ingredients. 
This strong temperature dependence raised the hope that neutrinos could be a good thermometer of the center of the Sun. 

 In parallel, helioseismology was being strongly developed on both the theoretical and experimental sides \citep{duvall79,grec80,claverie81,duvall82,christensen82,christensen85,gough85} with a real potential to properly describe the radiative zone and, maybe, the deep core of the Sun. Considering  how these two probes (helioseismology and the neutrinos) complement each other, 
S. Turck-Chi\`eze (1988) decided to use the solar sound speed obtained from helioseismic instruments in space to go beyond a purely theoretical prediction of the neutrino fluxes. Also, she contributed to build the GOLF \citep{gabriel95} instrument, onboard the SoHO spacecraft, dedicated to research on the solar core. With her collaborators, she developed more and more accurate standard and then seismic solar models. For more details, the reader is refered to the review of \cite{turck93}
describing the different aspects of the neutrino puzzle and of the solar research. The SoHO spacecraft, launched in December 1995, 
provided solar physicists with a wonderful opportunity to determine the solar sound speed down to 0.06 R$_{\odot}$ (where R$_{\odot}$ is the solar radius) with a good accuracy, 
thus reaching the region of neutrino emission \citep{turck01b}.
In the meantime, most of the physics of 
the standard solar model was improved (see sections 4 and 5). 
Today, the solar central temperature is estimated with a precision level of about 0.1 $\times 10^6$ K, much better than 1 \%, thanks to the combined efforts of astrophysicists and the SNO (Sudbury Neutrino Observatory) collaboration.

Since the middle of the nineteen-eighties, the amount of information available dramatically increased on both the solar physics and neutrino sides. 
Different neutrino detection techniques were developed to complete the remarkable results of 
the Homestake experiment. Twenty three years after the first result of Davis, a real-time experiment began at Kamiokande in
 Japan, with pure water in an installation dedicated to the determination of the proton lifetime. Kamiokande
confirmed the solar neutrino deficit \citep{hirata89}.
Two radiochemical detectors using gallium were ready to run at 
the beginning of the nineteen-nineties, in Gran Sasso (Italy) under the responsibility of T. Kirsten \citep{anselmann92, hampel98},
and in the Balkans under the responsibility of V. Gavrin \citep{abazov91,gavrin92,abdurashitov94,abdurashitov99a,abdurashitov99b}, 
thus allowing a comparison of different neutrino detection techniques. For the first time, the neutrino 
flux issued from the proton-proton (pp) reaction was measured in addition to the other neutrino sources. Strangely enough, the first flux announcements were values just above the predicted 
pp neutrino fluxes for one experiment, and an absence of detection for the other. With an increase in the sample size, both experiments showed a global deficit. Then, on April 1$^{th}$ 1996, it was announced that the detector of 
Superkamiokande started to detect solar neutrinos \citep{fukuda98,fukuda00}.
With Superkamiokande, the statistical accuracy was dramatically improved with more than 
5000 neutrinos detected  per year, coming from the Sun. A dream come true for R. Davis. Superkamiokande was 
the first experiment to unambiguously confirm the existence of neutrino oscillations, with the additional measurement 
of the atmospheric neutrinos. 
The neutrino oscillation in matter as solution to the solar neutrino puzzle was proposed by particle theoreticians as early as the first detections 
by Davis, and has been mainly developed by \cite{wolfenstein78},
and \cite{mikheyev86}.
Therefore, it has been coined the ``MSW'' effect.
 Finally, the Canadian heavy water detector, SNO, unambiguously proved its existence. SNO 
measures the interaction of different flavors of neutrinos \citep{ahmad01,ahmad02,ahmed04}
and allowed, 
for the 
first time, a determination of all the neutrinos (all flavors) coming from the Sun. 
The total neutrino flux thus measured immediately agreed with the predictions of solar models: standard and seismic models of the Sun, 
whose predictions were close to each other \citep{bahcall01, turck01b,couvidat03}. 

Undeniably, 2001 was a second very important year for solar neutrino problem.
Such a wonderful agreement strongly validates the measurements of the pioneer experiments. 
Their authors, R. Davis and M. Koshiba, received   
a joint Nobel prize in 2002 for their seminal contribution to this field. Recently, improvements on solar CNO (Carbon, Nitrogen, Oxygen) abundance determination put a renewed pressure on the standard solar model neutrino predictions and emphasized the need to recourse to a seismic solar model for these predictions. It is clear that solar observations benefit from two probes (neutrinos and helioseismology), and that
the Large Mixing Angle (LMA) solution is the favored solution to the solar neutrino oscillation 
\citep{bahcall04,inoue04,aharmim05}. 
Helioseismology and particle physics are now in agreement \citep{turck04a}
at a 10\% level, showing the limits of the  
neutrino-matter interaction effects, and the maturity of the different scientific approaches.
Borexino recently confirmed the impact of the MSW effect by measuring, for the first time, the $^7Be$ and $^8B$ neutrino fluxes with the same detector \citep{arpesella08}.
This experiment confirms the LMA solution of oscillations, and directly shows the reduction of neutrino fluxes at low energy (due to oscillations in vacuum) and at high energy (due to oscillations in matter). Physicists are now able to obtain individual measurements of the different sources of neutrinos. Borexino and KamLAND 
(the largest low-energy antineutrino detector presently measuring antineutrinos from nuclear reactors) confirm the reduction of neutrino flux at low energy due to neutrino oscillations. We will probably gain further constraints on CNO abundance in the solar radiative zone in a near future, thanks to, among others, the detection of CNO-cycle neutrinos. This could be an opportunity to look for magnetic interaction 
between neutrinos and matter.
In 
astrophysics, helioseismology opened the door to the study of extra phenomena not yet fully introduced in the standard solar model. For instance, the role of rotation and  magnetic field in the Sun is beginning to be considered in more detail than what was done in the past, which should help bridge the gap in the interpretation of the different solar activity data. In the meantime, the seismic solar model remains superior to the standard solar model  
to explain the different neutrino fluxes observed (see sections 5, 6, and 7). 

 The quest for solar neutrinos already produced fantastic results, and is far from over. New paths are being explored to answer the fundamental questions of neutrino physics: are they Majorana particles, what are their masses? New paths are also being explored regarding the Sun and other stars. These two research fields will contribute to the development of Neutrino-stellar Astronomy (section 8). 

In the next section we detail the physics of the neutrinos, and its development since the ``historical time'' 
of the first neutrino flux measurements by R. Davis in 1968.

\section{The weak interaction and the general neutrino properties}

The neutrino is a very peculiar particle: 

- neutrinos are generated in the Sun mainly by the transformation of a proton 
into a neutron, through weak interactions: either $\beta +$ disintegration (with production of a positron) or electron capture. The main solar energy source is the interaction between 
two protons which produces a deuteron plus a positron and a neutrino (table 1). 
It is remarkable that a neutron (mass of 939.56 MeV/$c^2$) can be generated  
from a proton (mass of 938.28 MeV/$c^2$) despite its smaller mass. The opposite case, the $\beta -$ disintegration  
$n \rightarrow p + e + \overline\nu_e$, is conceptually quite natural and free neutrons are unstable, with a mean 
life of $885.7$ s ($14.7$ mn). On the contrary, the long lifetime of the proton, estimated from the SuperKamiokande 
detector to be  $>$ 6.6 $\times 10^{33}$ yrs \citep{nishino09}, does not immediately favor the reaction: 
$p \rightarrow n + e^+ +\nu_e$. Indeed, it is noticeable that a free proton needs an initial 
velocity (kinetic energy) to produce a neutrino, or a reaction between two nuclei $A_1$=A(Z,N) and $A_2$=A(Z-1, N+1) 
with M(A$_1) < M(A_2$)+ $m_e$. In the Sun, $A_1$ is the sum of two protons and $A_2$ is a deuteron. 
This immediately shows that the cross section of such an interaction will be very small 
and that, consequently, the resulting neutrino energy will be low ($<0.4$ 
MeV). Hence the difficulty to detect neutrinos. 


- the absence of neutrino detection by Davis near nuclear reactors, combined to  the actual detection of solar neutrinos, seemed to favor the idea that \nue (particles emitted together with a positron in $\beta ^+$ decay) are 
different from  $\overline\nu_e$ neutrinos (particles associated to electron in $\beta ^-$ decay, near reactors). 
This premature conclusion was linked to the fact that the neutrino (resp. antineutrino) has a spin $1/2$, 
no electric charge, and a leptonic charge +1 (resp. -1). However this simple
picture was complicated by the demonstration of C. S. Wu (1957) regarding $\beta$ 
decay: the parity conservation is violated.  This followed the possible occurrence of such a violation, suggested by \cite{lee56}.
In the minimal standard model of strong and electroweak interactions, SU(3)*SU(2)*U(1), 
the consequences of this violation are the following: the neutrino is a Weyl particule, 
represented by a spinor at two components, and only the neutrinos of left chirality and antineutrinos of right chirality exist.  
The neutrinos can only exist in a specific state of polarisation: only left-handed helicity neutrinos are produced (that means that the spin and the momentum of a neutrino are opposite) and $\beta ^-$ 
produce antineutrinos of right helicity (same direction for the spin and momentum). 

- there are 3 families of neutrinos associated to the corresponding leptons: 
electron, muon, and tau. The discovery of neutral currents in 1973 showed that 
all the different neutrinos do not experience the same interactions. The different \nux may 
experience neutral interaction mediated by $Z_0$ (Z Exchange Process, ZEP), but only \nue can experience charged current 
interaction (W Exchange Process, WEP) mediated by the W particle, see \cite{bouchez05}.

The idea that neutrinos could oscillate 
between their different flavors was proposed by
\cite{gribov69}
soon after the first results      
of the chlorine experiment.
Indeed the nuclear reactions in the Sun produce only 
one flavor, \nue,
and the chlorine detector is sensitive only to this flavor.
If there are less neutrinos detected on Earth than neutrinos
produced in the Sun, this may be due to the transformation of \nue into
\numu or \nutau.
One condition for this to happen is that neutrinos are massive.
The propagation eigenstates are then different from the
flavor eigenstates and the phase change during the propagation
will induce a flavor change.
From the neutrino oscillation formalism
in vacuum and in matter, it was shown how this can be a natural
interpretation of the solar neutrino experiment results.

\noindent
\subsection{Vacuum neutrino oscillations}

In the standard model of particle physics, neutrinos are
assumed to be purely left handed and massless.
The standard model is based on arbitrary parameters and is
generally only considered as a first step toward a more complete theory.
In Grand Unified Theories (GUTs) neutrinos have a mass
and the mass eigenstates \nua, \nub,
and \nuc are different from the flavor eigenstates \nue,
\numu, and \nutau. The total lepton number is conserved, but
not the separate electron, muon, and tau numbers.
The transformation between the flavor
and the mass eigenstates can be written~:
\[ \rm \left( \begin{array}{c}  
   \nu_e \\ \nu_{\mu} \\ \nu_{\tau} \end{array}  \right)
      =\; U \; \left( \begin{array}{c} 
   \nu_1 \\ \nu_2 \\ \nu_3 \end{array} \right)  \]
where U is a 3x3 unitary matrix.
In the simple case where we consider only two flavors,
the transformation has only one parameter,
$\rm\theta$, which is called the mixing angle, 
and U is written :
\[ \rm \left( \begin{array}{c} \nu_e \\ \nu_{\mu} \end{array} \right) \; = \;
\left( \begin{array}{cc} \rm cos\theta & \rm sin\theta \\

\rm -sin\theta & \rm cos\theta \end{array} \right)
\left( \begin{array}{c} \nu_1 \\ \nu_2 \end{array} \right) \]

When neutrinos propagate in vacuum, the evolution equation takes the
following form:
\[ \rm i\frac{d}{dt}  \left( \begin{array}{c}     
     \nu_e \\ \nu_{\mu} \end{array} \right) \; = \;  
\left( \begin{array}{cc}
  \rm\frac{\Delta m^2}{4E_{\nu}}cos2\theta & 
  \rm -\frac{\Delta m^2}{4E_{\nu}}sin2\theta  \\
  \rm-\frac{\Delta m^2}{4E_{\nu}}sin2\theta & 
  \rm -\frac{\Delta m^2}{4E_{\nu}}cos2\theta 
\end{array} \right)  \left( \begin{array}{c}     
     \nu_e \\ \nu_{\mu} \end{array} \right)   \]
where \enu is the neutrino energy, and \dmd = $\rm m_2^2-m_1^2$ is the 
difference of the squared masses of \nua and \nub.
The probability P(\nue\rar\nue)
that a \nue 
produced at t=0 is still a \nue after a propagation time t 
(or a propagation distance l) is expressed as:
$$ \rm P(\nu_e\rightarrow \nu_e)\;=\;
|<\nu_e(t)|\nu_e(0)>|^2 \; = \;
1-\frac{1}{2}sin^22\theta\left[ 1-cos\frac{2\pi l}{l_v}\right]  \eqno(3.1) $$
where $\rm l_v=\frac{4\pi E_{\nu}}{\Delta m^2}$ is the vacuum oscillation
length. The amplitude of the oscillation depends on the mixing angle $\rm\theta$.

The general solution for three-neutrino flavor is~:
\[ \rm P(\nu_{\alpha}\rightarrow\nu_{\beta}) = 
\left|\sum_{i} U_{\beta i}\;exp(-iE_it)\;U^*_{\alpha i}\right|^2 \] 
\[ \rm = \; \sum_{i} |U_{\beta i}|^2|U_{\alpha i}|^2\;+\;
 Re\sum_{i\neq j}U_{\beta i} U_{\beta j}^* U_{\alpha i}^* U_{\alpha j}\,
  exp(-2\pi i \frac{l}{l_v}) \]

\vspace {5. mm}
\noindent
\subsection {Neutrino oscillations in matter} 

\cite{wolfenstein78}
observed that the presence of matter modifies
the propagation of neutrinos because of the effects of coherent
forward elastic scattering.
As previously mentioned, ZEP contributes to the
elastic scattering of all neutrinos, whereas WEP contributes
only to \nue scattering.
This implies a difference in the index of refraction for \nue
and for \numu or \nutau.
The propagation equation is then written~: 
\[ \rm i\frac{d}{dt}  \left( \begin{array}{c}     
     \nu_e \\ \nu_{\mu} \end{array} \right) \; = \;  
\left( \begin{array}{cc}
  \rm -\frac{\Delta m^2}{4E_{\nu}}cos2\theta+\frac{G\rho}{\sqrt{2}} 
  & \rm \frac{\Delta m^2}{4E_{\nu}}sin2\theta  \\
  \rm \frac{\Delta m^2}{4E_{\nu}}sin2\theta 
& \rm \frac{\Delta m^2}{4E_{\nu}}cos2\theta-\frac{G\rho}{\sqrt{2}}  
\end{array} \right)  \left( \begin{array}{c}     
     \nu_e \\ \nu_{\mu} \end{array} \right)   \]
where G is the weak interaction Fermi constant and $\rm\rho$ 
is the electron density of the medium.

If the density is constant, the previous system can easily be solved.
The probability P(\nue\rar\nue) has a form very similar to
the probability previously obtained in vacuum~:
$$ \rm P(\nu_e\rightarrow \nu_e)\;=\;
|<\nu_e(t)|\nu_e(0)>|^2 \; = \;
1-\frac{1}{2}sin^22\theta_m\left[ 1-cos\frac{2\pi l}{l_m}\right]  \eqno(3.2) $$
The  mixing angle in matter $\rm\theta_m$ and the oscillation length $\rm l_m$ depend on
the electron density of the matter through the relations~: \[ \rm tan2\theta_m =
\frac{sin2\theta}{cos2\theta - (l_v/l_o)} \] \[ \rm
l_m\,=\,\frac{l_v}{\sqrt{1-2(l_v/l_o)cos2\theta+(l_v/l_o)^2}} \] where $\rm
l_o\,=\,2\pi/(\sqrt{2}G\rho) $ is a characteristic length of the medium.
\cite{mikheyev86}
pointed out that, for a given set of
the neutrino oscillation parameters $\rm\theta_m$ and \dmd , and
for a given neutrino energy \enu, there is a value of the density
which induces a ``resonant'' mixing, i.e.
$\rm sin^22\theta_m=1$.

If the density of the matter is not constant, there is no general
solution to the evolution equation.
The mass eigenstates are no longer eigenstates of the Hamiltonian.
However, if the density has a slowly varying behaviour,
\cite{mikheyev86}
showed that 
the instantaneous mass eigenstates become eigenstates of the Hamiltonian.
This property, well known in quantum mechanics, is called
the adiabatic approximation.
The crossing of the resonant density may then induce a very important
phenomenon: the adiabatic transformation of a \nue, flavor eigenstate,
into a \nub, vacuum eigenstate.
This is called the Mikheyev, Smirnov,
and Wolfenstein (MSW) effect.

This situation may happen in the Sun where the
density decreases relatively slowly with the radial distance, with an exponential-like
behaviour, from about 150\,g/cm$^3$ at the center, where the neutrinos are produced, to almost zero at the surface.
In this case, a pure \nue can leave the 
Sun as a pure \nub, i.e. a pure state of propagation in vacuum.
Two conditions are necessary for this to happen: a) the density
where the neutrino is produced must be larger than the resonant
density corresponding to its energy and to the neutrino
oscillation parameters; and b) the adiabatic condition must be satisfied.
In this case:
a) the interval in which there is a \nue flux suppression
is large and its width decreases with \sindt;
and b) the minimum value of the flux is equal to sin$\rm ^2\theta_m$,
which means that the smaller the mixing angle, the larger
the \nue flux reduction.
The reduction factor depends on energy and a consequence is that,
depending on the oscillation parameters, the observed \nue spectrum may
be distorted compared to the theoretical one.  

Moreover, when neutrinos reach the detector on Earth
at night, there may partly be regeneration of the \nue which
``disappeared'' in the Sun, see \cite{bouchez86}.
Indeed if the neutrino arrives on Earth as a pure \nub, this \nub
is no longer an eigenstate of propagation in the Earth and some \nua
component may appear depending on the values of the parameters.

\begin{table*}
\caption{The nuclear network: nuclear energy produced by each reaction, maximal
value of the neutrino energy (E$_{\nu}$ max), its mean value deduced from
neutrino energy spectra ($\rm \overline E_ \nu$), for contributions to the total
luminosity greater than $0.1$\%. From \cite{turck88}.} 
\begin{center}
\begin{tabular}{|c|c|c|c|c|} 
\hline                                                     
Reaction  & nuclear energy  &  $ E_{\nu}$ max    &  
$\rm \overline E_{\nu}$      &
   luminosity \\
&  MeV  &  MeV  &  MeV &
\\ \hline p(p,e$^{+}\nu$)D & 1.442 & 0.420 & 0.265 & 8.26$\%$ \\
p(pe$^{-},\nu$)D & 1.442 & 1.442 & 1.442 & 0 \\
D(p,$\gamma$) $^{3}He$ & 5.494 &  &  & 38.29 $\%$ \\
$^{3}He$($^{3}He$,2p)$^{4}He$ & 12.860 &  &  & 42.84 $\%$ \\
$^{3}He$($\alpha, \gamma$) $^{7}Be$ &  1.586 &  &  & 0.77 $\%$  \\
$^{7}Be$(e$^{-}, \nu$)$^{7}Li$ & 0.862 \  0.324 &  & 0.862 \ 
0.324 & 0.  \\
$^{7}Li$(p, $\alpha$ ) $^{4}He$ & 17.348 &  &  & 8.07 $\%$ \\
$^{7}Be$(p, $\gamma$)$^{8}B$(e$^{+}\nu$) $^{8}Be
^{\ast}$($\alpha$)$^{4}He$ &  17.98 & 14.02 & 6.71 & 0.  \\
  &  &  &  &   \\
$^{12}C$(p,$\gamma$)$^{13}N$(e$^{+}\nu$)$^{13}C$ & 4.454 & 1.198  & 0.707 &
0.26$\%$ \\
$^{13}C$(p,$\gamma$)$^{14}N$ & 7.551 &  &  & 0.56 $\%$ \\
$^{14}N$(p, $\gamma$)$^{15}O$(e$^{+}\nu$)$^{15}N$ &10. 05 & 1.173 & 0.997 &
0.60 $\%$ \\
$^{15}N$(p, $\alpha$)$^{12}C$ & 4.966  &  &  & 0.33$\%$ \\
$^{15}N$(p, $\gamma$)$^{16}O$ &12.128 &  &  & 0.  \\
$^{16}O$(p,$\gamma$)$^{17}F$(e$^{+}\nu$)$^{17}O$  & 3.422 &   & 0.999 & 
0.  \\
$^{17}O$(p,$\alpha$)$^{14}N$ & 1.193  &  &  &  0. \\
\hline 
\end{tabular} 
\end{center}
\normalsize
\end{table*}
\subsection{Characteristics of the different solar neutrino sources}

Neutrinos are produced in the central part of the Sun by several
reactions described in tables 1  and 2: the
 pp reaction (with a small contribution, $0.25\%$, from the weaker pep
reaction), the  electronic capture on $^{7}Be$ which produces \nBe, and
the proton capture by $^{7}Be$,  which produces  excited
$^{8}B$ atoms which decay and generate the so-called \nB. The
reactions from the CNO cycle involving $^{13}N$ and $^{15}O$  are of 
minor importance for the luminosity of the Sun but are not
negligible for the neutrino flux. The 
$^{17}F$ source, which also comes from the CNO cycle, is much less
important. A reaction, energetically negligible, but which involves
$^3He$ and produces neutrinos with an energy up to 19 MeV (called hep
neutrinos) has been
identified by \cite{bahcall88}.
The  energy
spectrum  (figure 1) is crucial for neutrino detection. This spectrum 
is determined by
the kinematics and the energy released by the corresponding
nuclear reactions. It was calculated by  \cite{bahcall86}
and updated by \cite{bahcall88}
for other sources. 

All these
reactions generate electron neutrinos $\nu_{e}$, but these
neutrinos differ by their region of emission, their flux, and their
energy spectrum (figure 1). The main interest of the neutrino lies in its 
particularly low  cross section (about $10^{-44}$ to $10^{-41} \ 
\rm cm^{2}$ in the energy range considered: 0  to 14 MeV). This results
in a mean free path of  about $0.5$ parsec in the solar interior; 
accordingly, neutrinos might reveal the  thermodynamical state of
the emission region,  and  detecting different
sources of neutrinos might give information about 
different regions of the Sun.

The regions of emission are  determined by the  nuclear reaction rates and
their temperature and density dependence.  For the central conditions in the Sun, the three major neutrino fluxes
can be expressed in the following  way \citep{gough88}:
$$ { ¬†\Phi _ {pp}} \propto { \rho_{C} X_{C}^{2}¬† T_{C}^{4}} ; \;\;
{ \Phi_ {^{7}Be} } \propto {{{ (1-X_{C})} \over {(1+X_{C})}} T_{C}^{11.5}} ; \;\; 
 {\Phi_ {^{8} B}} \propto {\rho_{C} {{(1-X_{C})} \over {(1+X_{C})}} T_{C} ^{24.5}} \;\eqno (3.3) 
 $$
where $T_C$ is the central temperature, $\rho_C$ is the central density, and $X_C$ is the central mass fraction of hydrogen.
It should be noticed that these statements are, for all practical purpose, independent
of the solar model and only reflect the behaviour of each
nuclear reaction.

 Figure 2 shows the regions of emission: the \B neutrinos
are produced in the very central part of the Sun    (the inner 10\%
in radius or 20 \% in mass) because of the extreme dependence of this 
nuclear reaction rate  on the temperature; the \Be neutrinos appear
in a wider domain, and the pp neutrinos  originate  from the whole
nuclear region. 
The neutrino flux from each source must be integrated over the
relevant region  of emission, which corresponds typically to a region
where the temperature decreases by 20\% for
\B  neutrinos  and more than 30\% for pp neutrinos.  Thus, the real
temperature  dependence of the neutrino flux, which varies with
the solar model, is less steep  than quoted in Eq. (3.3). Moreover a
feedback is always necessary to adjust the luminosity of the solar models at present age to the observed value. It turns out that, owing to
partial compensations  due to the variation of the composition and the
density in different models,  the dependence of the solar neutrino
flux on the {\it central} temperature is in fact:
$$  { ¬†\Phi _ {pp}} \propto ¬†\  T_{C}^{-1.2} ; \;\;
{ \Phi_ {^{7}Be} } \propto  \  T_{C}^{8} ;\;\;
 {\Phi_ {^{8} B}} \propto   \ T_{C} ^{18} \; \eqno (3.4) $$

These expressions were deduced by \cite{bahcall88}
from 1000 different computations of a solar model with different input
parameters covering a large range of values.

\begin{figure}
\begin{center}
\vspace{-6. cm}
\includegraphics[width=14cm]{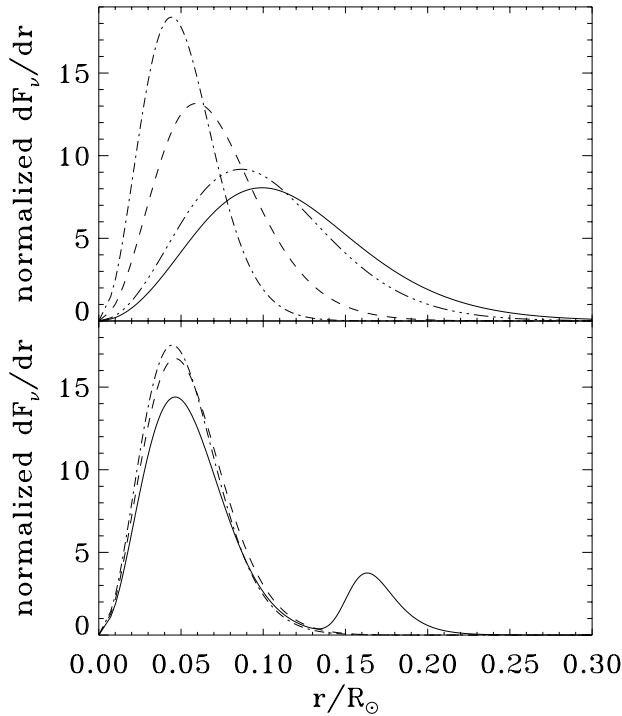}
\end{center}
\vspace{-1. cm}
 \small
{\caption{Spatial emission of the neutrino sources computed with the seismic
model. In the upper figure are drawn the p-p (plain curve), $^8$B (dot-dashed curve), $^7$Be (dashed curve), and the pep (dot-dot-dot-dashed curve) neutrinos. In the lower figure, the $^{13}$N (plain curve), $^{15}$O (dashed curve), and $^{17}$F (dot-dashed curve) neutrino production are shown. For each neutrino “type”, we have drawn (1/Ft) (dF/dr) where F is the flux in s$^{-1}$, r the fractional radius, and Ft the total flux for this neutrino type (integrated over the entire Sun).From \cite{couvidat03}.}}
 \normalsize
\end{figure}
\vspace {5. mm}

\section{The classical view of the Sun through the Standard Solar Model}
The time scale required for biological evolution on Earth is considerably greater than the Kelvin Helmholtz time 
of 30 Myr, corresponding to the estimated lifetime of the Sun if gravity alone compensates for the 
loss of luminosity at the surface: this fact was at the origin of public disagreements between Darwin and 
Lord Kelvin at the end of the 19th century. At the beginning of the 20th century, it became obvious 
that the Sun should be at least as old as the Earth. This is why Eddington suggested in 1920 that, 
perhaps, a subatomic source of energy may exist in stars. It then became evident that thermonuclear 
fusion had to play a fundamental role. Today, the origin of the Sun's stability is clearly 
established and it is reasonable to treat the long-term evolution of a star as a succession of static 
equilibrium models. 

\subsection{The fundamental equations}
The classical picture of the Sun (used to produce ``standard solar models'') was developed in the course of the 20th century with the following 
hypotheses: the star is spherical, described by a succession of hydrostatic equilibria, 
and without effects of rotation and magnetic field. We will show in 
sections 5 and 7 that we are currently expanding this picture to better describe the real Sun. 
The precise knowledge of the general characteristics like distance, mass, 
age, radius, and luminosity, gives the Sun a privileged position among the stars.
These characteristics are summarized in
Table 3. Their  
role is crucial in  solar modelling because they serve as
constraints: for a given model, a correct luminosity and radius must be reached at the correct age.
In this framework, the internal structure of the star is described by four structure equations 
which are the foundations of what is generally called the Standard Solar Model (SSM).

\small
\begin{table}
  \begin{center} 
    \caption{Evolution of the solar fundamental constants and their variability during the 11 year solar cycle. 
    The heavy element mass fractions have been estimated by \cite{anders89,grevesse92,asplund09}.}
    \begin{tabular}[h]{lcccc}
     \hline
     & Reference values Allen &  Present values &Time Variability  \\
 \hline
Luminosity  & 1360.488 ($\pm 2. 10^{-4}$)  &   1367.6 W/m$^2$ - 1361    W/m$^2$  &  1-4  W/m$^2$  \\
Age   & & $\rm 4.6 \pm 0.02\,\,Gyr$        & \\

Radius   &   695 990 km  &   693710 (min)   &     10-160 km \\  
Seismic radius &        -       &               695660 km             &          -            \\
Radius shape  &   -     &   oblateness 6 to 10 km    & 6-14 km  \\
Heavy element Z & 0.02 then 0.0173 &  0.0134  &  no evidence\\
Present mass loss & & $\rm {2 \times 10^{-14}} {M_\odot / yr} $ &   & \\    
  \hline
\end{tabular}
 \label{tab:fund}
    \end{center}
\end{table}
\normalsize 
  The 
first two equations assume hydrostatic equilibrium (each gas shell is balanced by the competition between the 
downward gravitational force and the outward pressure gradient) and mass conservation. P, T, and $\rho$, are respectively the pressure, temperature and density at the position r (radial distance from Sun center), and {\it M(r)} 
is the mass enclosed within a sphere of radius {\it r}: 
$$ \rm {dP \over dr} = -{{ M(r) G} \over{r^2}} \rho \;\; and \; \;{dM \over dr} = {4 \pi r^2 \rho} \;  \eqno (4.1; 4.2)$$ 
Thermal equilibrium is assumed. The 
energy produced by nuclear reactions (${ 4\pi r^2 \rho \epsilon }$, where $\epsilon$ is the nuclear energy production rate), balances the energy flux {\it 
L(r)} emerging from the sphere of radius {\it r}. It includes the energy loss by neutrinos (which is only about $0.4$ MeV for hydrogen burning, but larger for, say, a supernova explosion). 
Taking into account 
quasi-static gravitational readjustment and composition variation, a heat transfer 
term {\it TdS} is included, where {\it S} is the total entropy per gram of the gas and the energy conservation 
yields:$$ {dL \over dr} = {4 \pi r^2 \rho \left( {\epsilon - {T dS
\over  dt}} \right)} \;  \eqno (4.3)$$ 
The radial temperature gradient depends on the different  processes which contribute to the energy transport. In 
a radiative region of a star, the diffusion approximation is appropriate, and the relation between 
temperature gradient and luminosity is:  $$ {dT\over dr} ={
{{-3}\over {4 ac}} {{\kappa \rho }\over {T^{3}}} {L(r)\over{4 \pi r^2}}} \;\eqno (4.4 \rm a)$$ When the 
radiative opacity coefficient $\kappa$ increases too much, like in the Sun, or  when the energy production is very high (in the internal part of stars with mass $>$ 
$1.5 M_{\odot}$), the radiative gradient increases so much that matter becomes convectively unstable. The 
resulting temperature gradient is then nearly adiabatic: $$ {dT\over dr} = {\left({ dT
\over {dr}} \right)_{\rm ad}} ={{ \Gamma_{\rm 2}- 1 \over  {\Gamma_{\rm 2}}}{ T \over P} {dP \over dr}} 
\; \eqno (4.4 \rm b)$$  where ${\Gamma_{\rm 2}}$, and the other adiabatic exponents, are defined by: 
$$ {PV}^{{\Gamma}_{\rm 1} } = const \; ; 
 {P}^{1- \Gamma_{\rm 2}}T^{\Gamma_{\rm 2}} = const \; ;
 TV^{\Gamma_{\rm 3}-1} = const  \;  \eqno (4.5)$$
In  a monoatomic ideal gas, $\Gamma _{\rm 1}= \Gamma _{\rm 2} = \Gamma _{\rm 3} = 
{c_{P}\over{c_{V}}}$. In stars, these three quantities are not equal, especially because 
of the partial ionization of different elements. Moreover, these three quantities drop from 5/3 to 4/3 when 
a pure ideal gas is being replaced by pure radiation. The value $\Gamma_3 < 4/3$ is, for example, at the 
origin of pulsation of stars. $\Gamma_{\rm 1}$ determines the dynamic instability, while $\Gamma_{\rm 2}$  
governs the convective instability, and $\Gamma_{\rm3}$ governs the regime of pulsation instability. 

Boundary conditions (typically 
for pressure and  temperature at the stellar surface and for mass and luminosity at the stellar center) are 
included to solve these equations and to follow the solar  structure  evolution.

\subsubsection{The solar equation of state}
The total pressure P in a star is the sum of radiation pressure and gas pressure:
 $$P_{R} = {1 \over 3} {a T^{\rm 4}}  \; ; P_{G} = {N \over {V}} {kT} = {{\rho {kT}} \over \mu (r)} \eqno(4.6)  $$ where {\it a} is the radiation density 
constant  and {\it V} is the volume of the gas. The radiation pressure $P_{R}$ is negligible in the solar center (1/1000 of the gas pressure $P_{\rm 
G}$) but not near the  photosphere ($.6 P_{\rm G}$). When temperature is high and density low enough, 
interactions between the particles are negligible and the gas can be reasonably approximated by 
an ideal gas (equation 4.6). This approximation is quite good in the central part of the Sun for ions. However, the large range of solar temperatures ($5 \times  10^{\rm 3}$ to $ 10^{\rm 
7}$ K) and densities ($10^{\rm -12}$ to $ 10^{ 2}\; \rm {g/cm^3}$) requires a more detailed description of 
the equation of state \citep{rogers96,rogers02}. 
Since the star is constituted 
of a mixture of different chemical species, gas pressure and mean molecular weight $\mu$ are written as:
  $$P= \sum_{i} {{ {n_{\rm i}} \over {m_{\rm i}}} \rho kT} \;\;{\mu ^{-1}} = \sum_{i} { {n_{\rm i}} 
{(m_{\rm H}/ {m_{\rm i}})}} \;  \eqno (4.7)$$
 In these expressions, $n_{\rm i}$ is the number of free particles: Z electrons + 1 nucleus for each 
atom of atomic number Z.  More generally, this number depends on the degree of ionization of the 
species i considered. In astrophysics, X, Y, Z represent the mass fraction numbers (normalized to 1) of, respectively,  
hydrogen, helium and other elements (heavier than helium). For a completely ionized gas, 
$\mu$ is expressed as $\mu^{-1} = 2 X + (3/4) Y+ 0.5Z$  where it is assumed that the proton number 
equals the neutron number for heavy elements. The solar value $\mu$ increases outward due to the 
partial ionization of elements and the existence of molecules. As shown by their high ionization potential, He and Ne are 
difficult to ionize. 
Central solar density is sufficiently high to cause partial degeneracy of the electrons. Therefore, electron 
number $n_{\rm e}$ and electron  pressure $P_{\rm e}$ must be expressed as functions of the degeneracy parameter $\eta$: 
$$P_{\rm e} = {{8\pi kT} \over {3 h^{3}}} {{(2mkT)}^{\rm {3/2}}}  F_{\rm {3/2}}(-\eta)\; ; n_{\rm e} = {{4 \pi} \over h^{\rm 3}} {(2mkT)^{\rm {3/2}} F_{\rm {1/2}} 
(- \eta)}  \;  \eqno (4.8)$$ where $F_{1/2}$ and $F_{3/2}$ are the Fermi-Dirac functions. In the case 
of  the Sun, the effect of degeneracy is small ($\eta$ of the order of --1) but, if the star continues to evolve, it reaches a point where this effect becomes dominant. This leads to an increase in the  pressure and a quasi independence 
of the pressure on the temperature.  

\subsubsection{The thermonuclear source of energy}
The calculation of the nuclear energy rate $\epsilon$ requires a good knowledge of the reaction rates that take place in the solar interior.  \cite{gamow28}  
was the first to show that hydrogen is the most abundant element in stars, and is  the 
first element to be converted because it has the lowest Coulomb barrier (energy barrier that two nuclei need to overcome).  \cite{vonweizsacker38}
and \cite{bethe38}
clearly  showed that two different sets of reactions, the 
pp chain and the CNO cycles,  could provide the amount of energy explaining  the present solar 
luminosity. 
In the solar case, the pp
chain is  the most energetic one ($\epsilon \propto T^4$), but when temperature  increases (at the end of hydrogen burning), the CNO cycles 
 become the most efficient ($\rm \epsilon \propto T^8 \; or \; T^{12} $).
This is why low mass stars ($< 1.5 M_{\odot}$), driven by pp  reactions, have a 
slower evolution than more massive stars, driven by CNO cycles.

A typical nuclear reaction between two species {\it a} and {\it X} can be
described by: $${a+X} \rightarrow  {b+Y}\;  $$
The associated gain or loss of energy Q is given by the energy balance $Q={\Delta E}= {(M_{\rm X}+M_{ 
a}-M_{\rm Y}-M_{ b})c^2}$. If $\sigma (v)$ denotes the probability that a projectile  {\it a} 
collides with a fixed nucleus  {\it X} with relative speed $v$, the total number
of reactions per $\rm cm^3$ and per second is given by:
  $  r = {N_{ a} N_{\rm X} v  \sigma (v) } $
where $N_{ a}$ and $N_{\rm X}$ are the number densities of the species {\it a} and {\it X}.  Since 
matter in the stellar interior is in local thermodynamical equilibrium, the relative  velocities are 
given by a Maxwellian distribution f({\it v}), and the total reaction rate is  expressed
as: $$r= {N_{ a} N_{\rm X} \int  {v \sigma (v)f(v)dv}= N_{ a} N_{\rm X}   <{ \sigma v}> } \;  \eqno (4.9)$$ 
The total energy generated per unit
mass and per unit time is: $ \epsilon = {rQ \over {\rho }}$
where $\rho$ is the gas density. At typical stellar temperatures, the average  thermal energy of a 
particle (considering a Maxwell-Boltzmann distribution of velocities) is  several keV, which is about $10^{-3}$ times 
smaller than the Coulomb barrier ($\sim$ 1 MeV).
Therefore, from a classical point of view, no reaction could take place, and the pp interaction could only occur when 
the  internal temperature reached $10^{10}$ K (corresponding to the
Coulomb barrier). However, at that point, the ensuing nuclear reactions would be so strong  that the star would 
experience a catastrophic explosion. Of course,  nuclear reactions could also occur in the high energy tail 
of the  Maxwell-Boltzmann distribution, but the number of participating particles would be very small. 
This problem was solved by \cite{gamow28}
when he showed that, in quantum
mechanics, there is a  small but finite probability (1.7$ \times 10^{-4}$ at 5 keV) that the proton  
penetrates the Coulomb barrier. This tunneling probability is proportional to: 
$${\exp  \bigl[ {-2 \pi {\rm a}Z_{\rm 1} Z_{\rm 2} ({mc^2 \over 2E})^{1/2}}
\bigr] }= \exp{ \bigl ({-b \over{E^{1/2}}} \bigr) } \eqno (4.10)$$
\noindent 
where $Z_{\rm i}$ is the charge of the species i, and $\rm{a=e^2/hc}$.  The generation of energy is  expressed as:
$$ \epsilon = {N_{\rm a} N_{\rm X} Q \over {\rho}} \bigl( {8 \over {m \pi}
 } \bigr) ^{1/2}{ (
kT)^{-3/2}}{ \int_{0}^{\infty} { S(E) {\exp {({-E \over {kT}}-{ b \over {
\sqrt {E}}})}} dE}}   \eqno (4.11)$$
where S(E) gives the smooth variation of the cross section with energy $E$. In fact, the cross section for 
charged particle-induced  nuclear reaction drops rapidly for energies below
the Coulomb barrier and extrapolation is often needed to determine the factor 
$  \sigma (E)= {1 \over E} \exp {(-2 \pi \eta) S(E)} $.
The competition between these two trends, decrease of the Maxwellian distribution with $E$ and the increase of the 
penetration effect, leads to an increase in the cross section in a rather 
small range of energies in the vicinity of $E_{\rm o}= (bkT/2)^{2/3}$, known as the Gamow peak.
 $E_{\rm o}$ increases with the charge of the particle and is typically between 10 to 
50 keV for  hydrogen burning at a central temperature of 15 million K
(1.3 keV). Nevertheless, to calculate the production of  energy of (4.14), all coefficients S(E) must be
determined from  laboratory measurements for the reactions involved. In the case of 
non-resonant reaction rates in a non-degenerate, non-relativistic gas, the reaction rate r  
may be expressed as: $$ r = {{N_{a} N_{X}} \over {(1+ \delta_{12})}} ({2 \over
 M})^{1/2} {{\Delta E_{0}}\over{{(kT)}^{3/2}}} {S_{eff} \exp {\left [{{-3
E_{0}}\over {kT}}-({T \over T_{0}})^2 \right ]}}     \eqno (4.12)$$ where
$\Delta E_{0}$ is the full width at 1/e of the maximum value of the Gamow peak. The 
effective value $S_{eff}$ of the cross section factor, S(E), is given by:
\small
 $$  S_{eff}= S(0) {\left[ 1 + {{5
kT} \over {36 E_{0}}} + {S'(0) \over S(0)} {\left (E_{0} + {35\over 36} kT
\right )} + {1 \over2} {S''(0) \over S (0)} {\left ({E_{0}^{2}} + {{89\over
36} E_{0} kT }\right )} \right]} \rm{MeV barn} \eqno (4.13).$$ 
\normalsize
The extrapolations of $S(0), S'(0), S''(0)$ from laboratory measurements were performed by
\cite{fowler67},(1975),(1983) 
for many thermonuclear reactions.
They expressed the product $N_{A} < \sigma v>$ as: $$
N_{A} < \sigma v> = C_{1} T_{9}^{-2/3} \exp [C_{2}
T_{9}^{-1/3} -{(T_{9}/T_{0})}^{2}]   \ \ \eqno(4.14) $$
$$ \ \ \ \ \ \ \ {(1+ C_{3} T_{9}^{1/3} + C_{4}
T_{9}^{2/3} +C_{5}T_{9} + C_{6} T_{9}^{4/3} + C_{7} T_{9}^{5/3} )} \;
\rm{cm^{3} sec^{-1} (mole)^{-1}}$$ where all the coefficients $C_{i}$ depend
on $Z_{a}$, $Z_{X}$, and $S(0)$ or $S'(0)$ or $S''(0)$ and $T_{9}= T/
10^{9}$.
Most of the cross sections  relevant to the solar interior were measured
and extrapolated in the twentieth century. Their recommended values have been commonly defined in \cite{adelberger98}.
A few specific cross-sections have  recently been improved (see section 4.3).

\subsection{ Energy transport}
Two processes dominate the transport of energy in a solar-type star: radiation and convection. 
In the Sun, conductivity plays no role whatsoever (it has an effect in higher-density  stars and in 
supernova cores) and the energy transported by the neutrinos is very small. In the central 
part, radiation is the most efficient transport process.

\subsubsection{Radiative transport}

In the solar interior, the average energy per photon is in  the keV range, which is characteristic of X 
rays, whereas the light escaping from the surface is dominated by the visible spectrum, corresponding to an energy  
$10^{\rm 5}$ times lower. The cause of this sharp decrease in photon  energy is the coupling 
between radiation and matter. The effect of the interaction of photons with electrons, atoms, ions and  
molecules must be evaluated and is summarized by what is called the ``Rosseland mean opacity''  
$\kappa$, which is a function of temperature, density, and composition. The equation of radiative 
transfer adapted  to stellar interiors is given by equation (4.4a).
The radiation field is nearly isotropic 
and Planckian. 
The opacity of the gas really drives the rate at which energy escapes,  and plays a crucial role in 
determining the age of a star and its central temperature. The sources of opacity are the 
following: (1) electron scattering (Thompson and Compton scattering): scattering of the photon on free electrons;
(2) bound-bound transitions: the photon triggers a change of energy level of a bound electron; (3) bound-free (photoionization) transitions: a bound electron becomes free after interacting with the 
photon (or the opposite); (4) free-free transitions (inverse bremsstrahlung): a free electron stays free after the interaction with 
the photon; and (5) molecular transitions at the photospheric level. 

The computation of all these processes necessitates detailed calculations of  complex
atomic physics. Frequency-dependent opacities are needed for detailed stellar models:
$\kappa _{tot}(\nu ) = \sum _{i} \kappa _{i}(\nu)$, where i represents the 
different processes. It is noticeable that the Rosseland mean opacity $\kappa$ is not obtained by 
averaging the frequency-dependent opacities, but by averaging their reciprocals, which are the radiative 
conductivities (the weight function $f(\nu)$ is the temperature derivative of the Planck
function): $${1 \over {\kappa}} =  \sum _{i} \left[ \int {f(\nu) \over{\kappa_i(\nu)}} d\nu 
\right] \;  \eqno (4.15)$$
 The major difficulty with such calculations arises from the fact that a single source of opacity may 
exceed all the others. For example the bound-bound processes may introduce cross sections
which are 1 or 2 orders of  magnitude greater than the competing contributions.
 
\begin{figure}
\begin{center}
\includegraphics[width=9cm]{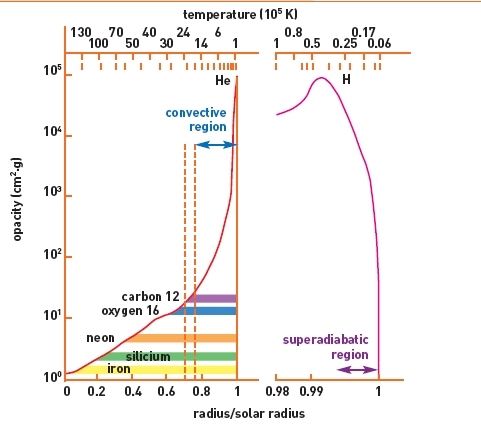}
\end{center}
\caption{
Total solar opacity coefficient as a function of the radial distance from Sun center. This figure
shows the impact of the different elements, mainly through bound processes,
 on the increase of the opacity. The most important elements are clearly
visible : C, N, O (at the limit of convective instability) and at low temperature
He and H. The appearance of neutral atoms or even molecules
produces the decrease of the opacity near the surface.  From \cite{turck93}.}
 \normalsize
\end{figure}

Therefore, even a trace element may produce a non negligible  contribution to the Rosseland mean opacity. 
 In the case of the Sun 
or solar like stars, the detailed knowledge of radiative processes is crucial. Indeed, 
the contribution of heavy elements (everything heavier than helium, representing less than $2\%$ in mass fraction and  $0.14\%$ in  number fraction) to the 
equation of state does not exceed 1 or $2\%$, but their contribution to the opacity is about $30\%$ in 
the solar central region, and more than $70\%$ in the intermediate region. 
This is why the iron contribution
 reaches 20 to 30 $\%$ of the total opacity (see \cite{turck93,turck09b}
 for more details). In the latter reference, we discuss how these estimates can be evaluated by measuring the opacities, in some specific cases, with high-energy lasers. All the radiation-matter interaction processes must be calculated for 
all the chemical elements. Moreover, when peaks appear near the maximum of the 
Rosseland weighting scheme, the bound processes play an even more important role. At low energies, more levels contribute to the opacities, and the number of peaks greatly 
increases. Clearly, since bound-bound and free-bound processes vary as $Z^4$, instead of $Z^2$ for free processes,
the heavy elements play an important role far exceeding what could be expected from their low abundances.
Moreover, due to the non-additive character of the opacity coefficients,
the spectrum must be calculated for the correct chemical element mixture before deriving the
Rosseland mean value. This leads to extensive calculations which must be
done carefully for different mixtures. Figure 3 illustrates the specific contribution of different elements when they become partially ionized
and when bound processes start to significantly impact the opacities: first (in term of distance from Sun center) Fe, then Ne, Si, O, N, C... When oxygen
(the most abundant heavy element) becomes  partially
recombined, the temperature gradient greatly increases. Further away from Sun center,
nitrogen and carbon exhibit a similar behaviour and  
the resulting increase in $\kappa$ becomes so significant that the radiative
temperature gradient exceeds the adiabatic one:
 $$ {\left| dT \over {dr} \right|}_{\rm rad} > {\left| dT \over {dr} \right|}_ 
{\rm ad}, \;  \eqno (4.16) $$
This is the criterion for the onset of convective instability, refered to as
the Schwarzschild  criterion. Therefore in the Sun, there is a large convective solar
region below the photosphere. In this region conditions are satisfied for first (in term of radial distance from solar center) oxygen, then nitrogen, carbon, and finally helium and hydrogen, to significantly impact the opacities through
bound-bound contributions. However, closer to the solar surface, at about $0.995 R_{\odot}$, hydrogen and helium
become neutral, and the opacity decreases, prompting the radiative flux to become significant again: we enter in a superadiabatic regime.

\subsubsection{Convective transport}

In any case, the energy produced at Sun center must escape from the star. When the opacity
increases too much, a very steep temperature gradient is required to
maintain the energy flow, leading to an unstable situation. When
the instability criterion is fulfilled, the real temperature  gradient
exceeds the adiabatic gradient and the difference is called the
superadiabatic  gradient: $${\Delta  dT \over dr} \equiv {{dT \over
{dr}} -{\bigl({ dT \over {dr}}
 \bigr)_{\rm ad}}} \;\;
\mathrm{with} \;\; {\bigl({ dT \over {dr}} \bigr)_{\rm ad}} ={{ \Gamma_{\rm 2}- 1 \over 
{\Gamma_{\rm 2}}}{ T \over P} {dP \over dr}}  \eqno (4.17)$$
Convection is a very efficient means of energy transport.
In fact, near the base of the convective envelope, the
radiative flux is so small that we can ignore the superadiabatic gradient
and use the adiabatic one as the temperature gradient (this is also the case
in the central part of massive stars). Unfortunately, this is not the case
in the outer layers of the Sun. In 1D solar models, the convection is usually treated by the ``mixing length approximation'' \citep{vitense53,bohm58}
which replaces  the real situation of ``plumes'',
convective eddies of different  sizes, etc... with an average situation where each
convective element travels a distance $\Lambda$, before
mixing with the surrounding matter. This distance is generally scaled to
the pressure scale height \footnote{The pressure scale height is the
distance corresponding to a decrease in pressure by a factor $e=2.718$.} by a
parameter $\alpha$ called the mixing length parameter: $$ \Lambda
= {\alpha \lambda} \equiv {\alpha {\bigl({ dlnP \over {dr}}  \bigr)}^{\rm
-1}}  \; \eqno (4.18)$$ The relative
amount of energy  carried by convection and by radiation must be estimated. 
The temperature of a convective element and the average temperature of the
surrounding matter are very different in the central and superficial solar regions. In the
central region, the temperature gradient is about $10^{-4}
\rm{K/cm}$  with a mean temperature of about $10^7 \ K$ and a
convective time of about 1 to 100 days. Therefore if a convective instability
exists in the core (for M$\ge$ 1.5 $M_{\odot}$), it is an extremely
efficient  process of energy transport and the temperature gradient can be approximated by the adiabatic one.  In the outer layers, the
temperature variation between shells is of the order of  100 to 1000 K (to be
compared with a gas temperature of 5000 to 10000 K), the convective time is
much shorter (1 to 100 minutes), and we really need a specific treatment of the 
convection. In that case, due to  partial ionization of hydrogen or helium,
the second adiabatic exponent $\Gamma_{2}$ drops to a value close to one and the adiabatic gradient
becomes relatively small. This favors convective transfer but the
radiative flux remains important and cannot be neglected.
Convective models have been greatly improved by the extension of the mixing length approach: some models take into account different Kolmogorov energy cascades \citep{canuto96}.
Moreover, 3D simulations of such
convective effects are now at a very advanced stage and reproduce realistically the surface and atmospheric turbulence \citep{nordlund09}.

The value of the
mixing length parameter is still calibrated
in solar models, so that these models
match the solar radius at the solar age. Therefore, in this specific
case the small superadiabatic region
just below the photosphere can be considered as correctly treated. The use of
this solar value for other stars  with rather different surface temperatures
was considered  more uncertain due to the fact that the mixing length parameter is largely dependent
on the opacity in the superadiabatic part of the convection zone.
It is now possible to test this approach by comparing precise radius measurements of red giants to models using different theories of convection (Piau et al. 2010). Significant progress 
 will also be achieved with stellar seismology (especially with the COROT and KEPLER missions), which allows the derivation of the convective 
depth for various solar-like stars \citep{ballot04a}.

\subsection{Improvements in the physics of the SSM. Evolution of the $^8B$ neutrino flux} 
Radial temperature, density, pressure and
 composition of the present Sun (and consequently the emitted neutrino
 fluxes) are obtained by solving the equations (4.1) to (4.4), at each time step, as a succession of hydrostatic equilibria.
The SSM is the most economical
 way of describing the Sun: it assumes that there is no important effect of rotation and magnetic field.
 In this framework, only 3 observed variables are needed: the luminosity of the Sun
 at an age of 4.55 Gyr (from the onset of hydrogen burning, or 4.6 Gyr since the solar formation), 
the solar radius,
 and the detailed element composition for nuclei greater than helium (determined at photospheric level 
 and compared to
 meteoritic compositions). The only free variable is the initial helium content 
 which is adjusted in order to reproduce these observations at the present time. 
Solving the equations (4.1) to (4.4) requires the knowledge of the physical inputs previously described, to derive $\rm P(\rho, T, X_i)(r),
\epsilon(\rho, T, X_i)(r) \, and \,
\kappa(\rho, T, X_i)(r)$ at each time step. Knowledge of these physical ingredients has been significantly improved during the last two decades.

Furthermore, several of these improvements have been accomplished since 1988, the year when the European Space Agency decided to favor helioseismology from space. Table 4 summarizes the physical processes 
introduced in the solar models which can be estimated thanks to one or two observable quantities, and the location in the Sun where a specific element reaches its peak impact.

\begin{table*}
\small
\caption{\label{Table 4}  Location in the Sun where a specific element can be used to shed light on a physical process
thanks to a specific observable quantity.}     			
\begin{tabular}{p{1.cm}*{5}{c}}
\hline
radius R$_{\odot}$ & element & physical process &  observables  & reference \\
\hline
0.98 & $\rm ^4He$   &microscopic diffusion & $\rm c^2$, $\Gamma_1$& \cite{vorontsov91} \\
0.71		&	$\rm ^{16}O$ &	transition radiation/convection &	$\rm c^2$, $\kappa$         	& \cite{christensen91}	\\				
0.70		&	$\rm ^7Li$		  &nuclear process, turbulence		& 	$c^2$, rotation	& \cite{brun99} \\	
		&			  & role of magnetic field	&  $\rm c^2$, rotation &  not solved \\ 
		0.57 		&	\rm $^9Be$		&  nuclear process, turbulence &	$c^2$			&  \cite{brun99} \\
0.25		&	$\rm ^3He$		&   presence of mixing ? (no) &	$\rm c^2, \rho $                        & \cite{turck01a}  \\
0.05		&	$\rm ^7Be$	&  presence of mixing ?	(no)		& \rm	$c^2$, neutrinos		&  \cite{turck01a}  \\
0.- 0.1		&	$\rm ^{56}Fe$	&  central temperature &	$\rm c^2$, neutrinos                &  \cite{turck01a} ;(2004)  \\
\hline
\end{tabular}
\normalsize
\end{table*}

\subsubsection {The detailed solar composition: Xi(t,r)}

The detailed composition of the plasma is a key ingredient of the SSM. It enters in the calculation of radial pressure, density, and opacity coefficients, as previously shown.
 It determines the duration of the proton burning phase which is largely influenced by
 the transparency of the star. The knowledge of the composition is required at various stages of the 
theoretical approach. The elements from hydrogen to oxygen determine the evolution 
of nuclear burning; the initial composition also influences the mean
molecular weight, which plays a crucial role in the determination of the
pressure.
Consequently, 
the \B neutrino flux is sensitive to the metal composition, as shown by \cite{bahcall68}.
The solar abundances were poorly known at the time \cite{bahcall68} was written, 
explaining why the SSM was calculated with different hypotheses
leading to a difference by a factor of two in the \B neutrino flux predictions. 

To determine the initial composition of the Sun,
three sources are used: the Earth, the solar photosphere, and the 
meteorites. Each source raises some specific issues. The Earth has lost
a large fraction of its volatile elements and there is a chemical fractionation
of the various elements in the different terrestrial layers. On the other hand, the Earth is very
useful in most cases to determine the {\it isotopic ratios} which are
not influenced by the chemical fractionation and can be very accurately determined.
Two other sources of 
information are currently favored: the direct spectroscopic
observation of the solar photosphere  and the chemical  analysis of C1
carbonaceous chondrite meteorites. Both are assumed to characterize  the
initial composition of the protosolar nebula. Therefore their agreement, or lack thereof, is extremely interesting. They inform us on the thermal and chemical
history of the early phases of the formation of the solar system: precise 
determination of the possibly biased meteoritic abundances has to be compared
with the reliable but often delicate photospheric abundances that depend on the knowledge of atomic
physics of the solar atmosphere. The
convective  zone does not  penetrate deeply enough to alter the surface composition, which is why the  
photospheric abundances have been considered for a long time to be the initial abundances of 
the Sun for most of the elements. In reality, a slow microscopic diffusion
leads to a decrease with time in heavy element abundances, relative to hydrogen at the surface, of typically 10\% in $4.6$ Gyr \citep{michaud93}.
This constraint on the solar models is added to obtain the present
photospheric abundances at the age of the Sun. 
Moreover, some
elements  are burned at very low temperature, such as deuterium, lithium,
and beryllium. For these elements, the photospheric values are affected by the Sun's
evolution. 

After years of studies and precise measurements of
atomic oscillator strengths \citep{grevesse92},
the  photospheric
determination of element abundances is very much in agreement with the meteoritic one, especially
for refractory elements (Table 5). This agreement is often better than 5 \% even if 
noticeable differences still persist on elements like Cl, Mn, Fe, Ga, and Ge. 
Following that compilation, iron received specific attention because its photospheric abundance 
 $\rm Fe_{high}= Fe/H=4.68 \pm 0.33 \   10^{-5}$
was 30\% greater than the meteoretic one. However,
\cite{holweger90}
based on ionized Fe (95$\%$) derived an
abundance much closer to the meteoritic value: $\rm Fe_{low}= Fe/H=3.24 \pm
0.075 \  10^{-5}$. The consequences of this improvement on solar neutrino fluxes
 have been discussed in \cite{turcklopes93}
 (see their table 4). 

\begin{table*}
\small
\caption{Abundances of the elements used in a solar model derived from the solar 
Photosphere (from \cite{anders89}:
AG,  from \cite{asplund05}:
AGS and from \cite{asplund09}
and compared to meteoritic values from \cite{lodders03}. They are given in fraction number normalised
such that log N$_ H$ = 12. The main changes are a strong reduction for the C, N, and O abundances (see text). Indirect solar estimates are mentioned in parentheses.}
\begin{tabular}{p{2.5cm}*{4}{c}}
\hline
\hline
Elements & Photosphere AG & Meteorites  & Photosphere AGS & Asplund et al. (2009)\\
\hline
$^1H$ & 12.00 & $8.25 \pm 0.05$  &   12.00 & 12.00 \\
$^2He$& $[10.99 \pm 0.035]$ & 1.2900              & $[10.93\pm 0.01]$&$[10.93\pm 0.01]$ \\
$^7Li$& $1.16 \pm 0.1$      &      $3.25\pm 0.06$ &$1.05\pm 0.1$ & $1.05\pm 0.1$\\
$^4Be$&$1.15 \pm 0.1$&$1.38\pm 0.08$& $1.38\pm 0.09$ & $1.38\pm 0.09$\\
$^6C$&$8.56 \pm 0.04$&$7.40\pm 0.06$& $8.39\pm 0.05$ & $8.43\pm 0.05$\\
$^7N$&$8.05 \pm 0.04$&$6.25\pm 0.07$& $7.78\pm 0.06$ & $7.83\pm 0.05$\\
$^8O$&$8.93 \pm 0.035$&$1.38\pm 0.08$&  $8.66\pm 0.09$ & $ 8.69 \pm 0.05$\\
$^{10}Ne$&$[8.09 \pm 0.1]$& -           & $[7.84 \pm 0.06]$ & $[7.93 \pm 0.1]$\\
$^{11}Na$&$6.33 \pm 0.03 $& $6.27 \pm 0.03$  & $6.17  \pm 0.04 $ &  $6.24  \pm 0.04 $\\
$^{12}Mg$&$7.58 \pm 0.05 $& $7.53 \pm 0.03 $ & $7.53  \pm 0.09 $ & $7.60  \pm 0.09 $ \\
$^{13}Al$&$6.47 \pm 0.07 $& $6.43 \pm 0.02$  & $6.37  \pm 0.06 $ & $6.45  \pm 0.06 $ \\
$^{14}Si$&$7.55 \pm 0.05 $& $7.51 \pm 0.02$  & $7.51  \pm 0.04 $ &  $7.51  \pm 0.03 $\\
$^{15}P $&$5.45 \pm 0.04 $& $5.40 \pm 0.04$  & $5.36  \pm 0.04 $ & $5.41  \pm 0.03 $\\
$^{16}S $&$7.21 \pm 0.06 $& $7.16 \pm 0.04$  & $7.14  \pm 0.05 $  & $7.12  \pm 0.05 $ \\
$^{17}Cl$&$5.5  \pm 0.3  $& $5.23 \pm 0.06$  & $5.50  \pm 0.3  $ &$5.50  \pm 0.3$ \\
$^{18}Ar$&$[6.56 \pm 0.1] $& -     & $[6.18\pm 0.08]$   & $[6.40\pm 0.08]$ \\
$^{19}K $ &$5.12 \pm 0.13 $& $5.06 \pm 0.05$  & $5.08  \pm 0.07 $  & $5.03  \pm 0.07 $\\
$^{20}Ca$&$6.36 \pm 0.02 $& $6.29 \pm 0.03$  & $6.31  \pm 0.04 $ & $6.34  \pm 0.04 $  \\
$^{22}Ti$&$4.99 \pm 0.02 $& $4.89 \pm 0.03$  & $4.90  \pm 0.06 $ & $4.95  \pm 0.06 $\\
$^{24}Cr$&$5.67 \pm 0.03 $& $5.63 \pm 0.05$  & $5.64  \pm 0.10 $ & $5.64  \pm 0.10 $ \\
$^{25}Mn$&$5.39 \pm 0.03 $& $5.47 \pm 0.03$  & $5.39  \pm 0.03 $  & $5.43  \pm 0.03 $\\
$^{26}Fe$&$7.67 \pm 0.03 $& $7.45 \pm 0.03$  & $7.45  \pm 0.05 $  & $7.50  \pm 0.04 $  \\
$^{28}Ni$&$6.25 \pm 0.04 $& $6.19 \pm 0.03$  & $6.23  \pm 0.04 $ & $6.22  \pm 0.04 $\\
\hline
\end{tabular}
\normalsize 
\end{table*}

The
lack of information on H, C, N, and O in  meteorites (because these elements do not enter in
the composition of condensable solids)  raises the  problem of the normalization
of the meteoritic abundances relative to hydrogen. This normalisation is in fact performed
using meteoritic and photospheric determination of the
refractory elements (Mg, Al, Si). For volatile elements the deviation is
smaller than 20 \%. Therefore it is hoped that, in the case of very heavy elements
(A$>$ 56) for which photospheric data are not  reliable, C1 meteoritic
abundances are representative of the solar  properties. These elements 
are not used in the calculation of the SSM or of seismic solar models. Therefore, we do not list them in table 5
but their fraction  number appears in \cite{asplund09}.
It is estimated that their influence is very small, but their impact on the opacity coefficients must be verified. 

The C, N, and O elements
are the main contributors to the ``heavy elements'' category. These major
volatile elements can only be studied in the solar photosphere. Their
abundances relative to hydrogen are extracted from molecules like CO, CH, OH and
NH, or from neutral lines.  Their abundance determination requires the
knowledge of accurate oscillator strengths. Several revisions were published in the last decade
\citep{grevesse90,grevesse91,biemont91},
and a recent decrease by 30\% in the abundances of these elements has been
advocated by several authors \citep{holweger01,asplund04}: such a decrease results, among others, from the use of better spectral lines. As a consequence of these successive updates in the heavy element abundances,
the solar Z value decreased from  $Z_{\odot}$ = 0.02 to the present value of 
$Z_{\odot}$= 0.013, in 20 years. Many papers focused on this update \citep{asplund05, caffau08,caffau09}, and the decrease is summarized in the last column of table 5 \citep{asplund09}. The consequences of these updates on the \B neutrino flux prediction is shown in table 6, as well as on the  sound speed in the radiative zone \citep{turck04a,guzik05},  rapidly followed by \cite{bahcall05}. This update also started a discussion on the ability to check solar composition by helioseismology \citep{basu04}. Today this promising idea remains difficult to implement, due to the variation between the existing equation of state (for the external layers) and an insufficient radial accuracy on the sound speed in the radiative zone. The detection of gravity modes may lead to some progress and to a better constraint on the opacity coefficients. Still, the most promising way to determine the inner composition remains the use of both helioseismology and the CNO neutrino detection. New MHD calculations put in evidence the role of the magnetic field on the broadening of the lines and could lead to a slight  increase of the C,N, O, Fe photospheric lines \citep{fabbian10}
 
\begin{table*}
\begin{center}
  \caption{\label{Table 6}  Evolution with time of the SSM or seismic predictions of the $^8B$ neutrino flux  in $10^6cm^{-2}s^{-1}$. Added are
  the central temperature Tc in $10^6K$, the initial helium abundance Y in mass fraction, and a specific problem that was solved. From \cite{turck10a}. }
    \vspace{5mm}   			
\begin{tabular}{p{2.5cm}*{5}{c}}
\hline
\hline
 $^8B$  flux  & Tc &  Y initial & problem solved & reference\\
3.8 $\pm$ 1.1& 15.6  &  0.276 & CNO opacity, $\rm ^7Be(p,\gamma)$ & TC88\\
4.4 $\pm$ 1.1 &	15.43 &	0. 271        &	-30\% Fe abundance, screening& TCL93	\\				
4.82		&	15.67	& 0.273	& microscopic diffusion		&	BTCM98\\
4.82   & 15.71   & 0.272   & turbulence tachocline  & BTCZ99\\
4.98 $\pm$ 0.73		&	15.74	&	0.276		  & seimic model	&    TC01b \\	
5.07 $\pm$ 0.76 		&	15.75 	&	0.277	&  seismic model, magnetic field& Cou2003 \\
3.98 $\pm$ 1.1	&  15.54    &  0.262   &   -30 \% CNO composition  & TC2004 \\
5.31 $\pm$ 0.6	&   15.75   &  0.277  &   seismic model+ $\rm ^7Be \; and  \; ^{14} N(p,\gamma)$ & TC2004 \\
4.21   $\pm$ 1.2 &  15.51 & 0.262  & SSM Asplund 2009 & TC2010\\
\hline
\end{tabular}
\end{center}
\end{table*}
\subsubsection{Solar composition and our Galaxy evolution}
An important result of solar
modelling is the determination of the pre-solar helium
abundance, derived by forcing the solar model to reach the present luminosity at the present age. This value is also reported in table 4 and is compared to nearby HII
interstellar medium (interstellar gas of temperature around $10^{4} K$  where
hydrogen is ionised) and to the composition of hot stars to determine whether the
Sun is typical of our neighbourhood or peculiarly rich in heavy
elements \citep{peimbert93}. 
Helium abundance cannot be directly measured in the photosphere due to its low temperature, 
but seismic derivation of the $\Gamma_1$ profile in the region  where helium is partially ionized
allowed a determination of its abundance immediately below the photosphere. The photospheric value of 
Y = $0.249$ obtained by \cite{vorontsov91} 
 puts a strong constraint on the gravitational diffusion of elements during the life of the star (see below).
This phenomenon of diffusion explains the difference between the helioseismic value and the initial value derived from a solar model (see its evolution in table 6). 
The solar model compatible with seismic data favors an initial Y = $0.277$. This value, combined with the recent O/X determination, suggests that the Sun was not born in an enriched environment resulting from a supernova explosion \citep{holweger01,asplund04}, contrary to what was previously thought. It is also the conclusion of \cite{peimbert07}. A new galactic helium abundance law can be derived:  
He/H= 0.075 + 44.6 O/H in fraction number. This law is now compatible with
all the indicators \citep{turck04a} called TC2004. 
This conclusion is also supported by other studies on radionucleides
\citep{gounelle01,gounelle03,gounelle08}. 

\subsubsection{The interaction between photons and matter $\kappa(X_i, \rho, T)$:}

Photons escape from the nuclear region and interact with matter through the phenomena described in section 4.2. The Rosseland opacity coefficient,
 $\kappa(X_i, \rho, T)$, has a strong influence on the central temperature. The opacity calculations require an accurate determination of the composition,
especially the helium, carbon, oxygen, and iron composition, and a detailed calculation of
the photon-matter interaction, obtained from a thorough knowledge in atomic physics. 
Astrophysicists commonly use tabulated values obtained for specific compositions. 
Most of the solar models use, for temperatures larger than 6000 K,
the opacity compilation of \cite{iglesias96} who considered 21 elements: H, He, C, N, O, Ne, Na, Mg, Al, Si, P, S, Cl, Ar, K, Ca, Ti, Cr, Mn, Fe, and Ni.
At lower temperatures, the compilation by \cite{alexander94} is preferred. Recent opacity tables were recalculated based on the new abundances (see http://webs.wichita.edu/physics/opacity/).

It is noticeable that the successive composition updates (Fe, C, N, and O) have systematically deteriorated the agreement between SSM prediction of the sound speed in the radiative zone and the "observed" (helioseismic) sound speed  \citep{turcklopes93,turck04a}. It is the case also for the location of the base of the convective zone and for the photospheric helium.  Three directions of investigation appear today: the magnetic effect on the photospheric lines, the detailed understanding of the photon interaction, some effort is under way  to validate these complex calculations through laboratory experiments \citep{bailey07,bailey09,loisel09,turck09b,turck10c}, the third is to go beyond the SSM (section 7).

\subsubsection{Temporal evolution of the composition: $X_i(\rho, T)$ }

Heavy elements diffuse toward the center during the solar life. This microscopic diffusion, mainly gravitational settling (GS) in the solar case, is a very slow process \citep{proffitt91}. However, it is a crucial phenomenon that must be introduced in the equation describing the temporal evolution of the composition: 
 $$ 
\frac{\partial X_i}{\partial t}=  {\frac{\partial X_i}{\partial t}_{nucl}}-\frac{\partial 
\left[(4\pi\rho r^2 (D_i+D_T)\frac{\partial X_i}{\partial m}-v_i X_i) \right]}{\partial m} \eqno(4.19)$$
Microscopic diffusion explains the observed stellar photospheric abundances, and its introduction in solar models \citep{christensen93,thoul94,berthomieu97,brun98} improved the agreement between theoretical and observed  
sound speed (Fig. 6a). 
In fact we introduced two terms in equation (4.19) in addition to the nuclear term: the first one describes the migration of elements relatively to hydrogen (microscopic diffusion) and the second is a turbulence term in the region of transition from the radiative to convective energy transport, which partly inhibits this microscopic diffusion \citep{brun99}.

The  microscopic diffusion of the elements modifies the composition along the 
radial profile and reduces the hydrogen burning lifetime by almost 1 Gyr. 
The photospheric composition (relative to hydrogen)
is reduced by about 12 \% during this lifetime.
The introduction of such a process increased 
the neutrino flux estimates for the chlorine 
and water detectors by about 20\% (table 6). The introduction of such a dynamical effect helps to better understand
the photospheric abundances of lithium and beryllium,  which burn at respectively 
$2.5 \times  10^6$K and $\rm 3.5 \times  10^6 K$ (see also section 8). Surprisingly, this change almost compensates for the impact of the change of composition on sound speed, temperature, and neutrino fluxes. 

\subsubsection {A detailed description of the nuclear interaction in the solar plasma} 
The knowledge of the reaction rates listed in Table 1 (except for the very weak pp reaction rate) are based on  measured cross sections in laboratory. Exceptionally, $(^3He, ^3He)$ has been measured down to the astrophysical range of energy \citep{junker98}. Updates were obtained on the ($^7Be, p$) cross section \citep{junghans03} and on ($^{14}N,p$), which was reduced by a factor 2 \citep{formicola04} compared to previous estimates. However,
the reaction rates described by equation (4.12) are modified by the solar plasma. The velocities of the reactants are indeed perturbed by 
the presence of free electrons and ion clouds  in their vicinity. A correction,
called ``the screening factor'', is needed and was further discussed in the framework of the neutrino problem \citep{dzitko95,gruzinov98,gruzinovbahcall98},
because the solar plasma is not the pure weak plasma described by the Debye theory. The current 
solar observations are compatible with the notion of intermediate plasma and any larger screening effect seems unwarranted.
The pp reaction rate is so small that it is only known theoretically. As the sound speed is extremely sensitive to this reaction rate,
 it is now possible to improve its determination through seismology \citep{turck01b}
 and also through the neutrino flux measurements.  Experimental efforts are pursued to improve the nuclear reaction rate predictions, and a new compilation, including a discussion on the various sources of uncertainty, was published this year \citep{adelberger10}. Laser measurements on LMJ or NIF represent a new challenge for the future in producing plasma in stellar conditions, this new technique has the advantage to measure directly the nuclear reaction rates.

The possibility that physical mechanisms such as 
diffusion, collision, or long-range coulomb interaction, could slightly deplete the Maxwellian tail 
at high energy has been evoked in the past. A detailed study showed that  a very small deviation of $0.5$ \% of this Maxwellian distribution leads to a change 
in reaction rate of 5\% in pp reaction, -36\% in ($^3He, ^3He$), -50\% in ($^7Be,p$), and -67\% in the CNO cycle. 
The adverse impact on the sound speed \citep{turck01a} coupled with the results from the SNO 
experiment, do not support the idea that the Maxwellian distribution of the particle velocities is deformed 
at high energy.

\subsubsection{Is there some mixing in the very central core?}

When the neutrino puzzle surfaced, the possibility was raised for mixing in the solar core to reduce the $^7Be$ composition
(which is extremely peaked) together with the $^3He$ content  
and consequently to reduce the central temperature and the emitted $^8B$ neutrino flux \citep{bahcall68b,schatzman81,lebreton87}. This idea was supported by the fact that the central solar conditions are not far from convective instability. Such an effect has an impact on the whole radiative zone far greater than what is suggested by the observed acoustic modes  \citep{turck01a}, see  figure 3 of \cite{turck01a}. However some small and localised mixing in the deep core could be attributed to an initial higher mass (see below).

A similar idea that the central solar core may be cooled down by the presence of 
WIMPs (Weakly Interacting Massive Particles) acting as a conductive medium, and transfering heat from the very center to the rest of the radiative zone  \citep{Faulkner}, was emitted. 
More that ten years ago, we indeed used seismology to show that such an idea is not favored \citep{kaplan91}.
At that time the constraints on the solar core were not as stringent. The limits imposed by the Sun are regularly re-estimated \citep{lopes02}. A revision could be justified by the recent cross-section updates and by the inclusion of the new abundances, but it is probably better to wait for the inclusion of the impact of the gravity modes on the density and sound speed profiles. The difficulty to separate the effect of mass loss from the effect of dark matter hampers progress on this issue (also see section 8).

In conclusion, it was shown in this section that the physics of the solar interior has been greatly improved during 
the last ten years and that at each stage the predicted neutrino fluxes were compared to detection results. The most obvious impact is on the $^8B$ neutrinos whose predicted flux changed by more than 50\% during that period, as shown in Table 6. In parallel,  the neutrino flux prediction of Bahcall and collaborators
varied between  7.5 and 5.8 $\rm \times 10^6cm^{-2}s^{-1}$. After the launch of the SoHO spacecraft, the predictions of the two teams converged to comparable numbers when using the same physics. Since this period, the improvements of the SSM are systematically confronted to helioseismic observations. This second probe was also useful to help rejecting several 
non-standard ideas impacting on the neutrino flux predictions, and to qualify the standard solar model which only introduces the main physical processes describing the evolution of stars.

\section{The seismic view of the solar interior and comparison with solar models}
In this section, we briefly describe the helioseismic tools. Other reviews can be read for details: \cite{turck93,leibacher99,christensen02,antia07}.
Acoustic waves are generated by the granulation at the solar surface and propagate
inside the Sun to a depth depending on their initial velocity. 
These waves generate very small motions in the solar atmosphere 
 that are  detectable.
 We formally treat this information through a perturbation theory because the amplitudes of these perturbations are small and the Sun is, to a good approximation, spherical. 

\subsection{The formalism}
The Sun, as a self-graviting sphere of compressible gas, oscillates around its 
equilibrium state with a period of about 5 min. These oscillations are 
interpreted as a superposition of waves propagating inside the star (acting as a 
resonant cavity), and forming standing waves: the eigenmodes of vibration. By 
projecting these modes onto spherical harmonics $Y_l^m$, we write any scalar 
perturbations as (in the case of the Eulerian pressure perturbation $p^{'}$) following \cite{christensen91}:
\[p^{'}(r,\theta,\varphi,t)=p^{'}(r)Y_l^m(\theta,\varphi)\exp i\omega_{n,l,m} t \]
and the $\vec{\xi}$ displacement vector as
$$\vec{\xi}(r,\theta,\varphi,t)=\left(\xi_r(r),\xi_h(r)\frac{\partial}{\partial 
\theta},\xi_h(r)\frac{\partial}{\sin \theta \partial 
\varphi}\right)Y_l^m(\theta,\varphi)\exp (i\omega_{n,l,m} t) \quad  \eqno(5.1) $$
where $\xi_h=1/(\omega^2 r)[p^{'}/\rho+ \Phi^{'}]$ is the horizontal 
displacement, $\Phi^{'}$ the gravitational potential perturbation, 
$\omega_{n,l,m}$ the eigenfrequency, and $\rho$ the gas density. The quantum 
numbers $n$, $l$, $m$ are respectively the radial order (number of nodes along 
the radius), the degree (the total horizontal wave number at the 
surface is $k_h\sim L/R_{\odot}$, with  $L=\sqrt{l(l+1)}$) and the azimuthal 
order (number of nodes along the equator with $|m| \le l$). Restricting the phenomenon to adiabatic 
oscillations within the Cowling approximation ($\Phi^{'}$ neglected) and 
considering only small radial wavelengths compared to $R_{\odot}$ (the solar radius), the 4th-order system equations are reduced to 
second-order wave equations, with the following 
dispersion relation:
$$k^2_r=\frac{1}{c^2_s}\left[F^2_l\left(\frac{N^2}{\omega_{n,l,m}^2}-1\right)+\omega_{n,l,m}^2-\omega^2_c\right] \quad \eqno(5.2) $$
where the squared norm of the wave vector is written as the sum of a radial and 
a horizontal component $|\vec{k}|=k^2_r+k^2_h$, $k^2_h=F^2_l/c^2_s$ is the 
horizontal wave number, $F^2_l=L^2c^2_s/r^2$ is the Lamb frequency, 
$N^2=g[1/\Gamma_1 d\ln p/dr - d\ln\rho/dr]$ is the Brunt-V\"ais\"al\"a frequency, 
$\omega^2_c=c^2_s(1-2dH_{\rho}/dr)/4H^2_{\rho}$ is the acoustic cut-off frequency ($\sim 
5.8$ mHz), $H^{-1}_{\rho}=-d\ln \rho/dr$ is the density scale height, $\Gamma_1$ 
is the first adiabatic exponent, and $c^2_s=\Gamma_1 p/\rho$ is the sound speed.

\begin{figure}
\begin{center}
\rotatebox{90}{\includegraphics[width=6.5cm]{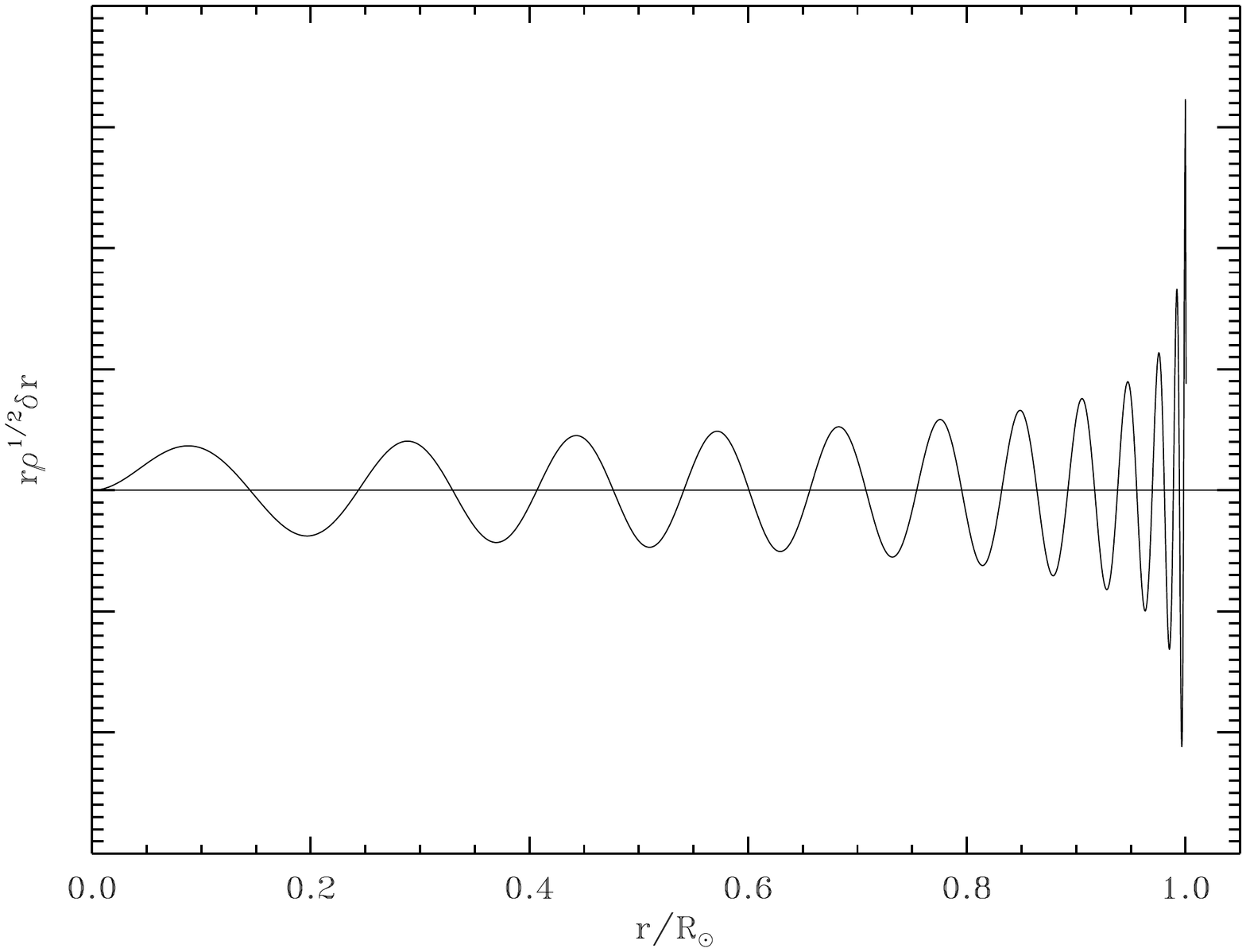}}\rotatebox{90}{\includegraphics[width=6.5cm]{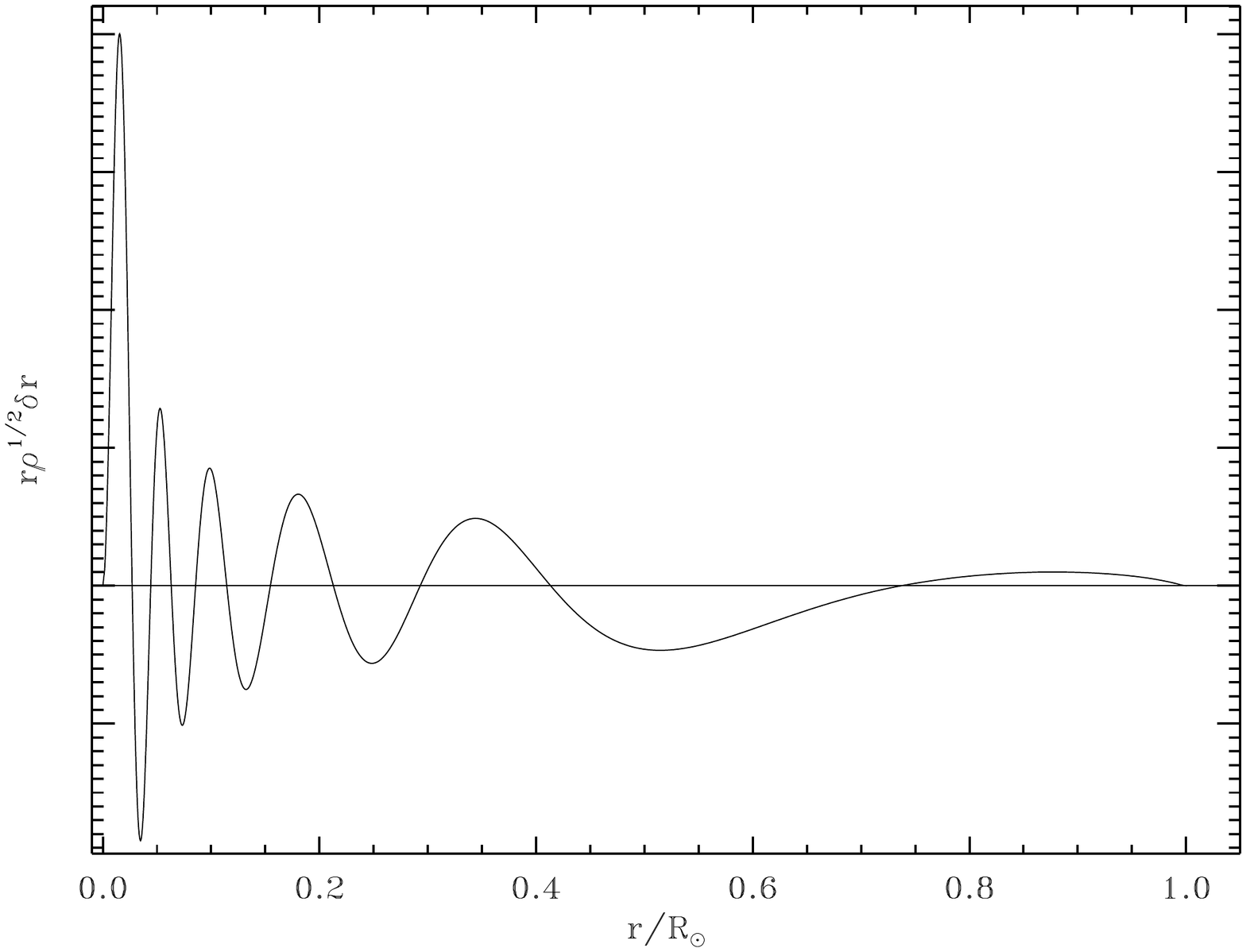}}
\caption{Eigenfunctions of an acoustic mode ($\ell$ =0 radial mode, n=23) (left panel), and of 
 a gravity mode ($\ell$ =2, n=-10) (right panel).}
  \label{fig:modes}
\end{center}  
\end{figure}

The oscillatory solutions to the wave equation define two types of waves (and modes, since some of these waves constructively interfere), namely 
acoustic ones (with $\omega_{n,l,m}> N, F_l$) and gravity ones (with $\omega_{n,l,m}< N, F_l$).
Figure 4 shows the properties of these waves: the acoustic modes have their maximum amplitude at the surface
(left side), while the gravity modes are excellent probes of the solar core and are evanescent at the surface
with a rather small amplitude (right side).

About 3500 acoustic modes (the so-called 5-min oscillations) have already been
observed. A refined analysis of their properties allows the obtention of a 
stratified information about the solar internal structure from the surface to the solar core through the sound speed profile. 
The gravity waves remain the best probes of the region of neutrino emission and also provide information on the rotation in the core.

There are two ways to apply the seismic data to probe the 
internal structure of stars:\\
a) The direct method: comparison of predicted with observed acoustic wave
frequencies;\\
b) The indirect method: using inversion procedures to deduce the solar radial profile of 
fundamental variables like the squared sound speed $c^2_s$, the density $\rho$, the 
adiabatic exponent $\Gamma_1$, or the rotation rate, and compare them to a computed 
model.

Since the launch of SoHO, we have been using two variables obtained from the space instruments GOLF: 
Global Oscillations at Low Frequency \citep{gabriel95}
and MDI: Michelson Doppler Imager \citep{scherrer95},
namely the sound 
speed profile and the internal rotation profile, 
and compared them to up-to-date solar models. It allowed an improvement of these models when the agreement was not satisfactory,
or the introduction in these models of some extra physical phenomena to check whether the agreement could be improved. Indeed, a lot of physical phenomena 
have very
specific signatures which are informative. We could also potentially access
data on the magnetic field, but until now we only managed to put upper limits on the field strength due to the difficulty 
to accurately derive such a quantity. 

\vspace{0.5cm} 
\begin{figure}
\begin{center}
\rotatebox{90}{\includegraphics[width=8cm]{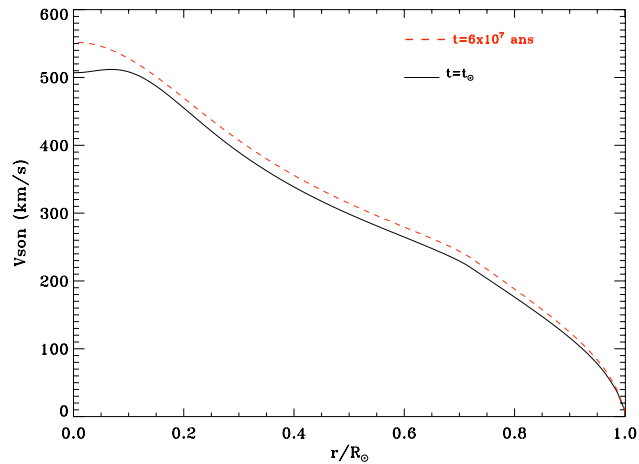}}
\vspace{-1cm}
\end{center}
\caption{Evolution of the sound speed during the hydrogen-burning phase}                
\end{figure} 
		
\subsection{The sound speed: a very useful but demanding quantity}
Since the launch of SoHO in 1995, the solar sound-speed profile has been derived with more and more accuracy from 
helioseismic data. Consequently, if it could be demonstrated that there is a significant discrepancy
between the observed Sun and the SSM, this would be an opportunity to reach beyond the classical
theoretical solar model. Moreover, in the meantime, this helioseismic information is used to
 deduce neutrino fluxes constrained by the seismic observation of the Sun. 

Before commenting on the role of the seismic probe,
 it is interesting to compare the sensitivity of the two probes at hand, the neutrino fluxes and 
 the sound-speed profile, 
 to the conditions at the center of the Sun. 
 We reminded the $^8B$ neutrino flux dependence on
  the central temperature in table 6. It is noteworthy that if the Sun was exactly 
 at the beginning of the hydrogen-burning phase, the prediction for the neutrino flux detectable by the chlorine experiment would be 0.57 SNU 
 (instead of the roughly 7 SNU
 currently estimated) while for the flux detectable by the gallium experiments it would be 67 SNU (instead of 127 SNU), meaning a factor, respectively, 15 and 2 smaller.

 On the other hand, as is shown on figure 5,  the central sound speed only varied 
 by 9 \%
 during the same timespan. In fact, as $$ \Delta c^2/c^2 = \Delta T/T - \Delta  \mu / \mu \eqno(5.3)$$ and 
 $\Delta T/T= 13.5 \%$ ($T_{init}$= $13.5 \times 10^6 K$),  we have $\Delta  \mu / \mu$ = 32 \% (the mean molecular weight varies from 0.31 to 0.41), 
 and therefore $\Delta c/c$= -9\%. Consequently 
the sound-speed profile needs to be known with
 a  high relative precision (better than $10^{-3}$) in order 
 to detect a temperature deviation of less than 1\%. This is exactly the precision required to 
 test the validity of the present solar 
 structure (see table 7 listing the sensitivity of the sound speed to different ingredients of the solar model). 
 An uncertainty of  2 \% on the central temperature leads to a reduction by a factor of 2 on the $^8B$
 neutrino flux. This challenge needs to be addressed by the SuperKamiokande and SNO experiments. 
 On the helioseismic side, the relative precision of the measurements is very high (about 10$^{-4}$) because those are metrological measurements. Therefore helioseismology offers very interesting constraints provided that the theoretical acoustic-mode characteristics can be validated. The accuracy of the oscillation frequency determination is given by the duration of the observation (which can be years or decades) if one can verify that these frequencies do not change with time. Their stability was demonstrated for frequencies below $1.6$ mHz \citep{couvidat03,garcia04}
 with the GOLF experiment.
  
  \begin{table*}
\begin{center}
\caption[]{\label{Table 7} Sensitivity of the sound speed to the physical processes }    	
\begin{tabular}{p{4cm}*{3}{c}}
\hline
Quantity 	& variation & $\Delta {c^2/c^2}$ variation  \\
\hline
T			& 1 \% 	&    1\% \\
$ \kappa$ 		&  1 \% 	& 0.1 \%\\
$X_c\,^{56}Fe$  &  4 \% &  0.1 \% \\
$X\,^{3}He$     	& 25\%	&  0.1 \%  \\
(p,p) reaction rate   		 & 1\%  &  $\pm$ 0.1\%  \\
($^{3}$He, $^{3}$He) reaction rate & - 25 \% &  - 0.1 \% \\
($^{3}$He, $^{4}$He) reaction rate & -25\% &  +0.2\%  \\
(p, $^{7}$Be) reaction rate & 10\% & none   \\
(p, $^{16}$O) reaction rate & -50\% &  - 0.1-0.2 \% just at the center  \\
\hline
\end{tabular}
\end{center}
\end{table*}

  Contrary to boron neutrinos, acoustic modes do not provide a direct determination of the temperature. $\Delta c^2/c^2$ does not depend only on
   $\Delta T/T$ as was previously shown. Nevertheless, if we have strong constraints 
   on the density and pressure, we also have an indirect but strong constraint on the central temperature.
In 2001, we investigated all the different issues 
mentioned in  table 6 and we reached the required precision to  predict neutrino fluxes constrained by helioseismology. These fluxes were compared to the very important SNO results \citep{turck01b, turck04a,couvidat03}.
During these years, 
 major efforts have been carried out to improve the quality
of the seismic indicators  because the acoustic mode frequencies 
are largely influenced by the physics of the outer solar layers.
The different techniques used contributed to identify biases in the oscillation frequency determination.  
 Three phenomena have been thoroughly studied for investigating the physics of the solar core: the stochastic excitation of acoustic waves, the influence of the solar activity on the absolute value of the mode frequencies, 
and the asymmetry of the mode distribution due to the interaction of the modes with the solar background 
\citep{basu00,couvidat03}. 
SoHO/GOLF detected modes of low frequency which have a long lifetime  and 
 propagate inside solar cavities for  which the noise coming from stochastic excitation of waves and the effects of the solar cycle are both reduced.
The relative accuracy of the sound-speed determination in the central solar region was improved by an order of magnitude and is now of the order of $10^{-5}$ in the core \citep{turck01b,couvidat03}. 
\begin{figure}
\centering
\rotatebox{90}{\includegraphics[width=7cm]{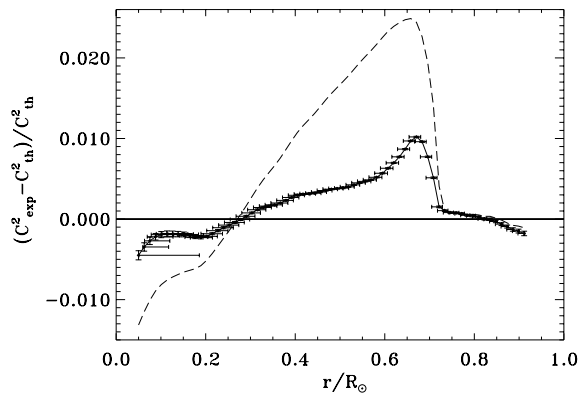}}
\caption{left: Relative difference between the squared sound speed in the Sun extracted from GOLF+MDI and the squared
sound speed of reference solar model in function of radius, compared with a model which does not contain the 
microscopic diffusion (dashed line). From \cite{brun98}.
}
\end{figure}

\begin{figure}
\centering
\vspace{-5cm}
\includegraphics[width=8.1cm]{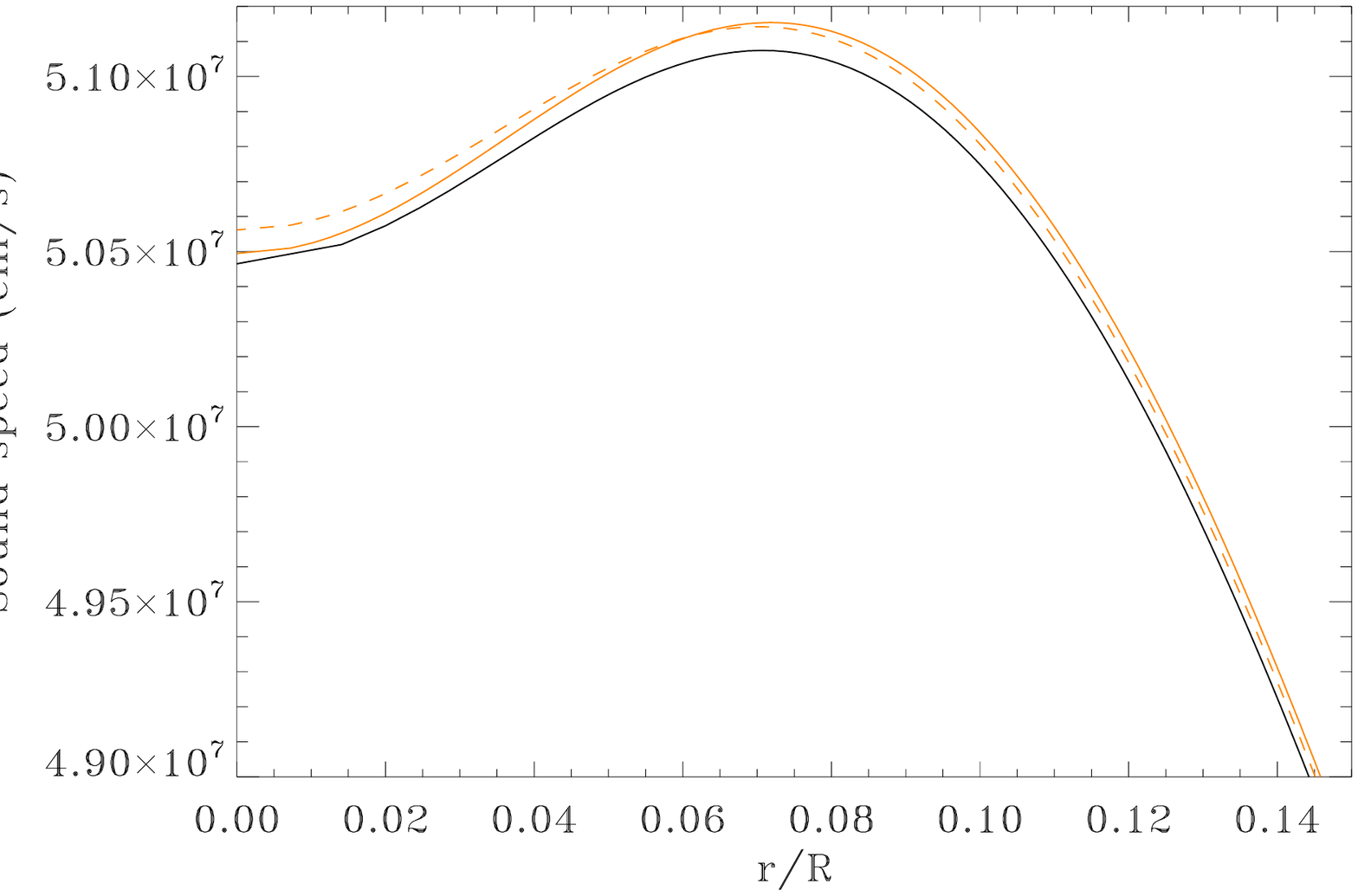}
\includegraphics[width=7.4cm]{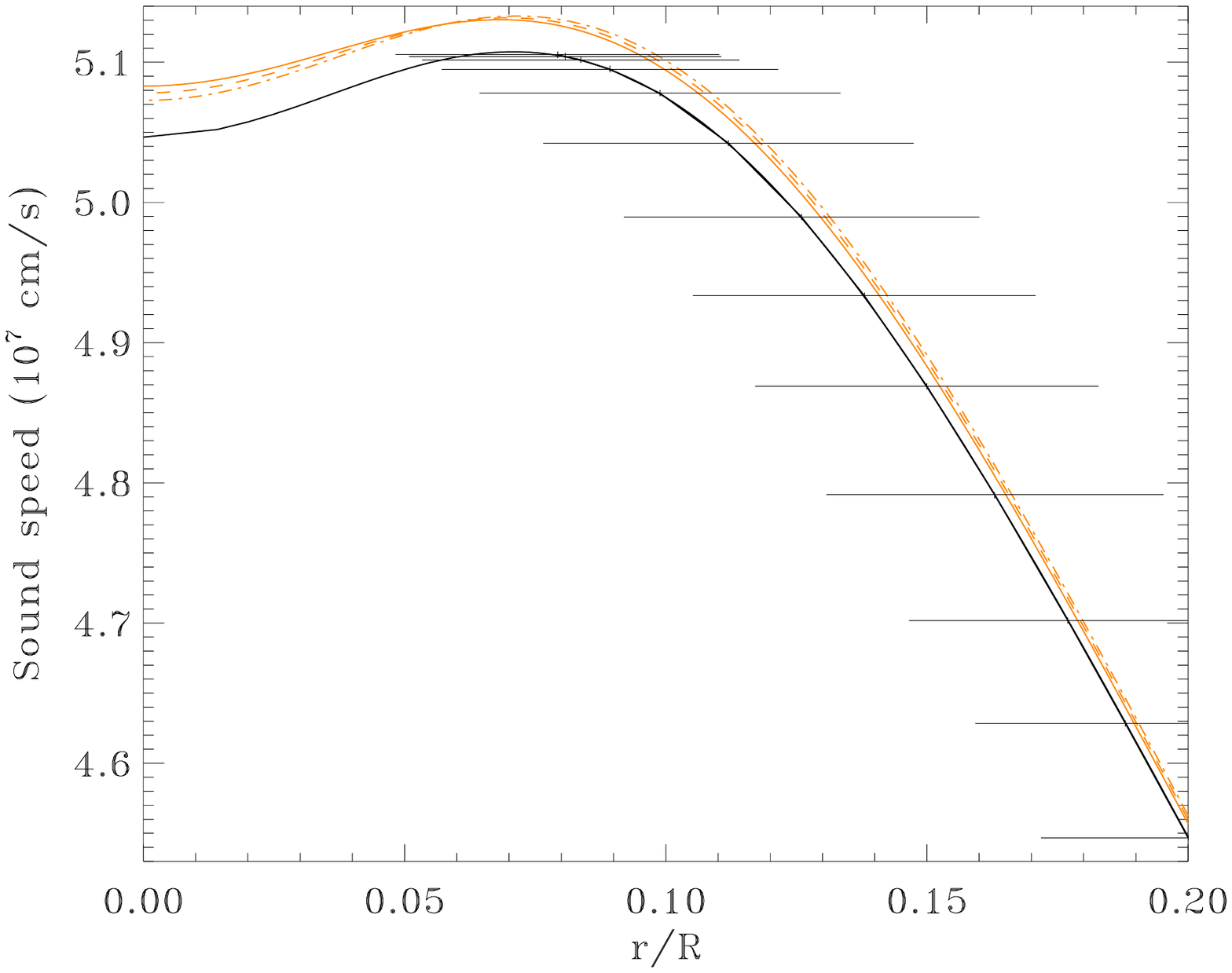}
\caption{\small Comparison of the sound speed in the core deduced from the acoustic modes and the sound speed of standard models and seismic models (in black). Acoustic modes explore the solar interior down to 6\% so reasonably well in the region of emission of boron neutrinos. Left: standard model of the nineties, Right: new standard model.
(points with error bars joined by straight lines).  Adapted from \cite{turck10b}.
 }
\end{figure}
\begin{figure}
\centering
\includegraphics[width=6.7cm]{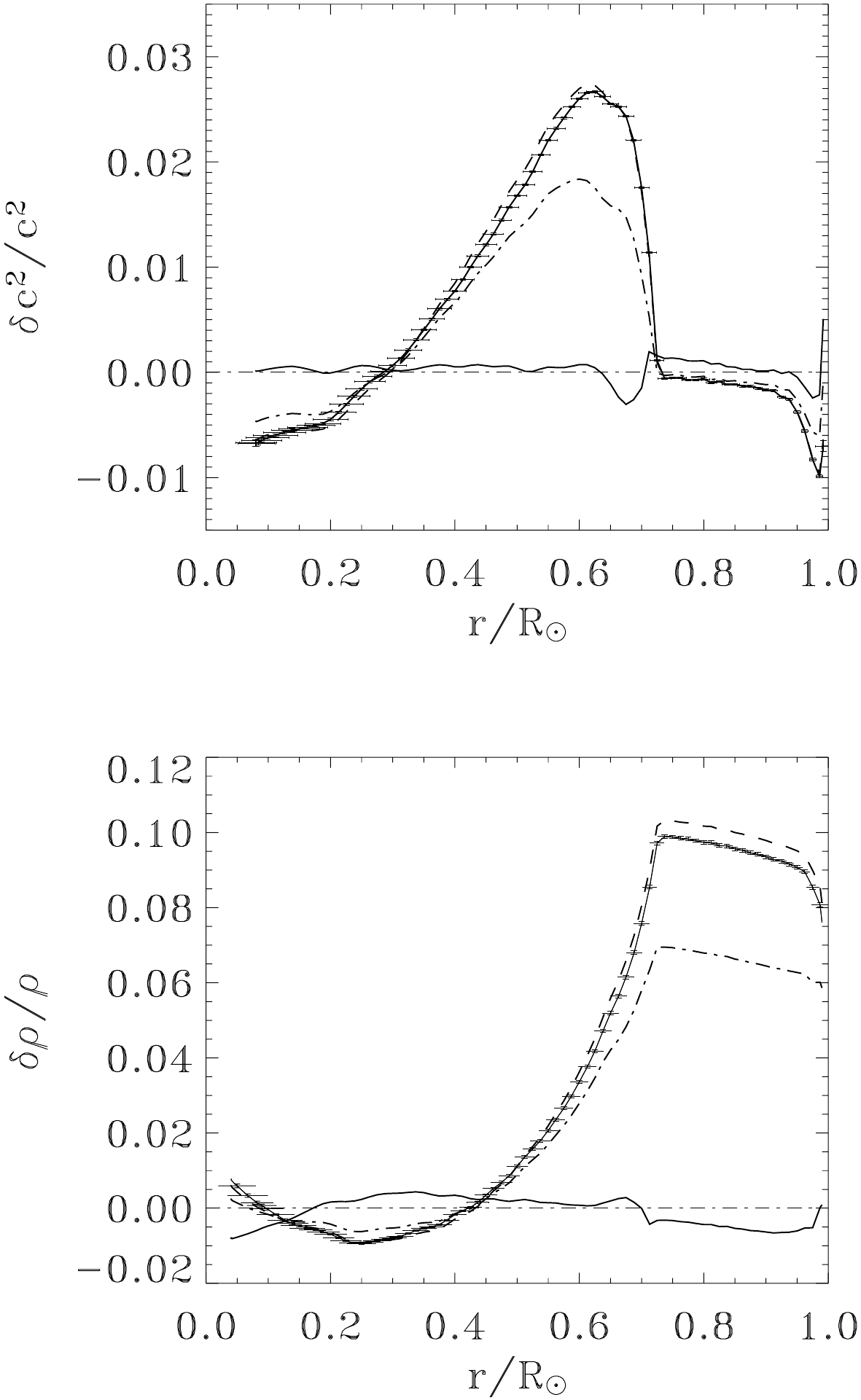}
\includegraphics[width=8.7cm]{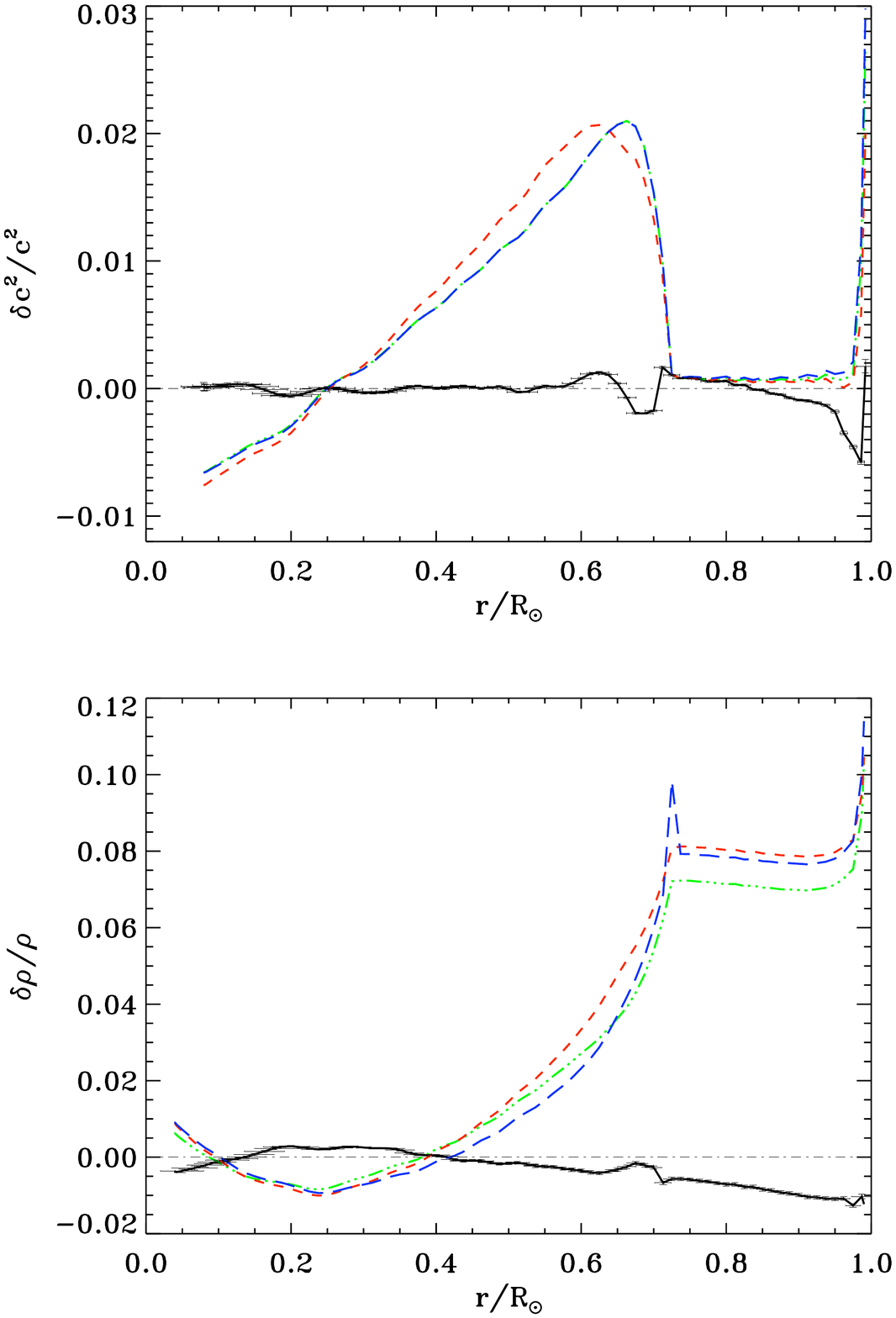}
\caption{\small Relative differences between (a) the square of the sound
speed and (b) the density deduced for the Sun using the GOLF/MDI 
frequencies 
and those of different Saclay 
models.  Left: Updated standard models including  new estimates of CNO abundances by \cite{asplund04} (full line with error bars coming from seismic 
observations: 
tac A model, dot line: tac L model, dot dashed line: tac H model) 
and the seismic model (full line). From \cite{turck04a}. Right: New SSM using \cite{asplund09}
 composition and CGM convective prescription. From \cite{turck10b}.}
\end{figure}

\subsection{The sound-speed and density profiles: observations and predictions}
               
Figure 6 shows the first squared sound-speed difference obtained using GOLF and MDI data aboard SOHO, between a standard model including or not the gravitational settling of elements.
Figure 7 shows a zoom on the solar core of the sound-speed profile of standard models (obtained with old and new abundances) and of the seismic model (see below). The vertical error bars are so small that they are almost invisible on the figure: such a high precision is due to the long temporal duration of the observation considered, and to the metrologic character of the Doppler velocity method (assuming that there is a good control of the instrumental noise). The horizontal error bars translate the sensitivity of each mode to the physics of the solar core. The large number of points compensate for these large error bars.
 The values of the sound speed  in the core have been recently confirmed with  ground-based observations \citep{basu09}. {\it The development of helioseismology (also see section 7) and the recent progress on CNO photospheric abundances, lead to the conclusion that the SSM reached its limits when it comes to representing the real Sun}.
 
  Figure 8 shows the evolution of the discrepancy between  the observed sound speed and density and the SSM predictions. Today, a difference between SSM, for \cite{caffau09} and \cite{asplund09} compositions, and seismic observations still exists, which is much larger than the seismic error bars.
  
\subsection{The gravity modes}
Gravity modes are the best probes of the solar core. They inform us on the density and rotation profiles of the core and consequently help us connect the current state of the central solar core to its past states and to the physical quantities characterizing them.
 Gravity modes are been searched for by the three helioseismic instruments onboard the SoHO satellite. The Doppler velocity technique, with measurements performed in the solar atmosphere where the noise is smaller than at the surface, proved the most adapted to this search. Indeed, the amplitudes of gravity modes are small and these modes are in the frequency range 10 to 400 $\mu$Hz, which  is dominated by granulation noise. In fact, two years after the launch of SoHO, the GOLF instrument, which looks at the Sun globally, appeared to be the most promising instrument for this search, thanks to its very low instrumental noise. Two publications presented positive detections of patterns in the gravity mode range: the first publication was dedicated to the range of mixed modes (modes that present characteristics of both gravity and acoustic waves) where two patterns were  followed over a period of 10 years \citep{turck04b,garcia08}. These patterns could favor an increase in the rotation rate in the solar core, but no clear determination of the central frequency of a potential quadrupole mode (n= -3) was obtained so far. The second paper is dedicated to the frequency domain where gravity modes are equally spaced in period. By cumulating the power of 20 oscillation modes, we highlighted a pattern attributed to dipole modes \citep{garcia07}. Some of these modes are now analyzed individually and at least 6 modes are properly identified (in the range n = -4 to n= -10). These results should be published this year. A definitive detection of these gravity modes improves the sound speed, density and rotation profiles in the core. The rotation rate of the very central region already seems larger, by a factor 5 to 7, than the rest of the radiative zone (figure 11) which is a very important result of the SoHO spacecraft. The frequencies of the detected modes are very close to the values predicted by the seismic model in \cite{couvidat03,mathur07}. 
 
\section{The detection of solar neutrinos and comparison with predictions}

Solar neutrino detectors focus on neutrino energy lower than 20
MeV, through one or more of the three different types of reactions~:

\noindent
- neutrino capture on a nucleus (A,Z) by
W Exchange Process (WEP)\footnote 
{The denominations
``charged current'' process (CC) and ``neutral current''
process (NC) have been replaced here by ``W exchange'' process
(WEP) and ``Z exchange'' process (ZEP) which better describe 
the physical mechanisms.}:
$$\rm \nu _x \,+\,(A,Z)\,\rightarrow\,X^-\,+\,(A,Z+1) \; \eqno(6.1) $$
Experiments based on this reaction are only sensitive to \nue.
When \nue oscillate and become \numu or \nutau,
they cannot be detected anymore by this process since the threshold for producing
a  muon (m\,=\,106 MeV) or a tau (m\,=\,1784 MeV) is well
above the maximum solar neutrino energy (\about 14\,MeV).
An electron is (almost) isotropically released by the reaction involving \nue, thus giving (almost) no
information on the neutrino incidence direction.
However the electron energy spectrum directly reflects  the
neutrino energy spectrum: E$\rm _e$ = E$\rm_{\nu}$ - E$\rm _{threshold}$
where E$\rm _{threshold}$ is the energy threshold of the reaction.

\noindent
- neutrino scattering by
Z exchange process (ZEP):
$$\rm \nu_x \, + \, A \, \rightarrow \, \nu_x \, + \, A^*  \; \eqno(6.2) $$
The reaction cross section does not depend on the neutrino flavor.
This type of reaction is  very useful  to distinguish neutrino
oscillations (constant flux but mixed flavors) from a genuine neutrino flux
deficiency. 

\noindent
- elastic scattering on electrons:
$$\rm \nu_x\, + \, e^-\,\rightarrow\, \nu_x\, + \, e^-  \;  \eqno(6.3) $$
This reaction occurs both via WEP and ZEP for \nue,
and only via ZEP for \numu and \nutau.
The resulting cross section for the \nue scattering is about 6 
times larger than for the \numu or \nutau scattering. This means that
the elastic scattering of \nue on electrons
proceeds mainly via WEP.
An advantage of this reaction is that, for kinematic reasons, 
the scattered electron direction is correlated with the direction of the neutrino.    
This property is very helpful for background noise reduction.
However the electron energy spectrum is not identical to
the neutrino spectrum, though scattering experiments give
significant information on this neutrino energy spectrum.

A summary of the solar neutrino experiments
is given in table 8. 

\begin{table*}
\small
\caption{\rm : Existing and funded solar neutrino detectors. }\label{Table 8}
\begin{center}
\begin{tabular}{|c|c|c|c|c|} 
\hline  
           &          & detection &  signature         &           \\             
experiment & reaction & threshold & main contributors & events/yr  \\
\hline
Homestake  & \nue + \Cl $\rightarrow$ \Ar + \emoins & 0.814 MeV & \Ar & 100 \\
 615 tons &                           &        &   \nB, \nBe, $\nu_{^{15}O}$, \npep   &    \\
1968-2003  &     &    &    &   \\
\hline 
Kamiokande   &  \nue \emoins \rar \nue \emoins & 5 MeV & E $>$ 7.5 MeV  &     \\
2140 tons \eau     &       &      &     \nB  &    \\
1987-1995    &      &     &   Sun direction   &    \\
\hline
SAGE (1990-		&  \nue + \Ga $\rightarrow$ \Ge + \emoins & 0.233 MeV & 30 \Ge&\\
50 tons      &     &    &    &      \\
Gallex (1991-1997) &   &     &  for the 3 detectors &  20\\
30 tons    &    &    &    &     \\
GNO (1998-2003)&   &     &  \npp, \nBe, \nB, &   20\\
radiochemical &   &    &  $\nu_{^{15}O}$, $\nu_{^{13}N}$, \npep &\\
\hline
Super-Kamiokande   & \nue \emoins \rar \nue \emoins & 5 MeV & E $>$ 5 MeV
                    & 24000 \\ 
50000 tons \eau  &                     &         &           \nB          &  \\
Cerenkov   &              &         & Sun direction      &  \\ 
1999-   &              &               &               &\\
\hline
Sudbury    & \nue D \rar \emoins p p & {\bf 6.5 MeV} & {\bf E $>$ 6.5 MeV  } \nB  & 9750 \\ 
1000 tons   &                         &         & $\rm(D_2O)-(H_2O)$ &  \\
           \cline{2-5}
\eaul      & \nux D \rar \nux p n    & 2.2 MeV & n capture on $\rm^{35}$Cl  \nB & 2800 \\
Cerenkov (1996-  &                         &         & $\rm(D_2O +NaCl)-(H_2O)$ &   \\
           \cline{2-5}
(2000- &      \nue \emoins \rar \nue \emoins & 5 MeV &   E $>$ 5 MeV  \nB &  1100   \\
       &                        &                  &   Sun direction   &  \\ 
       \hline    
       Kamland & $  \bar \nu_e$ + p \rar e$^+$ + n   &  1.8 MeV  &   reactors  & \\
       1000 t liquid scintillator &    &    &     & \\
    (2002-      &     &    &   &    \\
\hline
Borexino       & \nue \emoins \rar \nue \emoins & {\bf 0.2 MeV }&
                                      E $>$ 2.8 MeV      \nB      & 130 \\
 2000 tons  scintillator                                     &                    &     &   0.2$<$ E $<$ 0.8 MeV \nBe&  11000 \\
(278 t \bonze)           &                         &         & no Sun direction  &      \\ 
   \cline{2-5}
(2007-             & \nue \bonze \rar \conzex \emoins &  & E $>$ 3.5 MeV &    2300 \\
 &     \conzex \rar \conze $\gamma$ &    & no $\gamma$ or
                               $\gamma$(2,4.3,4.8 MeV)  &      \\
           \cline{2-5}
 & \nux \bonze \rar \bonzex \nux & 4.5 MeV & $\gamma$  &   \\
 &   \bonzex \rar \bonze $\gamma$ &    & (4.4 or 5 MeV) &  \\ 
\hline
Icarus I   & \nue \arq \rar \Kqx \emoins & 11 MeV & E $>$ 5 MeV  &  300 \\
600 tons    &  \Kqx \rar \Kq $\gamma$ &        & + $\gamma$ 2.1 MeV &   \\
           \cline{2-5}
liquid argon & \nux \arq \rar \arqx \nux & 6 MeV & $\gamma$       &   60 \\
drift chamber  &                         &       & 6.1,7.8,9.6 MeV &      \\
           \cline{2-5}
May 2010      & \nue \emoins \rar \nue \emoins & 5 MeV   & E $>$ 5 MeV
                                            &                       240 \\
           &                             &      & Sun direction  &     \\
\hline
\end{tabular} 
\end{center}
\end{table*} 
\noindent

\normalsize

\noindent
\subsection{Radiochemical experiments }
These experiments use the method of neutrino capture on a nucleus (A,Z) by
W exchange process (WEP) and detect only \nue.

\vspace{5mm}
\noindent
- Homestake experiment: 1968-1995

This historical experiment follows the initial idea of \cite{pontecorvo47}. It considers the reaction   
\begin{math}
\rm \nu_e\,+\,^{37}Cl\,\rightarrow\,^{37}Ar\,+\,e^-
\end{math}
to capture solar \nue.
The energy threshold is 0.814\,MeV, and the resulting radioactive
\Ar\,  isotopes decay by electron capture
with a half-life of 35\,days.
The cross section varies from  $5 \times$ 10$\rm^{-46}$ \cmd\,at 1\,MeV
to 10$\rm^{-41}$ \cmd\, at 14 MeV. Therefore, about 10$\rm^{31}$ atoms of chlorine are necessary to
observe one \Ar\,  atom a day. Consequently, from several hundred of
tons of a compound of chlorine, one atom of \Ar\, a day (or a few
atoms every other month) needs to be extracted, and its decay needs to be unambiguously observed.

A big tank containing 615 tons of perchlorethylene \CCl
was settled in 1967 in the Homestake gold mine (South Dakota, USA),
at a depth of 4100\,$\pm$\,200 hg/\cmd of standard rock 
\footnote{Underground detectors are located at different depths
of different kinds of rocks. In this field, one  
generally gives the depth with a unit called hg/\cmd 
of standard rock.
Standard rock is defined as rock with a density of 2.5 g/\cmt,
a mean atomic number Z=11, and a mean atomic mass A=22.}
to shield
against cosmic rays.    
The first results gave an upper limit of  3\,SNU \citep{davis68}, 
more than a factor 2 lower than the predictions from solar models.  


 A run consists of three steps: the exposure, the
argon extraction, and the counting of the \Ar\, atoms.
Before each run a small amount of carrier gas
(about 1\,\cmt  of non radioactive argon isotopes,
$\rm^{36}Ar\;or\;^{38}Ar$) is stirred and dissolved into the
\CCl.
The recovery of this carrier allows  
a measurement of the argon extraction efficiency.
The exposure to solar neutrinos lasts about two months.
Argon (the carrier and the \Ar\, produced by solar \nue or any
background), 
is then removed from the \CCl by circulating 
a few hundred cubic meters of helium through the tank.
A charcoal trap cooled at liquid-nitrogen temperature absorbs the argon
and lets the helium pass through.
The argon is then desorbed and pushed into a small proportional counter
(about 1\,\cmt). 10\,\% of methane is added to optimize the
working of the counter.
The counters are made of low-activity materials and put in
a lead shielding located in the mine itself.
\Ar\, decays by electron capture with a half-life of 35\,days
and emits 2.82\,keV Auger electrons.
The counter background noise is  reduced by a pulse-shape 
analysis of the signal: \Ar\, decays show a characteristic short
rising time.
The counter background noise
has been reduced to 0.01\,count/day after several years. The average \Ar
\, count has been of about 0.5 atoms/day.
The major source of background noise comes from cosmic ray muons (about
5\,/m$\rm^2$\,/day at this depth).  They produce low energy protons either
directly or  through the hadronic cascade which may follow. These 
protons produce \Ar\, through (p,n) reactions with \Cl.
Two indirect determinations of this background lead to a value
of (0.08$\rm\,\pm\,$0.03) \Ar\, /day (about 20\,\% of the signal
as will be seen below).
The other sources of background noise,
fast neutrons from surrounding rock and cosmic ray neutrinos, have been found
negligible.

More than one hundred runs have been performed since 1968, almost continuously
except in 1985-1986 \citep{davis94}.
The first two years of runs are no longer considered, due to
considerable improvements and modifications during this period. 
The final result for 25 years corresponds to a detection of 2200 atoms of \Ar.
The average solar neutrino flux of the 108 runs covering the period 1970-1995 
is $\rm 2.56 \pm 0.16 (stat.) \pm 0.16 (syst.)$ SNU \citep{cleveland98}. R. Davis received a joint Nobel Prize for this work, with M. Koshiba for the Japanese research on this subject (see below). 

\vspace{5mm}
\noindent
- Gallium experiments

The main objective of the radiochemical gallium experiments is the detection of
the \npp which are produced in the primary pp fusion
reaction using the fact that the reaction \nue\,+\,\Ga\,$\rightarrow$\Ge\,+\,e$^-$
has an energy threshold of 233\,keV, significantly below the
maximum value for \npp (420\,keV). This is an original idea of \cite{kuzmin66}.
The \Ge decays (half-life of 11.43\,days)
by electron capture 
(the probabilities of K, L, and M captures are respectively 87.7\,\%, 10.3\,\% and 2\,\%), emitting low-energy Auger electrons and X-rays.
Two gallium experiments have been designed: 
SAGE in Russia \citep{gavrin90}
and GALLEX in Italy \citep{anselmann92}.
Their principle is the same:  the gallium is exposed 
to solar neutrinos in a low-background environment,
 then
the \Ge atoms produced by the reaction are extracted by a chemical method and transformed into a counting gas
(germane \Germane). A proportional counter counts them.
The main difference between these two experiments is that GALLEX uses  a solution
of \Gacl as the  target while SAGE directly uses the gallium metal.

\medskip

- Gallex  and GNO:  1991-2003

The Gallex experiment, a Germany-Italy-France USA-Israel collaboration, 
was installed in the Gran Sasso Underground 
Laboratory (Italy), in a 
highway tunnel which crosses the
Apennine mountains.
The detector was a cylindrical tank (8\,m high and
3.8\,m diameter) containing 30.3\,tons of gallium in the form of \Gacl (8.2\,moles/liter of \Gacl and 1.9\,moles/liter of
HCl). The expected number of \Ge atoms was 1.2 per day according 
to theoretical predictions, but in fact it has been less than 1/day.

Every three weeks a large flow of nitrogen (150$\rm\,m^3/h$)
circulated through the
tank for about 20\,hours to remove the germanium. 
Germanium, in the form of \Gecl,
is very volatile in presence of HCl and was taken out with the
nitrogen.
It was absorbed into pure water
inside a system of exchange columns full of glass helices
which ensured a large surface of contact.
A concentration step was then followed by a transformation of the
\Gecl into germane \Germane.
A small amount of a given stable
isotope of germanium ($\rm ^{70}Ge,\,^{72}Ge,\,^{74}Ge,\,^{76}Ge$),
was added as a carrier in the tank before each run.
It allowed a measurement of the extraction efficiency and
was used as the counting gas.
The gas mixture in the proportional
counters was germane (30\%) and  xenon (70\%).
The small proportional counters (1\,\cmt) were made 
with ultrapure material. 
Only K (10.4\,keV) and L (1.25\,keV) electron-capture peaks could be observed, with an
energy resolution of about 20\,\%.
The first signature of a \Ge decay was a count observed in the K or L peaks.
Backgrounds as low as 0.06\,count per day (respectively 0.01) have been obtained for
the L-peak (respectively K-peak).
The extraction efficiency was larger than 95\,\% and the counting
efficiency was estimated  for each counter (about 65\,\%).
In a 21-day run, GALLEX observed about 3-4 \Ge decays taking into account 
all the efficiencies and natural decays.

The main source of background noise was the reaction \Ga\,(p,n)\,\Ge.
The protons came either from natural radioactivity of the liquid
solution or from cosmic ray muon flux 
(about 15\,/m$\rm^2$/day 
at a depth of about 3500\,hg/\cmd) 
or from neutrons from the surrounding environment
(building, rock, tank, etc...).
The purity specifications required for the solution have been fully 
satisfied (0.03\,pCi/kg of $\rm^{226}Ra$, less than 0.05\,ppb of U
and less than 0.1\,ppb of Th) and induced a contamination 
$\rm \leq\,0.1\,SNU$.
The protons coming from the interactions of the cosmic ray muons
induced a signal of about 3\,SNU.
The fast neutrons in the environment ($>$\,1\,MeV)
reacted through          
$\rm n\,+\,^{40}Ca\,\rightarrow\,^{37}Ar\,+\,\alpha$.
The induced background was estimated to be about 0.5\,SNU \citep{cribier95}.
The total background was $<$  4 \, SNU.
Solar neutrinos only produce \Ge from \Ga
but cosmic rays produce \Ger (half-life of only 1.63 days)
and also \Germ     
which decays to \Gall by electron capture with a half-life of 288 days
(\Gall, whose lifetime is 68\,minutes, decays mainly by positron emission).
Thus, its decay is identical to the \Ge decay. 
The gallium target was exposed to cosmic rays at
the ground level for several months but this problem was solved by heating the solution.

The results of the different runs were presented 
for different periods of time, Gallex 1:  \cite{anselmann92};  Gallex 2, 1998: \cite{hampel98};
GNO 2000: \cite{altmann00}. The final value of the joint 123 runs was $69.3 \pm 5.5$ SNU, confirming the deficit of neutrino detection \citep{altmann05}, see also table 9.
\medskip

- SAGE 1989-

The Soviet-American Gallium Experiment is located in the underground laboratory (Baksan Neutrino Observatory)
of the Baksan Valley in the Caucasus
mountains (ex-USSR), under about 4700\,hg/\cmd. The muon flux is
about 2 \,/m$\rm^2$/day, i.e. 7 times less than in the Gran Sasso
laboratory. SAGE started to take data in December 1989 with a 30-ton gallium target \citep{gavrin90}.
At that time, the counter
calibration was done with an \fer source. The data analysis is based on a maximum likelihood
method including \Ge decay and a constant background. Identifying no \Ge decay (after about two months of counting each
run), the first result of the SAGE collaboration was an upper limit on the solar neutrino flux but with a strong suspicion of deficiency.

The gallium 
is in the form of gallium metal (57\,tons at the beginning and then 49 tons since 1998) placed
in reactors of 7\,tons each, equipped with stirrers and
heaters that maintain the temperature just above the melting point
(29.8\,C).
The advantage of using the metal is its reduced sensitivity
to both internal and external backgrounds and a smaller volume.
The major disadvantage is the chemical treatment.
The chemistry for the extraction procedure is a little more
complicated than for a liquid \Gacl solution, \citep{gavrin92}
for details, and requires additional steps and somewhat
greater quantities of fresh reagents for the extraction. 
The efficiency is well known and of the order of 80-90\,\% instead of the 95\,\% reached in the Gallex experiment.
Low level radioactivity materials
have also been used in this experiment. The cosmic muon induced
background is negligible, being $<$ 0.5\, SNU.

 A test of the efficiency of the overall procedure has been performed \citep{abdurashitov96}.
An artificial neutrino source (using 200\,g of highly enriched
chromium : 86$\%$ of \Crc)  of radioactivity $\geq$ 1\,MCi
of \Cr, was used to calibrate the detector and to obtain an overall consistency check of
the whole experiment. 
\Cr is obtained by irradiating \Crc with thermal neutrons.
The source was produced from enriched chromium
(40\,\% of \Crc instead of 4.5\,\% in the natural chromium)
in the Silo\'e reactor located in Grenoble, France.

Assuming the solar neutrino production rate was constant during the period of data collection, the combination of 168 extractions from December 2007 until the end of 2008 returns a capture rate of solar neutrinos with energy higher than 233 keV of $65.4-3.0+3.1 (stat) -2.8+2.6 (syst)$ SNU. The weighted average of the results of all three Ga solar neutrino experiments, SAGE, Gallex, and GNO, is now $66.1\pm 3.1$ SNU, where statistical and systematic uncertainties have been combined in quadrature. During a recent period of data collection a new test of SAGE was made with a reactor-produced $^{37}$Ar neutrino source. The ratio of observed-to-calculated rates combined with the measured rates in the three prior $^{51}Cr$ neutrino-source experiments with Ga, is $0.87 \pm 0.05$. A proton-proton flux of $\rm 6.0 \pm 0.8 \;10^{10}/cm^2/s$ was derived.

\subsection{The real time experiments: Kamiokande, SuperKamiokande, SNO a,d KAMLAND}
These experiments have contributed to significantly increase the statistics of the solar neutrino detections. The first experiment, Kamiokande and, later, Superkamiokande, detect the elastic interaction on electrons.  The next generation of experiments, Sudbury Neutrino Observatory, Borexino, and Icarus, aim to measure
the  contributions of both 
the W and Z exchange processes and to detect the elastic interaction on electrons.

\noindent
 - KamiokaNDE, SuperkamiokaNDE experiments: 1986-1995 , 1999- , 2002-

The first real time experiment was
the Kamiokande experiment.
Its principle was based on elastic neutrino scattering:
\begin{math}     \rm
\nu_e\,+\,e^-\,\rightarrow\,\nu_e\,+\,e^-.
\end{math}
 The scattered electron is detected through the Cerenkov
light emitted and
its direction is strongly correlated with
the direction of the incoming neutrino.
The target consisted of a large cylindrical tank
(16\,m high and 15.6\,m diameter) containing 2140 tons 
of ultra-pure water.
The Cerenkov light was captured by about 1000 photomultipliers of
20\,inch diameter covering 20\,\% of the wall surface.
This detector, located in a mine at a depth of 2700\,hg/\cmd 
 of standard rock  \citep{hirata88}, 
 was primarily designed for observing the nucleon
decay with a lower limit of 2.6 $\times$ 10$\rm^{32}$ years. 
It was then surrounded by a water Cerenkov
anticoincidence layer of thickness 1.4\,m 
and started to collect solar neutrino
data at the end of 1986. Its great success was the observation of 12 neutrinos
from the supernova SN1987A on February 23, 1987.

The main background sources were 
the radioactivity in the water
($\rm\beta-decay \; of\; ^{214}Bi$) (E$<$ 9\,MeV), 
$\gamma$ rays from surrounding rock,
and unstable spallation products by cosmic muons (E$>$ 9\,MeV).
These backgrounds do not have any correlation with the
Sun's direction.
The water Cerenkov anticoincidence layer absorbed $\rm \gamma-rays$ from
surrounding rocks and monitored the cosmic muons.
The useful  volume (called fiducial volume)
for the solar neutrino detection was reduced
to 680 tons (about 2.5\,m from the wall of the detector), 
to enhance the signal-to-background ratio.
The electron detection efficiency was 50\,\% (respectively 90\,\%) at 7.6\,MeV 
(respectively 10\,MeV) in 1987 and has continuously improved since.
The selected event sample needed to have a directional  correlation with the Sun. 


 The enhancement around cos $\rm\theta$\,=\,1 (where $\theta$ is the angle between the trajectory of the
electron and the direction of the Sun) corresponded to
solar neutrinos. It constituted  a direct evidence for a detection of solar
neutrinos, but the neutrino energy
spectrum and the experimental electron spectrum both needed to be known to determine the neutrino flux.
 The theoretical 
prediction (including Monte-Carlo simulation and the energy spectrum
calculation) was deduced from \cite{bahcall88}.
The first result of Kamiokande indicated that the
observed signal was less than expected from solar models.

After nine months used to improve the detector,
new data were taken from January 1991 onwards.  The combined value accumulated from 1986-1991 is $0.49 \, \pm 0.05 \, \pm\, 0.06$ \citep{totsuka91} times the prediction of  \cite{bahcall88} and $0.74 \, \pm 0.07 \, \pm 0.09$
times the prediction of \cite{turck88}. 
The trigger rate was about 0.5 per second and the expected solar   
neutrino signal about 0.5 per day above 7.5\,MeV! Consequently, it was not
possible to assign an individual event to a solar neutrino interaction. 

SuperKamiokande is a significant extension of Kamiokande. It uses a tank of 50,000 tons of pure water
providing 65 solar neutrinos per day in a 22,000 tons
fiducial volume \citep{hirata88}. The SuperKamiokande experiment was ready in April 1996. 
Searches for possible day-night and semi-annual variations of
the \B solar neutrino flux have been carried out, using the 1040 days
of Kamiokande II data.
No such short time variations were observed \citep{hirata91}. Taking into account the various uncertainties, this results in
an upper limit of about 30$\%$ on the relative difference in the day/night neutrino fluxes.  The detection results obtained with SuperKamiokande contributed to the Nobel Prize of 2002, partly awarded for the demonstration of the existence of neutrino oscillations.

After full restoration of the detector in 2006 (following implosions of 6,600 photomultipliers in 2001), the energy threshold has been lowered down to 4.5 MeV. The mean values of the Superkamiokande measurements for all three experiment phases are respectively: $ \rm 2.35 \pm 0.08, 2.38 \pm 0.17, 2.31 \pm 0.05 10^6/cm^2/s$ \citep{hosaka06,cravens08}.
The day-night effect, which is a direct test of matter effect, does not clearly show any asymmetry within a 2-3\% error bar, and this error should be further reduced to 1.5\%.


\vspace{5mm}
\noindent
- The Sudbury Neutrino Observatory experiment: 2000-2006

The Sudbury Neutrino Observatory (SNO) experiment (a Canada-USA-UK collaboration) \citep{aardsma87} 
consists of 1000 tonnes of heavy water 
\eaul surrounded by 5\,m of purified light water \eau. 
The Cerenkov light
emitted by electrons
is detected by photomultipliers, like in the Kamiokande experiment.
The project was constructed in a deep mine
(2070\,m underground) near Sudbury, in Canada, and began taking data in
2000. The main advantage of this experiment is its ability to
detect both \nue (through the W exchange process on deuterium) and all
neutrino types (through a Z exchange process producing dissociation of
deuterium into a neutron and a proton). The WEP reaction is detectable via
Cerenkov light from electrons which are produced with an energy nearly equal
to the incoming \nue , thereby providing about 20\,\% energy resolution to
search for possible distortion (through the MSW effect) of the $\rm^8B$ energy
spectrum. The ZEP reaction is detected via the 8 MeV $\gamma$ rays produced
when the neutron is captured in $\rm ^{35}Cl$ from NaCl cycled in and out of
the heavy water. Discrete $\rm ^3He$ proportional counters have been
developed as an alternative neutron detection technique.

The major experimental challenge is to reduce the backgrounds to a very
low level. This requires  the use of low activity  materials: less
than $\rm 10^{-14}\,$ g/g of U and Th in the heavy water;  the high energy $\gamma$
rays from the U and Th chains can photodissociate the deuterium, emitting a
neutron which can simulate a Z exchange process.

This experiment is sensitive only to solar \nB coming from the $^8$B decay, but
presents a lot of advantages: measurement of the neutrino spectrum,
sensitivity to ZEP reactions, and sensitivity to a day-night variation of the solar
flux. SNO was key to proving the existence of neutrino oscillations by estimating the three neutrino fluxes (electron, muon, and tau neutrinos). 
SNO was measuring both neutrino-electron elastic scattering (ES) like Superkamiokande (that means $\nu_e$ + 0.16 $\nu_\tau$ +0.16 $\nu_\tau$) and  the charged-current interaction (CC) on deuterons which is sensitive to $\nu_e$. Therefore it measured for the fist time the sum of all three species, which is directly equal to the number of $\nu_e$ emitted by the Sun. Consequently a direct comparison between predictions of SSM or seismic solar model (SeSM) and detection results is possible. In June 2001, SNO obtained a detection of $\rm 4.95 \pm 0.72 \;10^6 cm^{-2}$ for a prediction of the SeSM of $5.44\pm 0.99$, which was a wonderful success for the two disciplines of neutrino physics and helioseismology. Such result triggered the attribution of the Nobel Prize to the pionnier experiments done in Japan and the USA. It also shed a new light on the SK results which include $0.16\%$ of the two other neutrino species (tau and muon), and explained the strong electronic neutrino deficit detected by the chlorine experiment.

 SNO measured the neutral-current (NC) disintegration of deuterium which is equally sensitive to all active neutrino favors:
$$ \rm \nu_X + D \rightarrow \nu_X' +  n + p $$ 
SNO captured the neutron produced by the NC reaction. In phase I, the neutron was captured mainly by the $D_2O$ of the solution. In phase II, the addition of 2 tonnes of NaCl to the 1000 tonnes of heavy water enhanced the efficiency of the neutron detection by at least a factor two. Moreover the threshold of electron detection for the boron has been reduced to 3.5 MeV. In phase III, the neutron detection  is also done with an $^3He$ proportional counter through the reaction $^3He + n \rightarrow p + ^3T$ (983 events per year instead of 267 for photomultiplier tubes). 
The final results published by \cite{aharmim08} for the SNO collaboration, after the end of the experiment, are :
$$\rm \Phi_{CC}= 1.67 _{-0.04}^{+0.05} (stat) _{-0.08}^{+0.07} ;\, \Phi_{ES}= 1.77 _{-0.21}^{+0.24} (stat) _{-0.10}^{+0.09} ;\, \Phi_{NC}= 5.54 _{-0.31}^{+0.33} (stat) _{-0.34}^{+0.36} \;10^6/cm^2/s$$
The charged current is only sensitive to electronic neutrinos, the elastic scattering is sensitive to all neutrinos but is dominated by the cross section of the electronic neutrinos which is 6 times greater than the cross section of $\nu_\mu$ and  $\nu_\tau$, and the neutral current allows a count of all the neutrino species.
These values are compared to predictions in table 6. Altogether, these seminal results demonstrating the detection of all the neutrino species and contributing to estimate the oscillation impact on the boron neutrinos, fostered a lot of prizes. They also put strong constraints on the neutrino oscillation parameters because SNO delivered different fluxes (electronic neutrinos alone or the sum of the three flavors, see section 6.5).  Finally the electronic-neutrinos boron flux corresponding to the total period of detection of SNO  is 5.046 $_{-0.152}^{+0.159} (stat) _{-0.123}^{+0.107} \;10^6/cm^2/s$
\cite{aharmim10b}. SNO was turned off on 28 November 2006 even though it also has the potential to detect atmospheric and supernova neutrinos, but with a lower rate than Super Kamiokande. SNOLAB operates other experiments and prepares the SNO+ detector. 

\vspace{5mm}
\noindent
 - The KAMLAND detector: 2002- 
 KAMLAND is the largest (1000 t) liquid scintillator detector. The objective of this detector is to measure not only the solar neutrino low energy spectrum but also to detect anti-neutrinos. It is located in the Kamiokande underground mine providing a cosmic-ray muon attenuation by a factor of 10$^{5}$ with respect to the earth surface level. The central spherical vessel of 13 m diameter  containing the liquid scintillator, is suspended in a transparent buffer oil. The photomultipliers (1325+554) cover 34$\%$ of  $\pi$ . An outside tank,  filled with 3200 tonnes of purified water, serves as a radiation shield against neutrons and $\gamma$ rays from surrounding rocks. Since 2002, the beginning of the data collection, KAMLAND has used nuclear reactors to perform neutrino oscillation measurements and geoneutrino detection. The first two years have led to 258$ \bar \nu$$_e$ candidate events with energies above 3.5 MeV, leading to a deficit of detection \citep{haraki05}. The distortion of the spectrum puts very strong constraints on the oscillation parameters (Figure 10). See section 6.5.

\subsection{Toward a neutrino spectrum: the BOREXINO experiment: 2007- }
\begin{figure}
\center
\includegraphics[width=6.7cm]{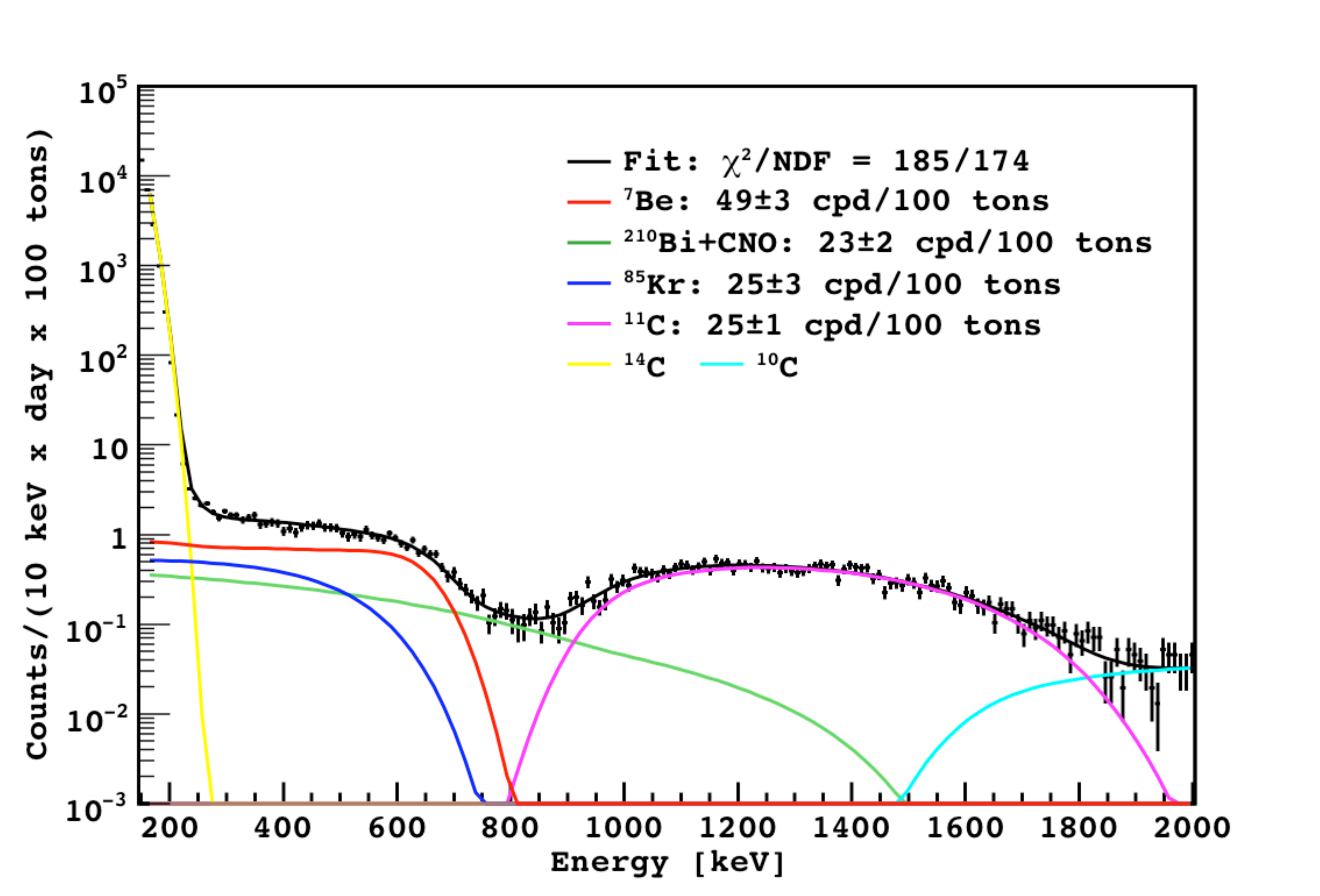}
\includegraphics[width=6.7cm]{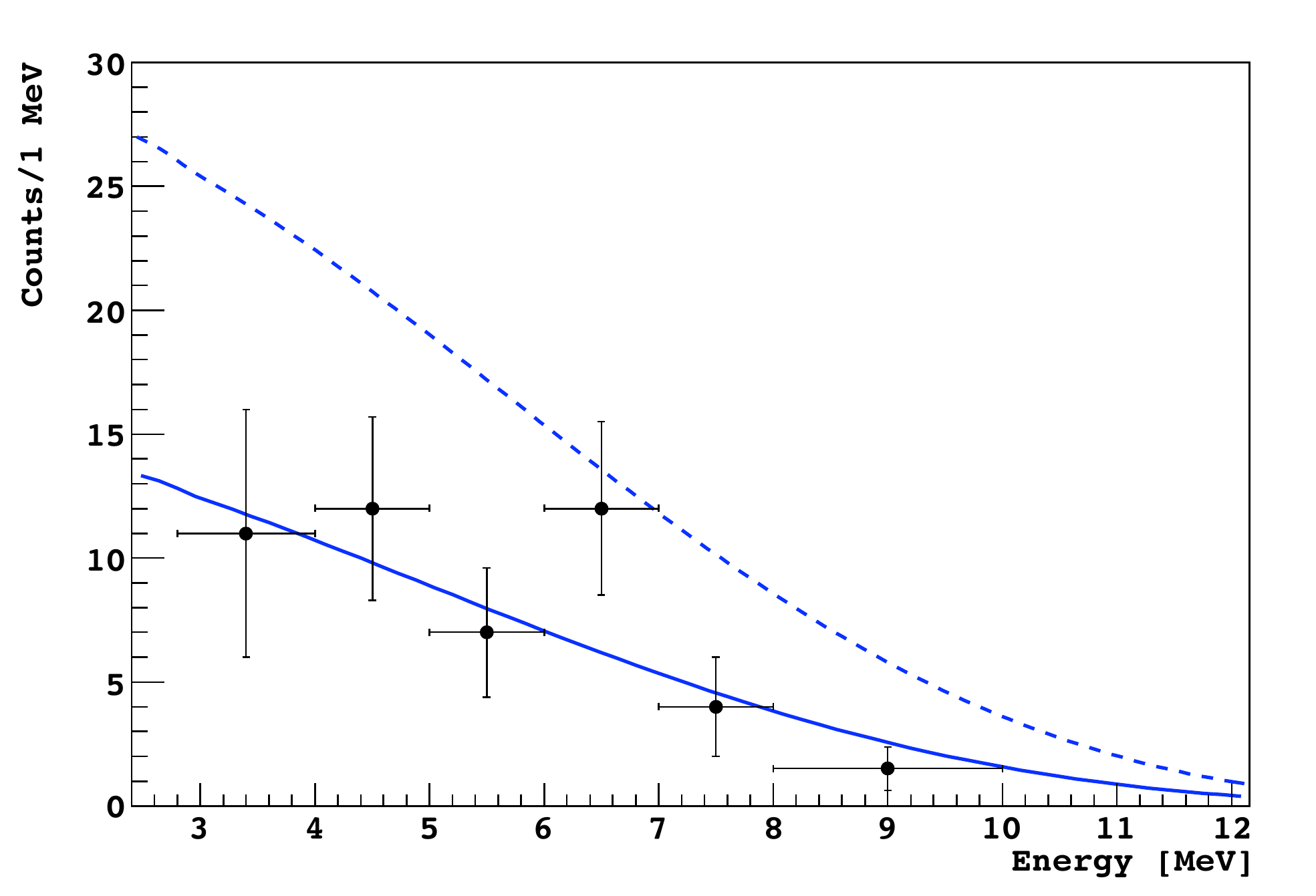}
\vspace {-4cm}
\includegraphics[width=10cm]{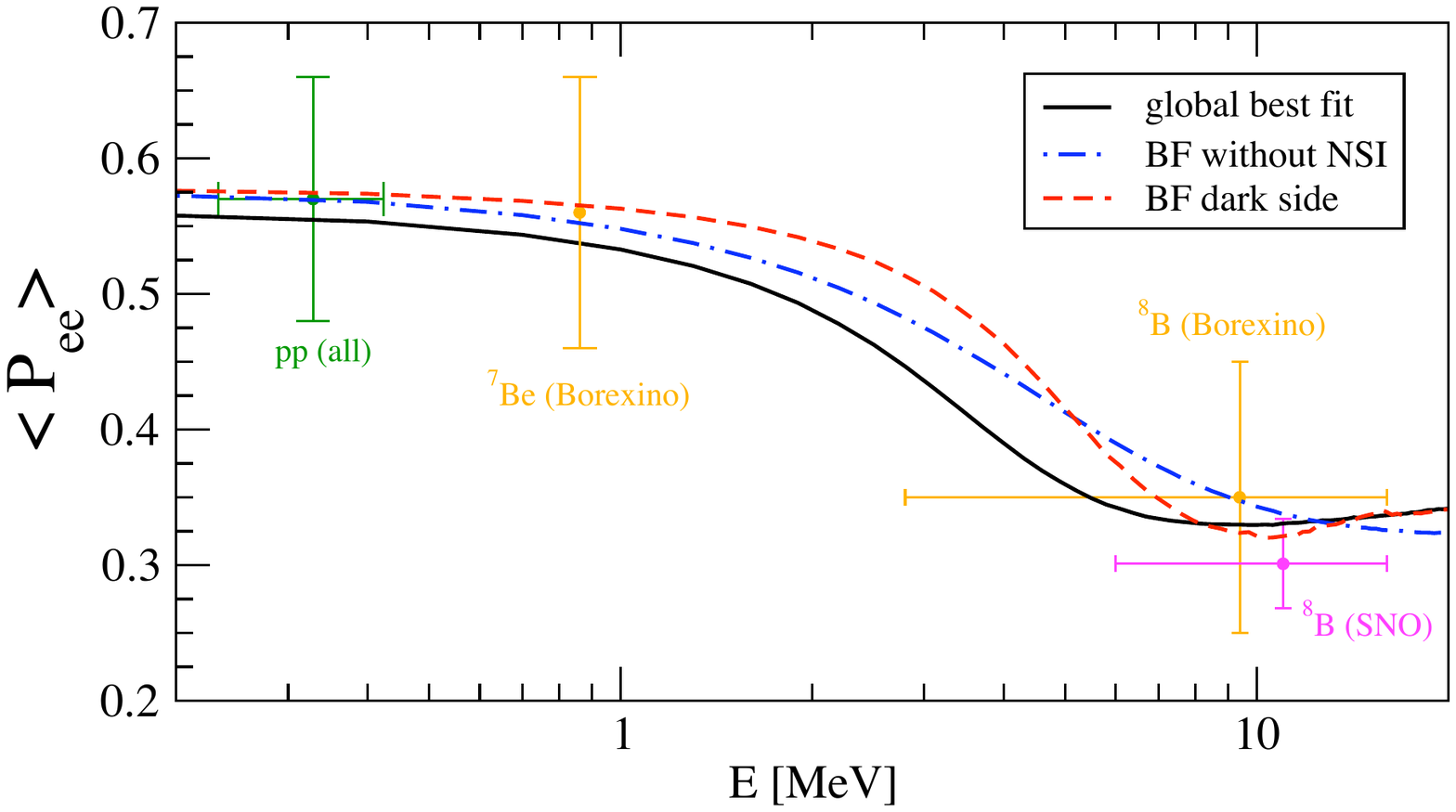}
\vspace {-1cm}
\small
\caption{
Top: Results of the first year of BOREXINO that shows the largest spectrum ever reached and the electron neutrino survival probability. From \cite{arpesella08,bellini08}. Bottom: Agreement between the measurement survival probabilities for the different sources of neutrinos and different predictions. From \cite{escrihuela09}.}
\normalsize
\end{figure}
The  Borex  project \citep{raghavan88}  
is another example of real time experiment. Nominally, it
involves a large tank containing 2000 tons of
borated liquid scintillator (200 tons of \bonze)  immersed in pure
water. It is supported
by a USA-Europe collaboration in the Gran Sasso underground laboratory
(about 3300 m of water equivalent) in Italy. A major difficulty is 
to obtain a very pure liquid scintillator. The WEP reaction
is marked by an electron in coincidence with a photon. 
A first step of Borex, called Borexino (``il piccolo Borex''), is currently
running  in the Gran Sasso laboratory with a detector of 278 tons of pseudocumene doped with 1.5g/l of a fluorescent dye of diphenyloxazole \citep{alimonti09}. 

Borexino aims to clarify the nature of the solar electronic neutrino oscillation by looking
at the \Be and \B neutrinos with the same detector.  The first objective of Borexino is to properly detect the \Be neutrinos because it is the first experiment to report a real-time observation of the low energy spectrum (between 0.2 to 4.5 MeV, see figure 9 top left). The nominal counting rate is of 1500 counts per day dominated by the residual muon flux.
These muons are rejected at the 99\% level by an outer detector. $^{208}$Tl and $^{214}$Bi are properly rejected. The pollution by neutrons is reduced to 10$^{-4}$  counts per day.   The 0.862 MeV $^7Be$ neutrinos signature is given by the Compton-like edge of the recoil electrons at 665 keV. 
After 192 days, the $^7Be$ flux is estimated at 49$ \pm$ 3 counts per day per 100 tons with a proper determination of all  the contaminants ($^{85}$Kr, $^{210}$Bi and $^{11}$C). This corresponds to a neutrino flux of $\rm 3.36 \pm 2 \pm 0.27 \times 10^6/cm^2/s$ \citep{arpesella08}.  Compared to the solar model predictions (see for example table 9), the neutrino-flux deficit is of only about 29\%, much less than the 50\% (se the same table) obtained for the boron flux. This confirmed the energy dependence of the survival probability of the electronic neutrinos (see below).
Moreover, Borexino also detects about 0.26 counts per day per 100 tons of boron neutrinos between 2.8 and 16.3 MeV, with 0.14 counts between 5 and 16.3 MeV (figure 9 top right). It is the first experiment which detects the boron neutrinos with a liquid scintillator detector and a threshold of 2.8 Mev, obtaining 75 $\pm$ 13 counts. The first analysis gives an elastic flux for boron  of $2.4 \pm 0.4 \pm 0.1$ \citep{bellini08,bellini10}, in good agreement with the previous measurements. Borexino also detects the other neutrino sources pp, pep, CNO  neutrinos and $^{11}C$ and $^{11}B$.

\subsection{Neutrino predictions from the standard and seismic models}
The excellent quality of the sound-speed data derived from GOLF+MDI allowed us to produce a Seismic Solar model (SSeM) that takes into account this sound-speed profile to avoid any possible theoretical bias in the neutrino predictions. Such biases result from the hypotheses applied to build the standard model. The seismic model was obtained by modifying some physical inputs (mainly reaction rates and opacity) within their error bars. The seismic model contributes  to stabilizing both the neutrino fluxes  and gravity-mode frequency predictions. Until a complete Solar Dynamical Model (SDM) can be built, the solar seismic model is the best way to deal with dynamical effects that are unaccounted for in 1D. This model is shown in Figures 7 and 8 and is defined in \cite{turck01b} and \cite{couvidat03}. Table 6 recalls how the boron neutrino prediction has evolved in parallel with the improved description of the microscopic physics of the solar models. Table 9 compares the seismic model predictions to  the results of all the detectors. It is noticeable that the most recent prediction of the SSM is not that close to the detected flux  \citep{turck04a,bahcall04}.

\begin{figure}
\centering
\includegraphics[width=7.5cm]{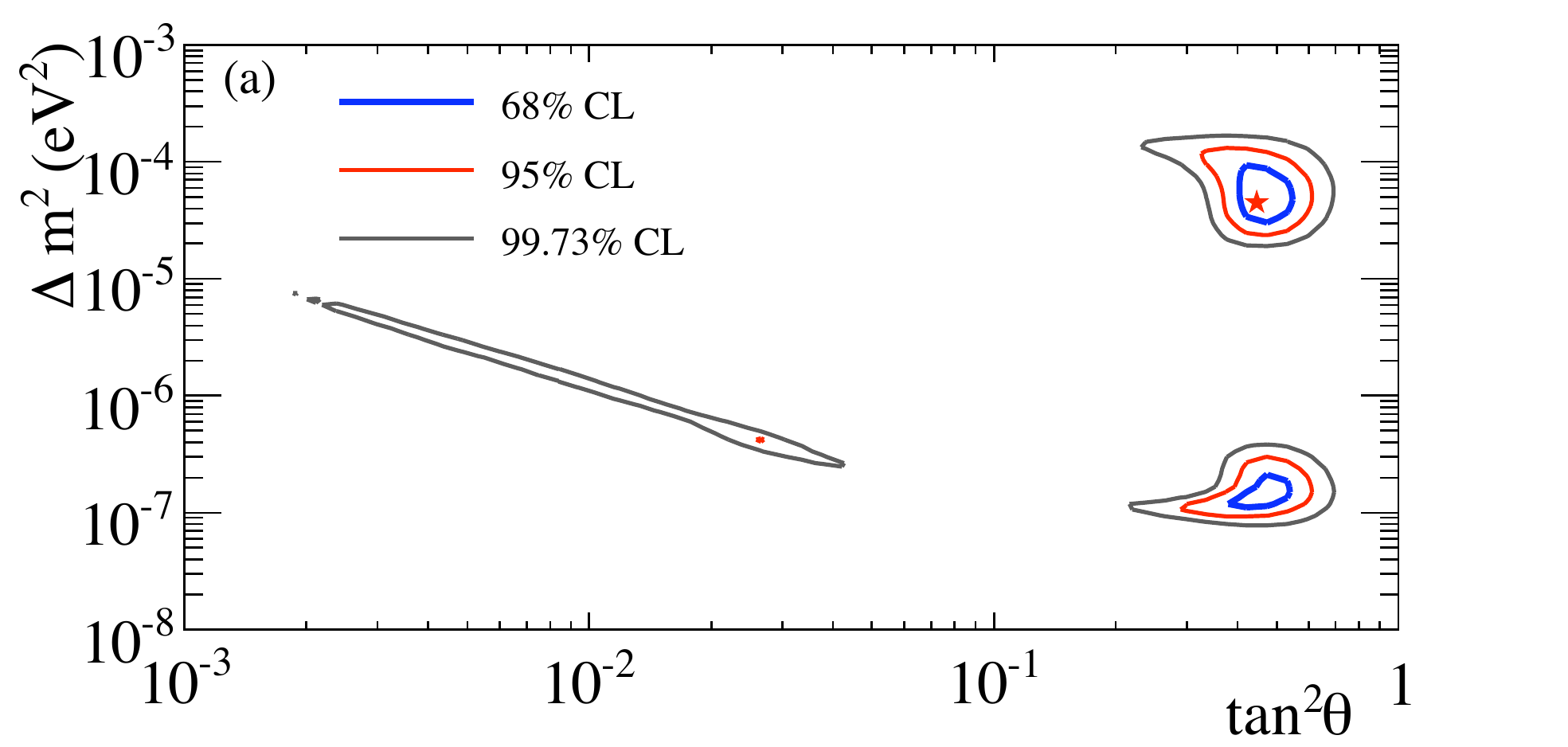}
\includegraphics[width=7.8cm]{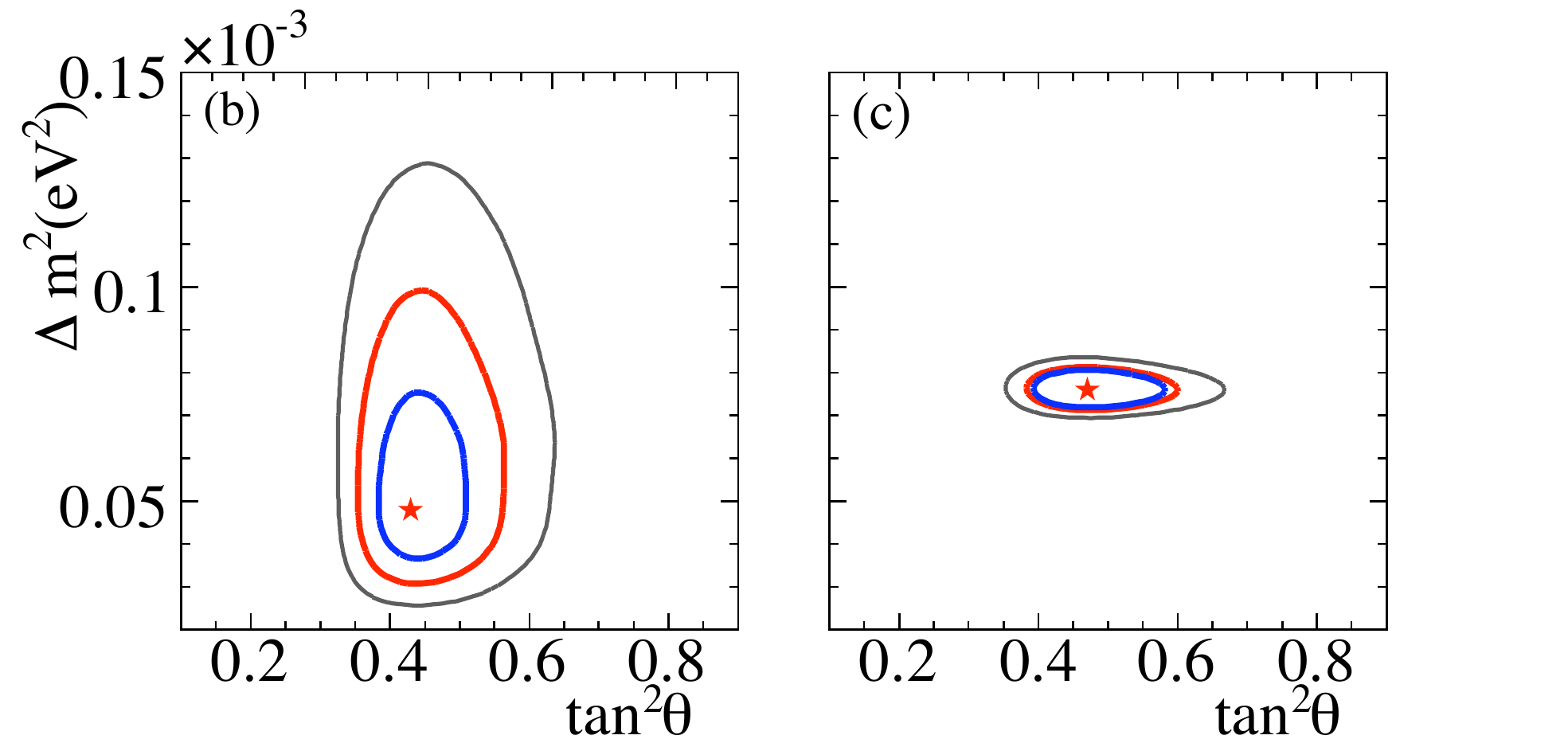}
\caption{\small Neutrino oscillation contour. Left SNO only;  middle: all the solar neutrino detectors: SNO, SK, Cl, Ga, Borexino; right: Solar neutrinos + KAMLAND. From \cite{aharmim08}.}
\end{figure}       

\begin{table*}[htb]
\vspace{-5mm}
\caption{Neutrino predictions of the present standard solar model and of the seismic solar model compared to radiochemical detections and real time experiments (including  BOREXINO) {\bf given in bold characters}. See text for the reference of the different experiments.
}
\label{table:7}
\newcommand{\m}{\hphantom{$-$}}
\newcommand{\cc}[1]{\multicolumn{1}{c}{#1}}
\renewcommand{\tabcolsep}{0.2pc} 
\renewcommand{\arraystretch}{1.} 
\begin{tabular}{@{}lll}
\hline
         & Predictions without & Predictions with \\
      &      neutrino oscillation & neutrino oscillation \\
\hline
{\bf HOMESTAKE }  &     &    {\bf 2.56 $\pm$ 0.23 SNU}\\
Standard model   2009      & 6.315 SNU & 2.24 SNU  \\
Seismic model   		  & $7.67 \pm 1.1$ SNU  & {\bf 2.76$\pm 0.4$ SNU}  \\
\hline
{\bf GALLIUM detectors}         &    GALLEX & 73.4 $\pm$ 7.2 SNU\\
					& GNO & 62.9 $\pm$ 5.4 $\pm$ 2.5 SNU\\ 
					& {\bf GALLEX + GNO}
& {\bf  67.6 $\pm$ 3.2 SNU}
\\
				& {\bf SAGE} &{\bf 65.4 $\pm$ 3.3 $\pm$ 2.7 SNU}\\
				& {\bf GALLEX+GNO+SAGE} &{\bf 66.1 $\pm$ 3. SNU}\\
Standard model   2009               & 120.9 SNU & 64.1 SNU \\
Seismic model   & $123.4 \pm 8.2$ SNU & {\bf 67.1 $\pm$ 4.4 SNU}  \\
\hline  
{\bf BOREXINO $^7$Be}   &     &   {\bf  3.36 $\pm$ 0.36 10$^{9}$cm$^{-2}$s$^{-1}$}\\  
Standard model &   & \\
Seismic model & 4.72 10$^{9}$cm$^{-2}$s$^{-1}$   &   {\bf 3.045 $\pm$ 0.35 \;10$^{9}$cm$^{-2}$s$^{-1}$} \\
\hline
\end{tabular}\\[2pt]
\begin{tabular}{@{}ll}
\hline
{\bf Water detectors}      & Predictions or Detections       $B^8$  electronic neutrino flux          \\
  {\bf SNO }& {\bf 5.045 $\pm$ 0.13 (stat)$ \pm$ 0.13 (syst) \;10$^{6}$cm$^{-2}$s$^{-1}$}\\
{\bf SNO +SK} &  {\bf \;5.27 $\pm$ 0.27 (stat)$\pm$ 0.38 (syst) \;10$^{6}$cm$^{-2}$s$^{-1}$}  \\
Standard model 2009  &    4.21   $\pm$ 1.2   \;10$^{6}$cm$^{-2}$s$^{-1}$     \\
Seismic model &  5.31 $\pm$ 0.6 \;10$^{6}$cm$^{-2}$s$^{-1}$\\
\hline
$B^8$ neutrino flux  &electronic + other flavors in 10$^{6}$cm$^{-2}$s$^{-1}$ \\ 
{\bf SK1 (5 MeV)}&{\bf 2.35 $\pm$ 0.02 (stat) $\pm$ 0.08 (syst)}\\ 
{\bf SNO D$_2$O (5 MeV)}  & {\bf 2.39 $\pm$ 0.23 (stat) $\pm$ 0.12 (syst)} \\ 
{\bf BOREXINO (2.8 MeV)}  &{\bf 2.65 $\pm$ 0.44 (stat) $\pm$ 0.18 (syst)}  \\
\hline
\end{tabular}\\[2pt]
\end{table*}

As shown in Tables 6 and 9, the seismic solar model produces flux predictions very close to the detected neutrino fluxes, and it has been prominently used as a guide in the search for gravity modes (\TC, 2004b, Garcia et al., 2007). It also contributes to estimating the quality of predictions of the standard solar model. We consider that this seismic solar model is the only one for which we can reasonably put some error bars on the neutrino fluxes. These error bars come from the measurements of the different physical inputs introduced in stellar equations. Any error bar attributed to a standard solar model prediction can only be considered as a minimum error because it cannot be excluded that this model neglects important additional processes. See section 6.5 for the oscillation parameters used.

\subsection{Comparison with predictions including mixing parameters}
Clearly, the comparison of all these neutrino detection results with the predictions of the seismic solar model or the standard solar model demonstrates the existence of the MSW effect (section 3). Figure 9 shows  the energy dependence of the neutrino oscillation due to matter or vacuum effects. Again, this is a very important result produced by the solar neutrino community. The predictions of Table 9 include the values of \cite{bahcall04} of $\rm \Delta m_{12}^2 = 7 10^{-5} eV^2$ and $tg^2 \theta_{12}= 0.45$ (neutrino mass differences and mixing angles). All the predictions agree well with the detection results, and this agreement is much better than the one obtained from the standard solar model including the new updated CNO abundances \citep{turck04a,bahcall06}.
Figure 10 \citep{aharmim08}
summarizes how the oscillation parameters for the neutrino mass-eigenstate couple (1,2) are more and more constrained by the different results, and how these results lead (independently of solar models) to the values: $\rm \Delta m_{12}^2 = 7.59 10^{-5} eV^2$ and $tg^2 \theta_{12}= 0.468$.


\section{Beyond the standard solar model: a dynamical view of the Sun}
The equations used in the SSM framework (sections 4-5) do not consider the internal rotation and magnetic field.  Twenty 
years ago, such a restriction was justified  and did not prevent the description of the main
phases of stellar evolution. However the description of the early and late stellar-evolution stages needs to be furthered. Young stars sustain a very strong magnetic field \citep{feigelson03}  and massive stars are strongly deformed by centrifugal forces and rapid rotation up to 50\% 
like in the case of Achernar \citep{desousa03,skumanich04}.  Moreover, the Sun  is clearly a magnetic star (see the variability of the fundamental quantities given in Table 3) and its activity impacts the Earth environment. Because the origin of solar magnetism is internal, it is necessary to introduce the effects connected to the internal
rotation and magnetic fields in  solar models, to first replace the SSM, then the SeSM,  by a more realistic DSM (Dynamical Solar Model). The most recent observations obtained with the SoHO satellite and the ground-based experiments contribute to the development of this new field. The Sun is rotating differentially at the surface, in roughly 24 days at the equator and 30 days at the poles. The internal rotation has a well known effect on the dynamo process which is characterized by variable activity cycles of about 11 years with large fluctuations in intensity and duration. Therefore the knowledge of the solar internal rotation profile is a crucial ingredient of the dynamics of the Sun which will help to understand the solar activity.
\begin{figure}
\centering
\hspace {2.5cm} \includegraphics[width=7cm]{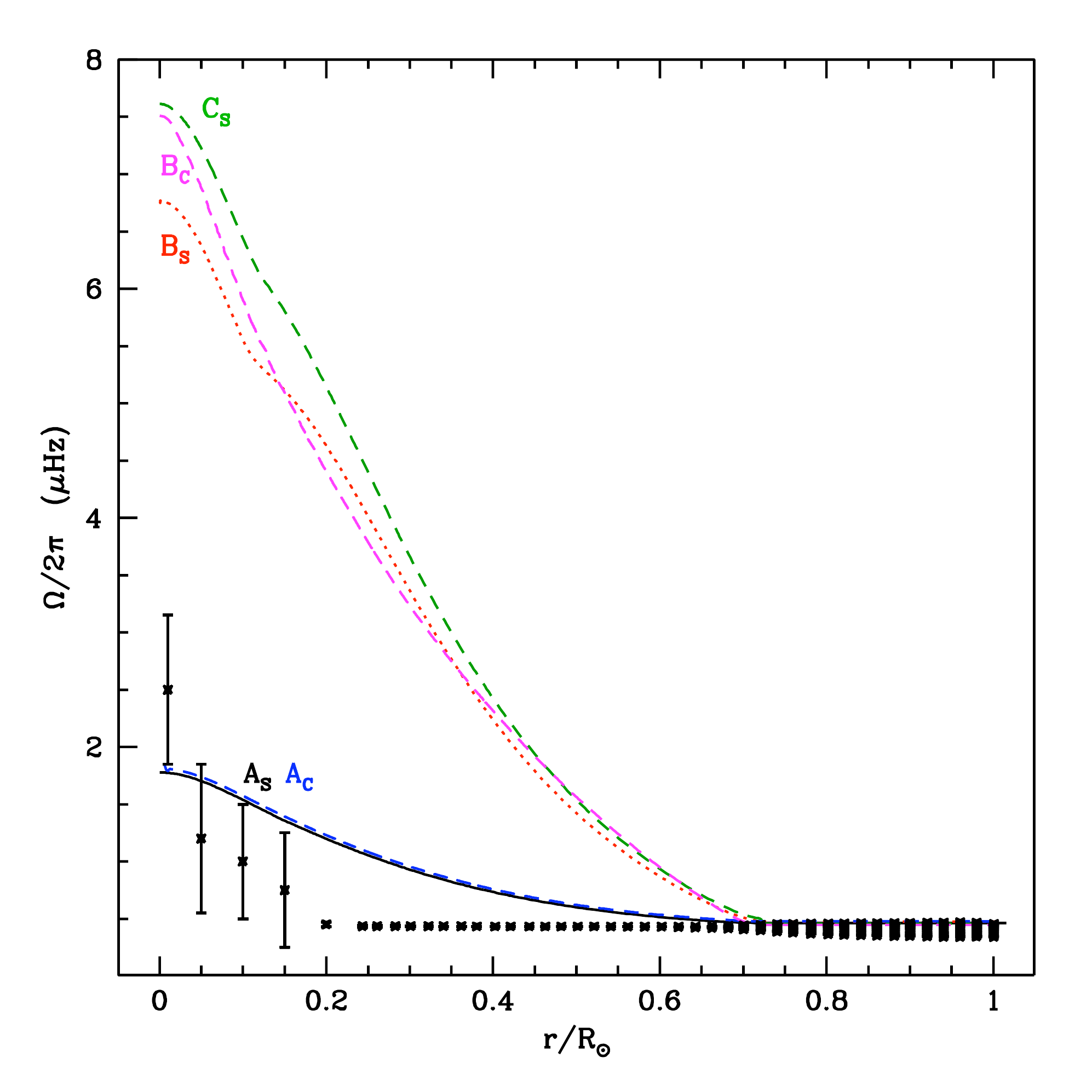} \hspace {2.5cm} 
\includegraphics[width=6.7cm]{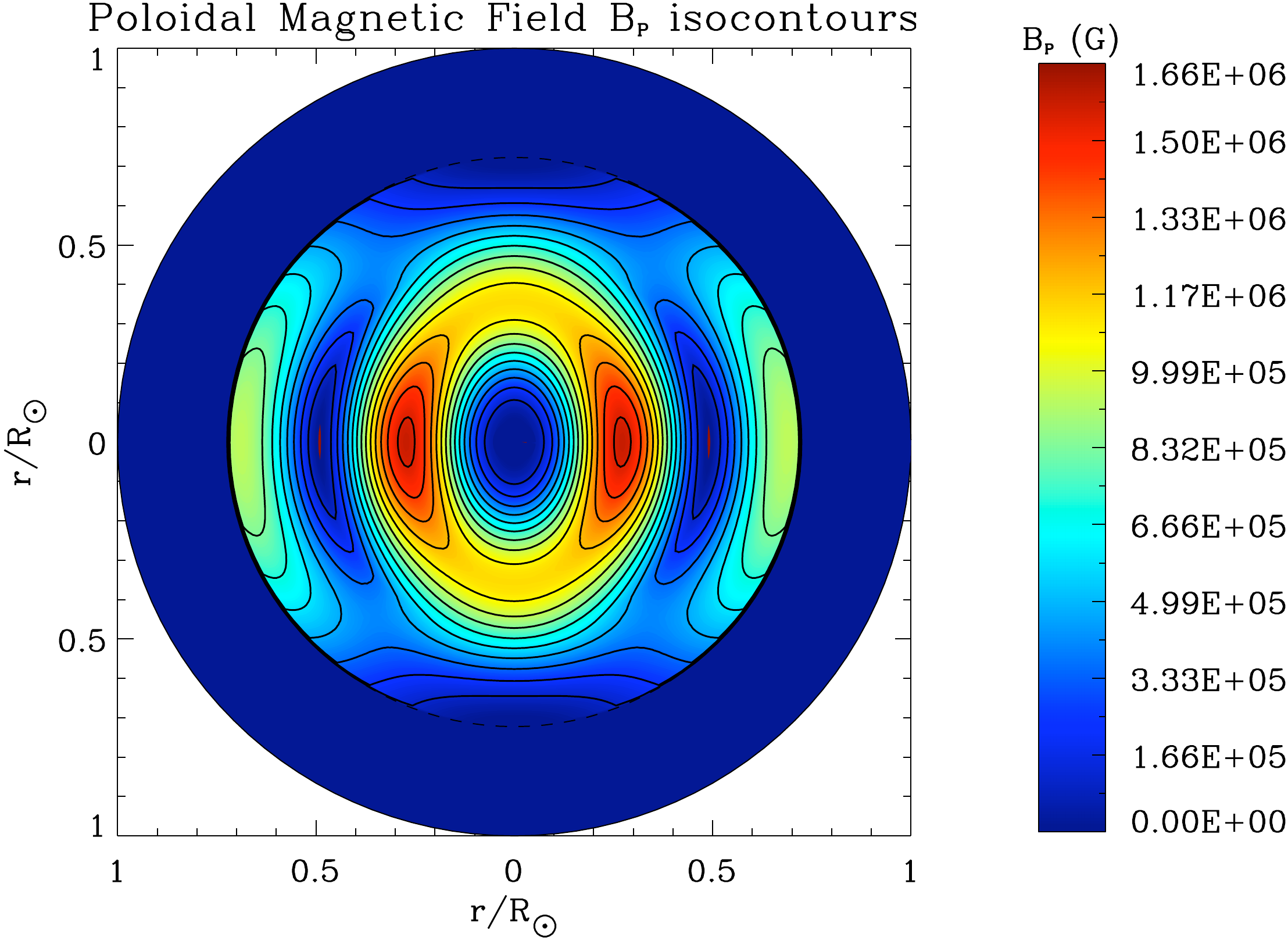}
\includegraphics[width=6.7cm]{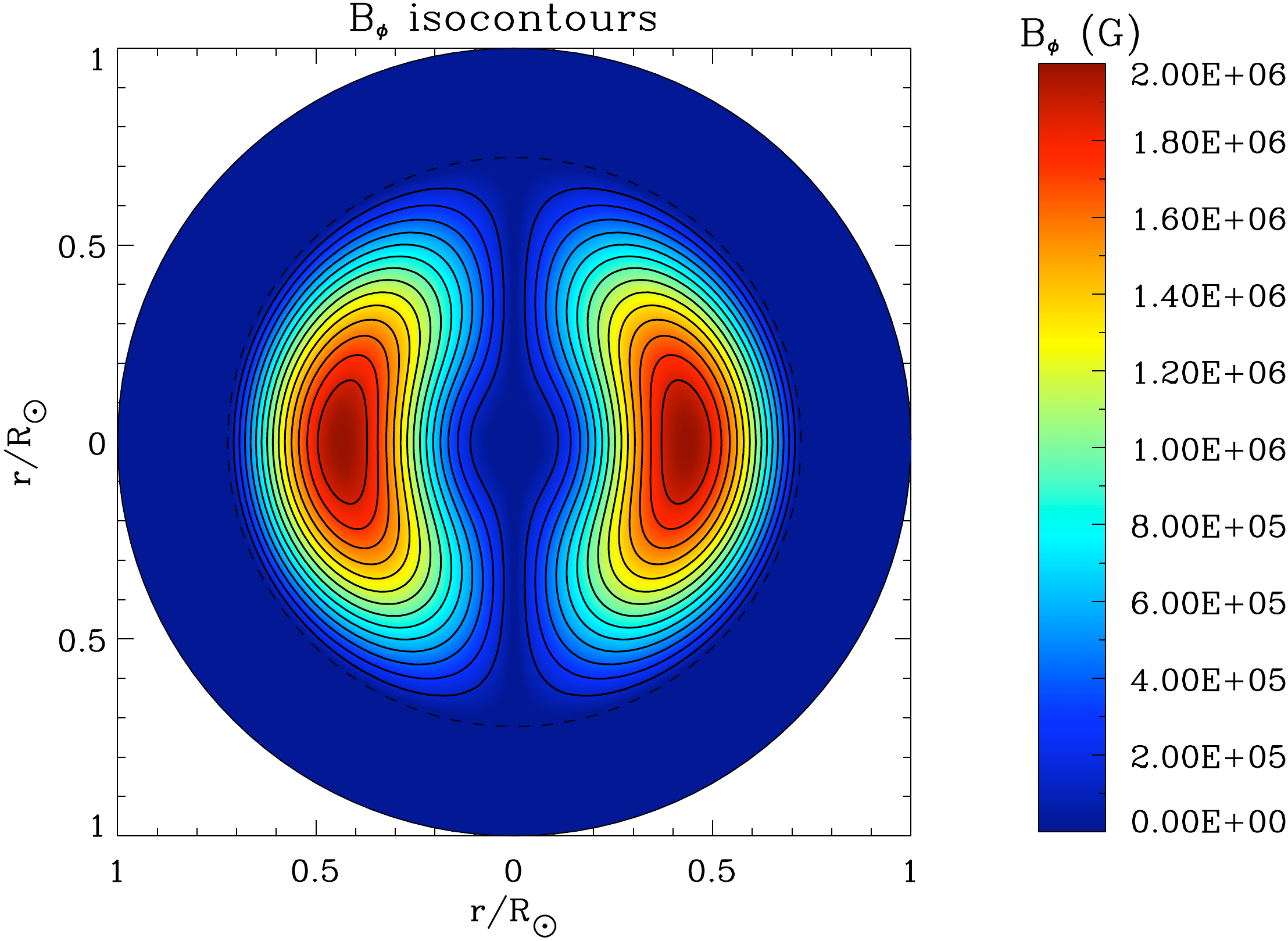}
\caption{\small {\bf Top}: Comparison between the solar internal rotation profile predicted by
    different solar models and the one deduced from helioseismology. From \cite{turck10a}. For models B$_S$ and 
C$_S$, the plotted value is $\Omega/2\pi$-0.2325 rather than 
$\Omega/2\pi$ in order for the surface value they reach to match the one 
reached by model B$_C$. The data points down to r/R$_\odot$ = 0.2 are deduced  from the acoustic mode splittings determined
by the observations of the GOLF, MDI and GONG instruments, from \cite{effdarwich08,garcia11}. The data points in the core correspond to the core rotation extracted from the potential
gravity modes observed. 
 Bottom: Potential configuration of a fossil field in the radiative zone of the Sun: left Isocontours of the poloidal magnetic field $\rm B_P(r, \theta)$ in G in meridional cut. c) Isocontours of the azimuthal magnetic field
$\rm B_\phi(r, \theta)$ in G in meridional cut. From \cite{duez10}.
}
\end{figure}

\subsection{The internal rotation and its consequences}
The Sun is the only star for which we can access the internal rotation in detail, thanks to the million of oscillation modes (million if we treat the different azimuthal orders $m$ as separate) observed at its surface.  The detection of the different components of the observed oscillation modes allows the determination of the solar interior rotation profile. 
$$ \Delta \nu_{n, \ell, m} = \rm {1 \over {2\, \pi}} \int_0^R \int_0^\pi K_{n, \ell, m} (r, \theta) \Omega(r, \theta) d r \,d \theta + \epsilon_{n, \ell, m} \eqno{7.1}$$

\subsubsection{Rotation profile in the radiative zone}
 Until recently, the only motion included in SSM was the 
gravitational settling (GS)  (see section 4.3.4). This phenomenon describes the temporal migration of the heavy elements in comparison with the lighter ones toward the solar center. It is a very slow process, and equation (4.19) uses a diffusivity coefficient $D_i \sim$ 10 cm$^2/s$. For comparison, the turbulent diffusivity $D_T$  reaches 10$^5$ cm$^2/s$ at the base of the convective zone \citep{brun99}. GS and turbulence operate during the entire life of the Sun and consequently the current solar
photospheric composition differs  from its initial value by about 10\%. Helioseismology, by measuring an helium content  at 0.9 R$_{\odot}$ of 0.243 that differs from the initial helium of the model of 0.277, validates this process. See table 6 for  the  
impact of GS on the neutrino predictions.

The rotation profile in the radiative zone has been difficult to establish. This is partly due to the small number of modes that reach this region (acoustic modes of degree $\ell < 40$) and the correction that needs to be applied  to reduce the influence of the external differential rotation. Figure 11 (top panel) summarizes some current estimates. The detection of asymptotic dipolar gravity modes \citep{garcia07} favors, like the first gravity-mode candidates of \cite{turck04b}, an increase of the rotation rate in the core. The very central region seems to rotate faster than the rest of the radiative zone by between 5 to 8 times the corresponding solid rotation rate of 430 nHz \citep{effdarwich08}. 430 nHz corresponds to the rotation rate of the convective zone at a latitude of 40 degrees. The transition between the radiative and convective zones generates an horizontal turbulent flow in a narrow region. Such hydrodynamic phenomenon occuring in a transition region from differential to solid-body rotation gave the name of tachocline to the corresponding layers \citep{spiegel92}.  
We studied in great details \citep{turck10a} the transport of angular momentum induced by the rotation, by following the evolution of the Sun since the initial phase of contraction and introducing in the angular-momentum equation a term of advection and a term of diffusion \citep{zahn92,mathis05}. We note that the rotation profile is mainly established during the early evolutionary phase and the present rotation rate favors the idea that the young Sun was a rather slow rotator. We also observe that there is a meridional flow in the radiative zone of the present Sun, but with a very low velocity of the order of 10$^{-6}$ 10$^{-7}$ cm/s, unlike the velocities observed at the top of the convective zone (order of a few m/s). This naturally produced interface layers (tachocline) which are both very turbulent and dominated by horizontal velocity. It was also shown that the presence of such a rotation profile has almost no impact on the sound-speed profile, but certainly sustains an internal deep magnetic field.
   
   \subsubsection{Meridional Circulation in the convective zone}
It is known since Galileo Galilei that the Sun rotates differentially in latitude and that the  equator rotates faster by 30\% than the poles, partly due to the thermal instability generated by the global rotation. Moreover it has been measured that such differential rotation is maintained in the entire convective zone \citep{kosovichev97,thompson03}.
The meridional circulation (MC) of a large-scale flow  velocity of 10-15 m/s, is observed at the solar surface and directed from the equator to the poles \citep{duvall79}.
A return flow (still unobserved) must take place deeper in the convective zone, from the poles to the equator.
This MC is probably playing an important role in the generation of magnetic field by the dynamo effect ({\it e.g.} \citep{brun09}.
Indeed, it transports not only matter but also magnetic flux (and angular momentum) with a timescale close to the solar cycle period (11 years), a crucial ingredient in flux-transport dynamo models \citep{choudhuri95}. Local helioseismology is conveniently used to invert the flow amplitude and direction in the upper solar layers, down to a few Mm below the surface.
For instance, ring-diagram analysis \citep{haber02} reached a 15 Mm depth. Time-distance analysis \citep{zhao04} confirmed their result but failed to detect any return flow.

\subsubsection{Zonal Flows in subsurface layers}

Zonal flows, or torsional oscillations \citep{howard80}, are latitudinal bands of modulation of the rotational velocity.
The mid-latitude bands move toward the equator during the activity cycle, while the high-latitude bands move toward the poles.
The amplitude of these velocity modulations is of the order of 5 m/s.
They are thought to be produced by feedback from the dynamo effect on differential rotation \citep{spruit03}.
\cite{kosovichev97} detected the zonal flows in helioseismic data using surface-gravity waves or f-modes.
Global helioseismology can only detect symmetrical patterns around the equator.
It was shown that the torsional-oscillation pattern reaches deep inside the convective zone.
Local helioseismology is needed to detect potential differences between the two solar hemispheres.
It gives us access to these torsional oscillations and their depth dependence.
The evolution of the zonal flow pattern during the solar minimum is being studied \citep{howe09} and  is now used to forecast the strength of the upcoming solar cycle. Indeed, unlike the butterfly diagram that was traditionally used to observe the evolution of the solar activity cycle, the torsional oscillations are clearly visible even in absence of sunspots (which do not always emerge). The solar magnetic activity never stops.

\subsection{The internal magnetic field}
The interaction between internal magnetic field and neutrinos would be very interesting to study \citep{couvidat03,rashba07}, but this phenomenon is probably small in the solar interior.

We believe that a fossil field generated during the early phases of the formation of the Sun, when it was totally convective, may have survived up to the present age. It is now well established that if such a field still exists, it diffused over a timescale of Gyrs and must be constituted of a mixture of poloidal and toroidal fields \citep{zahn07,duez10}. If it exists, we must determine its amplitude and its role on the variability of the solar activity, and on the potential magnetic interaction with neutrinos. 

The discovery of the so-called tachocline region emphasized the urgent need for the solar 
community to better understand the magneto-hydrodynamics of the radiative interior, and of the tachocline, and their
coupling to the upper solar layers. It now seems well established that the tachocline plays a crucial role in the 
solar dynamo, since it is most likely the layer where the mean toroidal magnetic field, 
thought to be at the origin of the surface sunspots and butterfly diagram, 
is streched, amplified (by at least a factor of 100) and stored until it
becomes magnetically buoyant \citep{parker93,rempel03}. 
However little is known about the dynamics of the solar tachocline: is it turbulent or laminar, what 
types of circulation are present, what is the dynamical influence of the magnetic fields, why is 
it so thin (extending at most over 5\% of the solar radius)? \cite{spiegel92} were the 
first to address directly some of those questions but in the purely hydrodynamical context. 
They showed that if no process were present to oppose its radiative spread, the solar tachocline 
would extend over 30\% of the solar radius after 4.6 Gyr, in complete contradiction with current helioseismic 
inversions. They demonstrated that the anisotropic turbulence could hinder the 
spread of the solar tachocline to only a few \% of the solar radius. 
\cite{elliott97} confirmed their results numerically with a 2D axisymmetric
hydrodynamic code. \cite{miesch03}, using a thin layer version of the ASH code, showed that 
the coupling between randomly forced turbulence and an imposed shear flow gives rise 
to Reynolds stresses that transport angular momentum such as to reduce the shear. 
However several authors \citep{rudiger97,gough98,macgregor99,garaud02,garaud08} proposed that the magnetic torques 
exerted by a weak internal fossil magnetic field (if it exists) could oppose the inward 
thermal hyperdiffusion (or viscous diffusion when thermal effects are neglected) 
of the solar tachocline. Such models favor a slow, rather laminar version of
the tachocline. Gilman and collaborators \citep{dikpati04} developed a series of models that showed that the tachocline could become unstable through
magnetic instability of toroidal structures embedded within it, resulting
in a latitudinal angular momentum transport that suppresses the shear and limits its inward 
diffusion. This increasing number of solar tachocline
models demonstrates how important it is to characterize the dynamical properties 
of the solar radiative interior. 

 With the advent of powerful supercomputers, non-linear
studies of the solar radiative interior and tachocline in full 3D MHD simulations became possible, 
but they lack the important and more detailed observational constraints that are required to 
progress in our understanding of the magnetohydrodynamics of the solar radiative interior.
A space project like DynaMICCS \citep{turck08}, constraining the solar radiative dynamics, is
clearly required to support and guide the 3D simulations.

\subsubsection{Time-distance helioseismology to study subsurface activity}
Distinct branches of local helioseismology have emerged: time-distance analysis \citep{duvall93}, ring-diagram analysis \citep{hill88}, and acoustic holography \citep{lindsey90} are the most widely used techniques. Local helioseismology is especially conducive to the study of the dynamics of the solar upper layers. For instance, it allows the analysis of meridional circulation, zonal flows, emergence of active regions, supergranular flows, and other phenomena closely related to the generation of magnetic fields by dynamo effect in the solar interior.
The potential impact of these large-scale magnetic fields on the propagation of solar neutrinos (through, for instance, the resonant spin-flavor precession process) makes local helioseismology directly relevant to this review.
For the sake of brevity, we will only introduce the time-distance analysis.

Time-distance analysis measures the travel times of wavepackets propagating between different points on the solar surface using a cross-covariance technique.
These wavepackets can be acoustic or surface-gravity waves. If the conditions on the photosphere were similar to the ones encountered on the Earth crust, it would be possible to follow the temporal evolution of a wave excited by a ``sunquake'' and measure its relevant travel time(s) without elaborate mathematical techniques.
However, the perpetual stochastic excitation of waves by millions of convective cells of the upper solar layers makes any attempt at directly measuring any travel time of a specific wavepacket quite challenging. Hence the need to compute cross-covariances.

We shall review a few results relevant to the potential interaction of neutrinos with the solar magnetic fields, because these results can (and are) used to discriminate between different theories of the formation of large-scale fields in the Sun.

\subsubsection{Structure and Dynamics of Sunspots}

Sunspots are regions of the solar surface that appear darker than their surrounding due to a lower temperature (see, {\it e.g.}, the review by \cite{solanki03}.
They are regions of intense magnetic fields (up to about 3000 Gauss in the umbra) and are directly connected to the generation of a toroidal field by the dynamo effect.
Although sunspots have been known for centuries, many details regarding their nature are still elusive.
They are thought to be the consequence of a magnetic-field flux raised by magnetic buyoancy (the density and pressure inside a flux tube are lower than the surrounding plasma because the magnetic pressure adds to the gas pressure to satisfy the hydrostatic equilibrium), and piercing the solar surface.
Local helioseismology proved a powerful tool to investigate their subsurface structure.
The first inversion of the sound-speed profile below an active region was presented in \cite{kosovichev00}. Since then, numerous articles have dealt with this topic ({\it e.g.} Jensen {\it et al.}, 2001; Couvidat {\it et al.}, 2004). The main result is the existence of a two-region structure: a decrease in the sound speed compared to the quiet Sun, immediately below the solar surface, followed by an increase in sound speed in deeper layers. The reality of this two-region structure is controversial, due to limitations and approximations made during the inversion process. In particular, the impact of a magnetic field on the propagation and conversion (into fast MHD waves) of acoustic waves {\citep{crouch05} is neglected.
Similarly \cite{zhao01} applied the time-distance formalism to derive the flow velocities underneath a sunspot (down to 30 Mm), and found results consistent with the Parker model of sunspots (cluster model).
Knowing the structure of sunspots can shed light on the interaction of the magnetic field fluxes with solar convection (strong magnetic fields are thought to impede convection), on the way these flux tubes arise and decay, where they are generated and anchored, and so on.

\subsubsection{Supergranulation as a Travelling Wave}

Using surface-gravity waves to measure the horizontal divergence of the flows in supergranules (convective cells of typical size 30 Mm) \cite{gizon03} found that supergranulation undergoes oscillations and supports waves with periods of six to nine days. These waves are predominantly prograde, which explains the apparent super-rotation of supergranules (supergranules appear to rotate faster than magnetic features at the solar surface).
The nature of this wave-like behaviour is not understood, but numerical simulations of convection in oblique magnetic fields showed the existence of traveling waves.
Therefore, local helioseismology studies of supergranulation could shed light on the subsurface solar magnetic field.
Indeed, \cite{gizon03} suggest that their result could be the beginning of the use of supergranular waves to probe the upper convection zone.

\subsection{Toward a dynamical model of the Sun}
 Building a dynamical model of the Sun is a difficult task, it is why we use today the seismic model to predict the different observed quantitites. The Dynamical Solar Model takes into account all the dynamical processes: the present magnetic variability of the external layers together with the slow evolution of the activity along the solar life.  The history of the momentum transport will include the time evolution of a potential fossil field (introduced to describe the very young solar analogs) in interaction with the dynamo field and a good understanding of the phase of accretion ejection which may have modified the initial solar mass and its impact on the formation of planets \citep{turck10b}. This model will include also  the  role of the gravity waves along time. It could take 10 years to  build it with a lot of consequences on different fields of physics so, building such model justifies the instrumental progress one can perform during the next decades.  

\newpage





\section{Conclusion, Open questions \& Perspectives}

The solar neutrino puzzle stimulated impressive instrumental developments on both the physics and astrophysics sides. This puzzle was solved by proving the existence of the neutrino flavor oscillations, thanks mainly to the SNO detector but also to SuperKamiokande and BOREXINO. These results help us put strong constraints on the oscillation parameters $\theta_{12}$ and $\delta m_{12}$ and on the survival probability Pee(E). All the neutrino detectors, including the gallium ones, show the deficit and support the LMA neutrino-oscillation solution. In parallel, the seismic investigation of the solar core with the very convincing GOLF and MDI instruments \citep{turck01b,turck08}, recently confirmed by the ground-based networks GONG and BiSON \citep{basu09}, shows the reliability of the neutrino flux predictions  deduced from  models taking into account the observed solar sound-speed profile confronted to the actual neutrino detections, on the contrary the SSM predictions vary with time and appear now in marginal agreement with all the experiments. In this review we focused on the main facts demonstrating how successfully this complex problem was tackled. 

Of course, in light of this success, new dramatic questions have been raised and agreement does not mean complete understanding of the solar interior. It is not totally established if the SSM is the appropriate picture or if dynamical effects play some crucial roles in the context described in this review. We have shown the different directions of improvements which require to take into account all the components of the solar magnetism and the development of new laboratory facilities which produce equivalent plasmas to check their microscopic properties.

\subsection{Secondary effects of the neutrino properties}
Variability has been searched for in the neutrino detections because they may reveal other properties of the neutrinos: day-night variability can unveil some oscillation characteristics, and longer time variability can be the signature of some sensitivity to the magnetic moment of the neutrino. The global fit to the SNO
day - night energy spectra, plus data from other solar
neutrino experiments, strongly favored the LMA solution
in a 2-flavor MSW neutrino oscillation analysis \citep{ahmad02}. Long term variability has not been convincingly evidenced in the SNO and SK detectors \citep{aharmim10}. It has been thoroughly investigated in the chlorine and gallium experiments \citep{gavryuseva01,sturrock01}, but but  a clear signal   has not been put definitively in evidence \citep{sturrock09}. The reason is threefold: (1) in real time experiments, the statistics is high but the solar neutrino  counts are the result of  a  complex analysis and of a strong reduction of  the initial statistics; (2) in radiochemical experiments the statistics is, on the contrary, very low, and some differences in the extraction procedure over time may have perturbed the time analysis of the signal; (3) the magnetic moment is probably too small. The upper limit is presently less than 10 $^{-10} \mu_B$  with BOREXINO and the internal solar magnetic field could be too weak, probably smaller than 3 MG even in the radiative zone. 
\citep{duez10}
Therefore,  a clear signature of magnetic interaction seems difficult to achieve. Nevertheless, more and more works are dedicated to this kind of interaction for the Sun  with the hope of detection of a signal in BOREXINO \citep{das09}, and with important consequences for more massive stars \citep{heger09}.

\subsection{The neutrino masses}
Solar neutrinos contributed to develop neutrino physics. The sensitivity to $\sin^2 \theta_{13}$ ($1\sigma$ error) of the solar neutrino detectors (including BOREXINO, LENS, SNO+) can be improved down to 0.01 - 0.02 by precise measurements of the pp-neutrino flux and the CC/NC ratio, as well as spectrum distortion at high energies. The combination of experimental results sensitive to the low and high energy parts of the solar neutrino spectrum resolves the degeneracy of angles $\theta_{13}$ and $\theta_{12}$ \citep{goswami05}. The comparison of the present $\sin^2 \theta_{13} = 0.017 \pm 0.026$   as well as $\sin^2 \theta_{12}$ measured in the solar neutrinos and in the reactor/accelerator experiments may reveal new effects which cannot be seen otherwise. This field has already benefited from atmospheric neutrinos and reactors like KAMLAND (see its role in Figure 10).   For the global analysis of the experiments, see \cite{fogli08,fogli10}, and  \cite{gonzalez10} who include also the cosmological results. 

Then come Double Chooz and the first long baseline with SK, ready to improve the present situation. Moreover the double $\beta$ decay at Mainz and Troitsk, MiniBooNE and MINOS add very important constraints on the effective electron neutrino mass  which is now smaller than 1 eV. A lot of discussions appears around the potential existence of a fourth neutrino, the sterile neutrinos  \citep{Mention} and other short baseline electron neutrino disappearance \citep{giunti}.
\begin{figure}
\includegraphics[width=15cm]{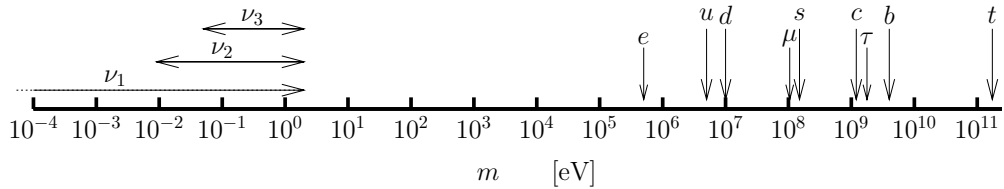}
\vspace{-17cm}
\caption{\small Order of magnitude of the leptons and quarks. From \cite{giunti09}.}
\end{figure}
If the neutrino were massless, neutrino and antineutrino would unambiguously be the same particle. However  neutrinos have a mass, which results in two possibilities: neutrinos are a Dirac-type particle (neutrinos and antiparticles differ by a leptonic internal charge); or they are Majorana 
particles, and the neutrinos are the self conjugate of the antineutrinos. Today we know for sure that neutrinos are massive, see Figure 12 to see the order of magnitude of their fundamental constituents. See also \cite{petcov09,giunti09,valle10} for general reviews on the present status of the field. The Large Hadron Collider, which just started running, could contribute to this success story by shedding light on the Majorana character of the neutrino.

\subsection{Future experimentations: Neutrino Astronomy  from  Sun to Supernovae}
The currently running solar neutrino experiments and the upcoming ones (ICARUS, \cite{arneodo06}, MiniLENS - LENS \cite{raghavan08}) are extending the energy threshold of neutrino detection toward the low energy (more statistics and less dependence on the theoretical spectrum for the boron neutrinos). Therefore they will be able to separate the pp contribution from the other neutrino sources, because  the boron and beryllium neutrinos are measured independently; they will also allow a detection of the neutrinos coming from the CNO cycles, thus heralding the beginning of neutrino astronomy. The liquid scintillator of the SNO+ experiment in the SNOLAB (located in the same Sudbury mine than SNO, but at a 2 km depth), containing dissolved neodymium, will measure neutrinoless double beta decay in that isotope to check whether or not the neutrino is a Majorana particle. All the current underground laboratories will also be able to detect supernova explosions. Moreover, a network of neutrino detectors is now being installed around the world. If a burst of neutrinos is detected, SNEWS (Supernova Early Warning System) will alert the astronomy community (and the amateur astronomer community as well).

On the astronomical side, the activity of the Sun is now at the center of the research effort, with the inclusion of all the different dynamical processes in solar models in 1, 2, or 3D. A new generation of missions are beginning to operate this year, SDO has been launched in February 2010 \citep{scherrer02}, PICARD in June 2010 \citep{thuillier07}. A new instrument  GOLF-NG \citep{turck06,turck08b,turck10d} has been qualified on ground to pursuit in space the gravity mode detection by a multichannel detection and add also some solar atmospheric investigation in an extended mission of formation flying \citep{turck09a} . The main goal of these missions is to better understand the Sun-Earth relationship in near real time (on a scale of days or less for the forecast of the Space Weather) or on a longer timescale (decade, to understand the space climate). Of course, this detailed information will be accompanied by the strong development of asteroseismology with the COROT \citep{baglin06} and KEPLER \citep{borucki10,chaplin10} missions, which will observe thousands to millions of pulsating stars. This dramatic evolution will give a new orientation to the Neutrino-Astronomic community: the detailed study of the explosion of different types of stars is bound to produce, once more, excitating times. 

 \subsection{Solar core and Dark Matter}
Despite the great success of the solar physics story, the central solar core below 0.10 R${_\odot}$ containing about  a quarter of the solar mass, is still not completely understood. Most of the acoustic modes have been detected but they do not allow us to properly describe the thermodynamics of this region of the Sun through the sound speed. Consequently the central temperature deduced from the seismic model is obtained by assuming that the temperature and density profiles follow the classical equations of stellar evolution. This hypothesis leads to a good agreement between prediction and detected boron neutrino fluxes but it has been known for 20 years \citep{giraud-heraud90,dearborn90,kaplan91} that these profiles could  be modified by the presence of dark matter. This matter exists and must be present in the stellar cores. It has not yet been identified but it shall change the transport of energy \citep{spergel85} and this transport depends on the scattering cross-sections between dark matter and stellar plasma  \citep{giraud-heraud90,lopes02, lopes02b}. This matter thermalizes the central core and, due to the fact that the temperature stays relatively constant in this region, it modifies the asymptotic behaviour of the gravity mode frequencies. This could be an interesting signature to look for it \citep{lopes10}. The first detection of dipole gravity modes with GOLF/SoHO excludes a large range  of WIMPs with mass $<$ 10 GeV \citep{Turck2011}. This interesting investigation is an additional argument between others \citep{turck05,turck09a} to pursue the gravity mode detections with the new generation of instruments mentioned above. 

\subsection{Last remarks}
The present  ongoing experiments and others in construction shows the real beginning of Neutrino Astronomy. The detection of CNO neutrinos and pure pp will add new constraints on the knowledge of the central CNO in addition to the present revised photospheric values. The on going opacity measurements will add crucial checks to separate the effects of microscopic physics from macroscopic physics in the radiative zone. The g-modes  will  reveal the order of magnitude of the fossil field and its interconnection with the dynamo field, this progress will open a new page on the knowledge of the variability of  our star. Double Chooz and Minos will constrain $\theta_{13}$ which could reveal some difference between matter and vacuum. 

These last three decades have been extremely productive.  It was not possible in this review to give credit to all the  works devoted to this subject. This review is certainly not totally exhaustive but we describe the main highlights and the direction of new works. We limited the number of tables, so we encourage the reader to return to our first review of 1993 and to the tables published by J. Bahcall and his collaborators. We  hope,  for the youngest members of our community, that stars (including Sun) and supernovae will trigger studies as exciting as the ones summarized in this review.

\section*{Acknowledgments}
This work was partly  supported by the space agencies ESA and CNES. We would like to  thank our collaborators J.C. Barri\`ere, P.H. Carton, J. Cr\' etolle, P. Daniel Thomas, R. Duc, V. Duez, H. Dzitko, R. A.
Garcia, R. Granelli, I. Lopes, S. Mathis, S. Mathur, P. Nghiem, F. Nunio,  L. Piau and Y. Piret, who have actively participated to some of the works described here and to the technical development of GOLF or of the prototype of GOLF-NG.  SOHO is an international collaboration between ESA and NASA. We are indebted toward CEA and CNES for financial support.
S\' ebastien Couvidat was funded these last years by NASA grant NAS5-02139 (HMI). We also thank our numerous colleagues from Particle Physics: ``we are sharing together a real fascination for these mysterious neutrinos''. 
\newpage

\References
\bibitem[Aardsma et al.(1987)]{aardsma87} Aardsma G. et al. 1987 {\it  Phys. Lett. B}{\bf  194}  321
\bibitem[Abazov et al.(1991)]{abazov91} Abazov A I {\it et al} 1991 {\it Phys. Rev. Letters} {\bf 67} 3332 
\bibitem[Abdurashitov et al.(1994)]{abdurashitov94} Abdurashitov J N {\it et al} ; the SAGE collaboration 1994 {\it Phys. Lett. B}, {\bf 328} 234
\bibitem[Abdurashitov et al.(1996)]{abdurashitov96}Abdurashitov J N, Gavrin V N, Giron S V ; the SAGE collaboration 1996 {\it Phys. Rev. Lett. B.} {\bf 77} 4708
\bibitem[Abdurashitov et al.(1999a)]{abdurashitov99a} Abdurashitov J N, Bowles T J, Cherry M L ; the SAGE collaboration 1999a {\it Phys. Rev. Lett} {\bf 83} 4686
\bibitem[Abdurashitov et al.(1999b)]{abdurashitov99b}Abdurashitov J N, Gavrin V N; the SAGE collaboration 1999b {\it Phys. Rev. C.} {\bf 60} 5801
\bibitem[Abdurashitov et al.(2009)]{abdurashitov09}Abdurashitov J N, Gavrin V N ; the SAGE collaboration 2009 {\it Phys. Rev. C.} {\bf 80} 15807
\bibitem[{{Adelberger} {et al.}(1998)}]{adelberger98} Adelberger E  {\it et al} 1998 {\it Rev. Mod. Phys.} {\bf 70} 1265
\bibitem[{{Adelberger} {et al.}(2011)}]{adelberger10} Adelberger E  {\it et al} 2011 {\it Rev. Mod. Phys.} {\bf 83} 195
\bibitem[{{Aharmim} {et al.}(2005)}]{aharmim05} Aharmim B  {\it et al} 2005 {\it Phys. Rev. C} {\bf 72} 055502 and references therein
\bibitem[{{Aharmim} {et al.}(2008)}]{aharmim08} Aharmim B {\it et al}; the SNO collaboration 2008 {\it Phys. Rev. Lett.} {\bf 101} 111301
\bibitem[{{Aharmim} {et al.}(2010a)}]{aharmim10} Aharmim B {\it et al}; the SNO collaboration 2010 {\it ApJ} {\bf 710} 540
\bibitem[{{Aharmim} {et al.}(2010b)}]{aharmim10b} Aharmim B {\it et al}; the SNO collaboration 2010 {\it Phys. Rev. C} {\bf 81} 055504
\bibitem[{{Ahmad} {et al.}(2001)}]{ahmad01} Ahmad Q R {\it et al} ; the SNO collaboration 2001 {\it Phys. Rev. Lett.} {\bf 87} 071301
\bibitem[{{Ahmad} {et al.}(2002)}]{ahmad02} Ahmad Q R {\it et al}; the SNO collaboration 2002 {\it Phys. Rev. Lett.} {\bf 89} 11301 
\bibitem[{{Ahmed et al.}(2004)}]{ahmed04} Ahmed S N; the SNO collaboration 2004 {\it Phys. Rev. Lett.} {\bf 92} 181301 
\bibitem[{{Alexander \& Ferguson}(1994)}]{alexander94} Alexander D R and Ferguson J W 1994 {\it Astrophys. J} {\bf 437} 879 
\bibitem[{{Alimonti et al.}(2009)}]{alimonti09} Alimonti G {\it et al} ; the Borexino collaboration 2009 {\it NIMP A} {\bf 600} 568 
\bibitem[{{Altmann et al.}(2000)}]{altmann00} Altmann M {\it et al} ; the GNO Collaboration 2000 {\it Phys. Lett. B} {\bf 490} 16
\bibitem[{{Altmann et al.}(2005)}]{altmann05} Altmann M {\it et al} ; the GNO collaboration 2005 {\it Phys. Lett. B} {\bf 616} 174
\bibitem[{{Anders \& Grevesse}(1989)}]{anders89} Anders E and Grevesse N 1989 {\it Geochim. Cosmochim. Acta} {\bf 53} 197
\bibitem[{{Anselmann et al.}(1992)}]{anselmann92} Anselmann P {\it et al} ; the GALLEX collaboration 1992 {\it Phys. Lett. B} {\bf 285} 376
\bibitem[{{Antia \& Basu}(2007)}]{antia07} Antia H M and Basu S 2007 {\it Astro. Nach.} {\bf 328} 257
\bibitem[{{Arneodo et al.}(2006)}]{arneodo06} Arneodo F {\it et al} ; the ICARUS collaboration 2006 {\it Phys. Rev.} {\bf 74} 11 
\bibitem[{{Arpesella et al.}(2008)}]{arpesella08} Arpesella C {\it et al} ; the BOREXINO collaboration 2008 {\it Ph. Lett. B} {\bf 658} 101;
{\it Phys. Rev. Lett.} {\bf 101} 091302
\bibitem[{{Asplund et al.}(2004)}]{asplund04} Asplund M, Grevesse N, Sauval A J, Allende Prieto C, and Kiselman D 2004 {\it A\&A} {\bf 417} 751 
\bibitem[{{Asplund et al.}(2005)}]{asplund05} Asplund  M {\it et al} 2005 {\it A\&A} {\bf 435} 339
\bibitem[{{Asplund et al.}(2009)}]{asplund09} Asplund M, Grevesse N, and Scott P 2009 {\it ARA\&A} {\bf 47} 481
\bibitem[{{Baglin et al.}(2006)}]{baglin06} Baglin {\it et al} 2006 {\it ESA} {\bf 624} 34; {\bf 1306} 33

\bibitem[{{Bahcall}(1962)}]{bahcall62} Bahcall J N 1962 {\it Phys. Rev.} {\bf 128} 1297
\bibitem[{{Bahcall et al.}(1963)}]{bahcall63} Bahcall J N, Fowler W A, Iben I Jr \& Sears R L 1963 {\it ApJ} {\bf 137} 344
\bibitem[{{Bahcall}(1964a)}]{bahcall64} Bahcall J N 1964 {\it Phys. Rev. Lett.} {\bf 12} 300
\bibitem[{{Bahcall}(1964b)}]{bahcall64b} Bahcall J N 1964 {\it Phys. Rev. } {\bf 135} 137
\bibitem[{{Bahcall}(1989)}]{bahcall89} Bahcall J N 1989 Neutrino Astrophysics (Cambridge University Press, Cambridge)
\bibitem[{{Bahcall \& Shaviv}(1968)}]{bahcall68}Bahcall J N and Shaviv G 1968 {\it ApJ} {\bf 153} 113
\bibitem[{{Bahcall, Bahcall \& Ulrich}(1968)}]{bahcall68b}Bahcall J N, Bahcall N A  and Ulrich R 1968 {\it ApL} {\bf 2} 91
\bibitem[{{Bahcall \& Holstein}(1986)}]{bahcall86} Bahcall J N and Holstein B R 1986  {\it Phys Rev C} {\bf 33} 2121
\bibitem[{{Bahcall \& Ulrich}(1988)}]{bahcall88} Bahcall J N and Ulrich R 1988 {\it Rev. Mod. Phys} {\bf 60} 297
\bibitem[{{Bahcall, Pinsonneault \& Basu}(2001)}]{bahcall01} Bahcall J N, Pinsonneault M H and Basu S 2001 {\it ApJ} {\bf 555} 990
\bibitem[{{Bahcall \& Pena-Garay}(2004)}]{bahcall04} Bahcall J N and Pena-Garay C 2004 {\it New J. of Phys.} {\bf 6} 63 
\bibitem[{{Bahcall et al.}(2005)}]{bahcall05}Bahcall J N, Basu S, Pinsonneault M \& Serenelli A M 2005 {\it ApJ} {\bf 618} 1049
\bibitem[{{Bahcall et al.}(2006)}]{bahcall06}Bahcall J N, Serenelli A M \& Basu S 2006 {\it ApJ Supp} {\bf 165} 400
\bibitem[{{Bailey et al.}(2007)}]{bailey07} Bailey J {\it et al} 2007 {\it Phys. Rev. Lett. } {\bf 99} 5002
\bibitem[{{Bailey et al.}(2009)}]{bailey09}Bailey J {\it et al}  2009 {\it Phys. of Plasmas} {\bf 16} 058101
\bibitem[{{Ballot, Turck-Chi\`eze \& Piau}(2004)}]{ballot04a} Ballot J, Turck-Chi\`eze S and Piau L 2004 {\it EDP Science} {\bf 319}
\bibitem[{{Basu et al.}(2000)}]{basu00} Basu S {\it et al} 2000 {\it Astro. Journal} {\bf 535} 1078
\bibitem[{{Basu \& Antia}(2004)}]{basu04} Basu S \& Antia H M 2004 {\it Astrophys. J} {\bf 606} L85
\bibitem[{{Basu et al.}(2009)}]{basu09} Basu S {\it et al} 2009 {\it Astrophys. J} {\bf 699} 1403
\bibitem[{{Bellini et al.}(2008)}]{bellini08} Bellini G{\it et al} 2008 {\it Journal Phys: Conf Series} {\bf 120} 052006
\bibitem[{{Bellini et al.}(2010)}]{bellini10} Bellini G{\it et al}, the Borexino collaboration 2010 {\it Phys. Rev. D} {\bf 82} 033006
\bibitem[{{Berthomieu, Provost \& Morel}(1997)}]{berthomieu97} Berthomieu G, Provost J and Morel P 1997 {\it Astronomy and Astrophysics} {\bf 327} 349
\bibitem[{{Bethe}(1938)}]{bethe38}Bethe H A and  Critchfield, C L 1938 {\it Phys. Rev.} {\bf 54} (1938) 248; Bethe, H A 1939 {\it Phys. Rev.} {\bf 55} 434
\bibitem[{{Biemont et al.}(1991)}]{biemont91}Biemont E, Baudoux M,  Kurucz R L, Ansbacher W and Pinnington E H 1991
{\it Astron. Astrophys.} {\bf 249} 539
\bibitem[{{Bohm}(1958)}]{bohm58}Bohm  K H 1958 {\it Zeit. Fur Astro.} {\bf 46}  245
\bibitem[{{Bouchez}(1986)}]{bouchez86}Bouchez J 1986 {\it Neutrino Physics and Astrophysics} / Singapore: {\it World Scientific} {\bf 194}
\bibitem[{{Bouchez}(2005)}]{bouchez05}Bouchez J 2005 {\it CR Acad√©mie des Sciences} {\bf 6} 706
\bibitem[{{Koch, Borucki et al.}(2010)}]{borucki10}Koch D G, Borucki W J  et al. 2010  {\it  ApJ lett} {\bf 713} L79
\bibitem[{{Brun \& Rempel}(2009)}]{brun09}Brun A S, Rempel M 2009 {\it Space Science Reviews} {\bf 144} 151
\bibitem[{{Brun, Turck-Chi\`eze \& Morel}(1998)}]{brun98}
Brun A S, Turck-Chi\`eze S and Morel P 1998 {\it ApJ} {\bf 506} 913 
\bibitem[{{Brun, Turck-Chi\`eze \& Zahn}(1999)}]{brun99}
Brun A S, Turck-Chi\`eze S and Zahn J P 1999 {\it ApJ}  {\bf 525} 1032
\bibitem[{{Caffau et al.}(2008)}]{caffau08} 
Caffau E et al.
2008 {\it A\&A} {\bf 488} 1031 
\bibitem[{{Caffau et al.}(2009)}]{caffau09} 
Caffau E {\it et al} 2009 {\it A\&A} {\bf 498} 877
\bibitem[{{Canuto, Goldman \& Mazitelli}(1996)}]{canuto96} Canuto V M, Goldman I and Mazitelli I 1996 {\it ApJ} {\bf 473} 550
\bibitem[{{Chadwick}(1932)}]{chadwick32} Chadwick J 1932 {\it \PRS} {\bf 136} 830 
\bibitem[{{Chaplin et al.}(2010)}]{chaplin10} Chaplin W J  et al. 2010  {\it  ApJ lett} {\bf 713} L169
\bibitem[{{Choudhuri, Schussler \& Dikpati}(1995)}]{choudhuri95} Choudhuri A R, Schussler M, Dikpati M 1995 {\it Astronomy and Astrophysics} {\bf 303} 29
\bibitem[{{Christensen-Dalsgaard}(1982)}]{christensen82} Christensen-Dalsgaard J 1982 {\it MNRAS} {\bf 199} 735
\bibitem[{{Christensen-Dalsgaard et al.}(1985)}]{christensen85} Christensen-Dalsgaard J  {\it et al}  1985{\it Nature} {\bf 315} 378
\bibitem[{{Christensen-Dalsgaard \& Berthomieu}(1991)}]{christensen91} Christensen-Dalsgaard J and Berthomieu G 1991 {\it Solar Interior and Atmosphere} {\bf 401} 478
\bibitem[{{Christensen-Dalsgaard, Proffitt \& Thompson}(1993)}]{christensen93} Christensen-Dalsgaard J, Proffitt C R and Thompson M J  1993 {\it  Astrophys. J} {\bf 403} L75
\bibitem[{{Christensen-Dalsgaard} (2002)}]{christensen02} Christensen-Dalsgaard J 2002 {\it Rev Mod Phys} {\bf 74} 1073
\bibitem[{{Claverie} {et al.}(1981)}]{claverie81} Claverie A {\it et al} 1981 {\it Sol. Phys.} {\bf 74} 51
\bibitem[{{Cleveland} {et al.}(1998)}]{cleveland98} Cleveland B T {\it et al} 1998 {\it Astr. Journal} {\bf 496} 505
\bibitem[{{Cowan} {et al.}(1956)}]{cowan56} Cowan C L {\it et al} 1956 {\it Science} {\bf 124} 103 
\bibitem[{{Cox \& Guzik} (2004)}]{cox04} Cox, A N and Guzik, J A 2004, {\it ApJ} {\bf 613} L169 
\bibitem[{{Crane} (1948)}]{crane48} Crane H.R. 1948, {\it Rev. Mod. Phys}., {\bf 20}, 278
\bibitem[{{Cravens} {et al.}(2008)}]{cravens08} Cravens J P {\it et al}, the Superkamiokande collaboration 2008, {\it Phys. Rev. D}, {\bf 78}, 032002
\bibitem[{{Cribier} {et al.}(1995)}]{cribier95} Cribier {\it et al} 1995, {\it Astroparticle Physics}, {\bf 4}, 23
\bibitem[{{Couvidat, Turck-Chi\`eze \& Kosovichev}(2003)}]{couvidat03} Couvidat S, Turck-Chi\`eze S and Kosovichev A 2003 {\it ApJ} {\bf 599} 1434
\bibitem[{{Couvidat, Birch \& Kosovichev} (2004)}]{couvidat04} Couvidat S, Birch A C and Kosovichev A G {\it et al} 2004 {\it Astrophysical Journal} {\bf 607} 554
\bibitem[{{Crouch \& Cally}(2005)}]{crouch05} Crouch A D and Cally P S 2005 {\it Solar Physics} {\bf 227} 1
\bibitem[{{Das, Pulido \& Picariello}(2009)}]{das09} Das C R, Pulido J and Picariello M 2009 {\it Phys Rev D} {\bf 79} 073010
\bibitem[{{Davis}(1955)}]{davis55} Davis R Jr 1955 {\it \PR} {\bf 97} 766
\bibitem[{{Davis}(1964)}]{davis64}Davis R Jr 1964 {\it Phys. Rev. Lett.} {\bf 12} 303 
\bibitem[{{Davis, Harmer \& Hoffman}(1968)}]{davis68}Davis R Jr, Harmer D S and Hoffman K C 1968 {\it Phys. Rev. Lett. } {\bf 20} 1205
\bibitem[{{Davis}(1994)}]{davis94}Davis 1994 {\it Intern.School Nucl. Phys. 15} {\bf Neutrinos in cosmology, astrophysics, particle and nuclear physics} 13
\bibitem[{{Davis}(2002)}]{davis02}Davis 2002, Nobel prize lecture
\bibitem[{{Dearborn et al.}(1990)}]{dearborn90}Dearborn, D., Raffelt, G., Salati, P., Silk, J.,  Bouquet, A. 1990 {\it ApJ} {\bf 354} 568
\bibitem[{{de Souza, Kervella \& Jankov}(2003)}]{desousa03}De Souza A, Kervella P \& Jankov S 2003 {\it A\&A} {\bf 407} L47
\bibitem[{{Dikpati, Cally \& Gilman}(2004)}]{dikpati04}Dikpati M, Cally P S and Gilman P A 2004 {\it Astrophys. J} {\bf 610} 597
\bibitem[{{Duez, Mathis \& Turck-Chi\`eze}(2010)}]{duez10}Duez V, Mathis S and Turck-Chi\`eze S 2010 {\it MNRAS} {\bf 402} 271
\bibitem[{{Duvall}(1979)}]{duvall79}Duvall T L Jr 1979 {\it Solar Physics} {\bf 63} 3
\bibitem[{{Duvall}(1982)}]{duvall82}Duvall T L Jr 1982 {\it Nature} {\bf 300} 242
\bibitem[{{Duvall et al.}(1993)}]{duvall93}Duvall T L Jr, Jefferies S M, Harvey J W, Pomerantz M A 1993 {\it Nature} {\bf 362} 430
\bibitem[{{Dzitko et al.}(1995)}]{dzitko95}Dzitko H, Turck-Chi\`eze S, Delbourgo-Salvador P and Lagrange G 1995 {\it ApJ} {\bf 447} 428
\bibitem[{{Eddington}(1920)}]{eddington20}Eddington A S 1920 {\it The observatory} {\bf 43} 341
\bibitem[{{Eddington}(1926)}]{eddington26}Eddington A S 1926 {\it Nature} {\bf 117} 25;  {\it The internal constitution of the stars, (Cambridge University Press)}
\bibitem[{{Eff-Darwich et al.}(2008)}]{effdarwich08}Eff-Darwich A {\it et al} 2008 {\it Astrophys. J} {\bf 679} 1636    
\bibitem[{{Elliott}(1997)}]{elliott97}Elliott J R 1997 {\it A\&A } {\bf 327} 1222
\bibitem[{{Escrihuela et al.}(2009)}]{escrihuela09} Escrihuela F J,Miranda, F O, Tortola M A and Valle J W F 2009   {\it Phys.\ Rev.\  D} {\bf 80} 105009 [Erratum-ibid.\  D {\bf 80}  129908]   
\bibitem[{{Fabbian, Khomenko, Moreno-Insertis \& Nordlhund}(2010)}]{fabbian10} Fabbian D, Khomenko E, Moreno-Insentis F \& Nordlhund A 2010,  {\it ApJ} {\bf 724} 1536
\bibitem[{{Faulkner, Gough and Vahia}(1986)}]{Faulkner}Faulkner J, Gough D O and Vahia M N 1986 {\it Nature} {\bf 321}  226   
\bibitem[{{Feigelson}(2003)}]{feigelson03}Feigelson 2003 {\it EAS Publ. series} {\bf 9} 317
\bibitem[{{Fogli et al.}(2008)}]{fogli08} Fogli G L {\it et al} 2008   {\it Phys.\ Rev.\  Lett.} {\bf 101} 1801 \bibitem[{{Fogli et al.}(2010)}]{fogli10} Fogli G L  {\it et al} 2010   {\it J Ph CS} {\bf 203} 2061 
\bibitem[{{Formicola et al.}(2004)}]{formicola04}Formicola A {\it et al.} 2004 {\it Phys. Lett. B} {\bf 591} 61 
\bibitem[{{Fowler, Caughlan G R \& Zimmerman}(1967)}]{fowler67}Fowler W A, Caughlan G R and Zimmerman B A 1967 {\it An. Rev. Of Astronomy and Astrophysics} {\bf 5} 525
\bibitem[{{Fowler, Caughlan G R \& Zimmerman}(1975)}]{fowler75}Fowler W A, Caughlan G R and Zimmerman B A 1975 {\it An. Rev. Of Astronomy and Astrophysics} {\bf 13} 69
\bibitem[{{Fowler, Caughlan G R \& Zimmerman}(1983)}]{fowler83}Fowler W A, Caughlan G R, Zimmerman B A and Harris M J 1983 {\it An. Rev. Of Astronomy and Astrophysics} {\bf 21} 165
\bibitem[{{Fukuda et al.}(1998)}]{fukuda98}Fukuda S {\it et al} ; the Superkamiokande collaboration 1998 {\it Phys. Rev. lett. } {\bf 81} 1158
\bibitem[{{Fukuda et al.}(2000)}]{fukuda00}Fukuda S {\it et al} Superkamiokande collaboration, 2000, {\it Phys. Rev. 
  Lett}, {\bf 85}, 3999
  \bibitem[{{Gabriel et al.}(1995)}]{gabriel95}Gabriel {\it et al} 1995 {\it Solar Physics} {\bf 162} 61
\bibitem[{{Gamow}(1928)}]{gamow28}Gamow 1928 {\it Zeits. f. Physik} {\bf 52} 510
\bibitem[{{Garaud}(2002)}]{garaud02}Garaud 2002 {\it MNRAS} {\bf 329} 1
\bibitem[{{Garaud} \& {Garaud} (2008)}]{garaud08} Garaud P \& Garaud J-D 2008, {\it MNRAS}, {\bf 391} 1239
\bibitem[{{Garcia et al.}(2004)}]{garcia04}Garcia R A {\it et al}  2004 {\it Sol. Phys.} {\bf 220} 269
\bibitem[{{Garcia et al.}(2007)}]{garcia07}Garcia R A {\it et al} 2007 {\it Science} {\bf 316} 1537
\bibitem[{{Garcia et al.}(2008)}]{garcia08}Garcia R A {\it et al} 2008 {\it Astron. Notes} {\bf 329} 476
\bibitem[{{Garcia et al.}(2011)}]{garcia11}Garcia R A {\it et al}  2011 {\it Journal Phys Conf Ser} {\bf 271} 2046
\bibitem[{{Gavryuseva \& Gavryusev}(2001)}]{gavryuseva01}Gavryuseva E, Gavryusev V 2001 {\it Memorie della Societ√† Astronomica Italiana} {\bf 72} 532
\bibitem[{{Gavrin et al.}(1990)}]{gavrin90}Gavrin V N ; the SAGE collaboration 1990 {\it Cosmic Ray Conference} {\bf 7} 179 
\bibitem[Gavrin et al.(1992)]{gavrin92}Gavrin V N {\it et al} 1992 {\it Nucl. Phys. B (Proc. Suppl.)} {\bf 28} 75 
\bibitem[{{Giunti \& Studenikin}(2009)}]{giunti09}Giunti C and Studenikin A 2009 {\it Phys Atom Nuclei} {\bf 72}  2089
\bibitem[{{Giunti \& Laveder}(2009)}]{giunti}Giunti C and Laveder M 2010 {\it Phys Rev D} {\bf 82}  53005
\bibitem[Giraud-Heraud et al.(1990)]{giraud-heraud90}Giraud-Heraud, Y., Kaplan, J., de Volnay, F. M., Tao, C., Turck-Chieze, S. 1990 {\it Sol. Phys.}{\bf 128} 21
\bibitem[{{Gizon, Duvall \& Schou}(2003)}]{gizon03}Gizon L, Duvall T L Jr, Schou J 2003 {\it Nature} {\bf 421} 43
\bibitem[{{Gonzalez-Garcia, Maltoni \& Salvado}(2011)}]{gonzalez10}Gonzalez-Garcia M C, Maltoni, M and Salvado J 2011 {\it JHEP} {\bf 5} 75
\bibitem[{{Goswami \& Smirnov}(2005)}]{goswami05}Goswami S and Smirnov A Y 2005 {\it Phys. Rev. D} {\bf 5} 053011
\bibitem[{{Gough}(1985)}]{gough85}Gough D O 1985 {\it Sol. Phys.} {\bf 100} 65
\bibitem[{{Gough}(1988)}]{gough88}Gough D O 1988 {\it in Solar-Terrestrial Relationships and the Earth Environment in the Millennia}, Varenna, eds G. Cini Castagnoli 90
\bibitem[{{Gough \& Thompson}(1990)}]{gough90}Gough D O and Thompson 1990 {\it MNRAS.} {\bf 242} 25
\bibitem[{{Gough \& McIntyre}(1998)}]{gough98}Gough D O and McIntyre M E 1998 {\it Nature} {\bf 394} 755
\bibitem[{{Gounelle et al.}(2001)}]{gounelle01}Gounelle M {\it et al} 2001 {\it Astrophys. J} {\bf 548} 1051
\bibitem[{{Gounelle et al.}(2003)}]{gounelle03}Gounelle M {\it et al} 2003 {\it \PR C} {\bf 68}
\bibitem[{{Gounelle \& Melbom et al.}(2008)}]{gounelle08}Gounelle M and Meibom A 2008 {\it ApJ} {\bf 680} 781
\bibitem[{{Grec, Fossat \& Pomerantz}(1980)}]{grec80}Grec G, Fossat, E and Pomerantz M 1980 {\it Nature} {\bf 288} 541
\bibitem[{{Grevesse et al.}(1990)}]{grevesse90}Grevesse N,  Lambert D L, Sauval A J, van Dishoeck E F, Farmer C B and  Norton R H
1990 {\it Astron. Astrophys.} {\bf 232} 225
\bibitem[{{Grevesse et al.}(1991)}]{grevesse91}Grevesse N,  Lambert D L, Sauval A J, van Dishoeck E F, Farmer C B and  Norton R H 1991 {\it A \& A} {242} 488
\bibitem[{{Grevesse et al.}(1992)}]{grevesse92}Grevesse N,  Lambert D L, Sauval A J, van
Grevesse N, Noel A and Sauval A J 1992 {\it ESA} {\bf 305} 308
\bibitem[{{Grevesse \& Noels}(1993)}]{grevesse93}Grevesse N and Noels A 1993 {\it in Origin and Evolution of the Elements}, Symposium in Honour of Hubert Reeves' 60th birthday, 15 
\bibitem[{{Gribov \& Pontecorvo}(1969)}]{gribov69}Gribov V and Pontecorvo B 1969 {\it \PL B} {\bf 28} 493
\bibitem[{{Gruzinov}(1998)}]{gruzinov98}Gruzinov A V 1998 {\it ApJ} {\bf 469} 503 
\bibitem[{{Gruzinov \& Bahcall}(1998)}]{gruzinovbahcall98}Gruzinov A V and Bahcall J 1998 {\it ApJ} {\bf 504} 996 
\bibitem[{{Guzik, Watson \& Cox}(2005)}]{guzik05}Guzik J, Watson L S and Cox A N 2005 {\it ApJ} {\bf 627} 1049
\bibitem[{{Haber et al.}(2002)}]{haber02}Haber D A, Hindman B W, Toomre J {\it et al} 2002 {\it Astrophysical Journal} {\bf 570} 855
\bibitem[{{Hampel et al.}(1998)}]{hampel98}Hampel W ; the GALLEX collaboration 1998 {\it} {\bf}
Hampel W ; the GALLEX collaboration 1999 {\it Phys. Lett. B} {\bf 447} 127
\bibitem[{{Haraki et al.}(2005)}]{haraki05}Haraki  {\it et al}: KAMLAND collaboration 2005 {\it Phys. Rev. Lett} {\bf 9} 081801
\bibitem[{{Heger et al.}(2009)}]{heger09}Heger  {\it et al} 2009 {\it Astrophysical Journal} {\bf 696} 608
\bibitem[{{Hill}(1988)}]{hill88}Hill F 1988 {\it Astrophysical Journal} {\bf 333} 996
\bibitem[{{Hirata et al.}(1988)}]{hirata88}Hirata K S {\it et al} 1988 {\it Phys. Rev. D} {\bf 63}16 
\bibitem[{{Hirata et al.}(1989)}]{hirata89}Hirata K S {\it et al} 1989 {\it Phys. Rev. Lett.}{\bf  63} 16; 1990  {\it Phys. Rev. Lett.} {\bf 65} 1297
\bibitem[{{Hirata et al.}(1991)}]{hirata91}Hirata K S {\it et al} Phys. Rev. D44 (1991) 2241; Phys. Rev. Lett. 66 (1991) 9. 
\bibitem[{{Holmgren \& Johnston}(1958)}]{holmgren58}Holmgren H D and Johnston R L 1958 {\it Phys. Rev.} {\bf 113} 1556
\bibitem[{{Holweger}(2001)}]{holweger01}Holweger H 2001 {\it AIP Conf. Proc.} {\bf 598} 23
\bibitem[{{Holweger, Heise \& Kock}(1990)}]{holweger90}Holweger H, Heise C, Kock M 1990 {\it Astron. Astrophys.} {\bf 232} 510
\bibitem[{{Holweger et al.}(1991)}]{holweger91}Holweger, H.,  Bard, A., Kock  A. and  Kock, M.
Astron. Astrophys. 249, (1991), 545
\bibitem[{{Hosaka} {et al.}(2006)}]{hosaka06} Hosaka J {\it et al}: the SuperKamiokande collaboration 2006 {\it Phys. Rev. D} {\bf 73} 112001
\bibitem[{{Howard \& Labonte}(1980)}]{howard80}Howard R, Labonte B J 1980 {\it Solar Physics} {\bf 239} 33
\bibitem[{{Howe et al.}(2009)}]{howe09}Howe R, Christensen-Dalsgaard J, Hill F {\it et al} 2009 {\it Astrophysical Journal} {\bf 701} L87
\bibitem[{{Iglesias \& Rogers}(1996)}]{iglesias96}Iglesias  C A and Rogers F J 1996 {\it Am. Astro. Soc. } {\bf 28} 915
\bibitem[{{Inoue}(2004)}]{inoue04}Inoue K 2004 {\it New Journal of Physics} {\bf 6} 147 and references therein
\bibitem[{{Jensen et al.}(2001)}]{jensen01}Jensen J M, Duvall T L Jr, Jacobsen B H {\it et al} 2001 {\it Astrophysical Journal} {\bf 553} 193
\bibitem[{{Junghans et al.}(2003)}]{junghans03} Junghans A R et al. 2003 {\it Phys. Rev. C}{\bf  68} 065803
\bibitem[{{Junker et al.}(1998)}]{junker98}Junker and LUNA collaboration 1998 {\it Phys. Rev. C} {\bf 57} 2700
\bibitem[{{Kaplan et al.}(1991)}]{kaplan91}Kaplan, Martin de Volnay, F., Tao, C.,  Turck-Chieze, S. 1991 {\it ApJ} {\bf 378} 351
\bibitem[{{Kavanagh}(1960)}]{kavanagh60}Kavanagh R W 1960 {\it Nucl. Phys.} {\bf 15} 411
\bibitem[{{Kirsten}(1993)}]{kirsten93}Kirsten, T 1993 {\it Nucl. Phys. S.} {\bf 31} 117
\bibitem[{{Kosovichev et al.}(1997)}]{kosovichev97}Kosovichev A G {\it et al} 1997 {\it Sol Phys.} {\bf 170} 43; Kosovichev A G, Schou J 1997 {\it Astrophysical Journal} {\bf 482} L207
\bibitem[{{Kosovichev, Duvall \& Scherrer}(2000)}]{kosovichev00}Kosovichev A G, Duvall T L Jr, Scherrer P H 2000 {\it Solar Physics} {\bf 192} 159
\bibitem[{{Kuzmin}(1966)}]{kuzmin66}Kuzmin V A 1966 {\it Soviet Astronomy} {\bf 9} 953
\bibitem[{{Lebreton \& Maeder}(1987)}]{lebreton87}Lebreton Y and Maeder A  1987 {\it A\&A } {\bf 175} 99
\bibitem[{{Lee \& Yang}(1956)}]{lee56}Lee T D and Yang C N 1956 {\it Phys. Rev.} {\bf 104} 254
\bibitem[{{Leibacher et al.}(1999)}]{leibacher99}Leibacher J {\it et al} 1999 {\it Adv Space Res.} {\bf 24} 173
\bibitem[{{Lindsey \& Braun}(1990)}]{lindsey90}Lindsey C, Braun D C 1990 {\it Solar Physics} {\bf 126} 101
\bibitem[{{Lodders}(2003)}]{lodders03}Lodders K. 2003 {\it ApJ} {\bf 591} 1220
\bibitem[{{Loisel et al.}(2009)}]{loisel09}Loisel G {\it et al} 2009 {\it HEDP} {\bf 5} 173
\bibitem[{{Lopes, Bertone \& Silk}(2002)}]{lopes02}Lopes I, Bertone G and Silk J 2002 {\it MNRAS} {\bf 337} 1179
\bibitem[{{Lopes \& Silk}(2002)}]{lopes02b}Lopes, I. P. and Silk, J. 2002 {\it Phys. Rev. Lett.}  {\bf 88} 151303
\bibitem[{{Lopes \& Silk}(2010)}]{lopes10}Lopes, I. P. and Silk, J. 2010 {\it Astrophys. J. Lett.}  {\bf 722} L95
\bibitem[{{MacGregor \& Charbonneau}(1999)}]{macgregor99}MacGregor and Charbonneau 1999 {\it Astrophys. J} {\bf 519} 911
 \bibitem[{{Mathis \& Zahn}(2005)}]{mathis05}Mathis S and Zahn J P 2005 {\it A\&} {\bf 440} 653
\bibitem[{{Mathur et al.}(2007)}]{mathur07}Mathur S, Turck-Chi\`eze S, Couvidat S \& Garcia R A 2007 {\it ApJ} {\bf 668} 594      
 \bibitem[{{Mention et al.}(2011)}]{Mention}Mention G et al 2011 {\it Phys Rev D} {\bf 83} 3006    
\bibitem[{{Michaud \& Proffitt}(1993)}]{michaud93}Michaud G J and Proffitt C R 1993 {\it IAU Colloq 138}
 Inside the Stars, ed. W. W. Weiss \& A. Baglin (San Francisco: ASP), 246
\bibitem[{{Miesch}(2003)}]{miesch03}Miesch M 2003 {\it ApJ} {\bf 586}  663
\bibitem[{{Mikheyev \& Smirnov}(1986)}]{mikheyev86}Mikheyev S P and Smirnov A 1986 {\it Nuovo Cimento}  {\bf 9C} 17 
\bibitem[{{Nishino et al.}(2009)}]{nishino09}Nishino A G {\it et al} 1997 {\it Sol Phys.} {\bf 170} 43
\bibitem[{{Nordlund, Stein \& Asplund}(2009)}]{nordlund09}Nordlund A, Stein R F and Asplund M 2009 {\it LRSP} {\bf 6} 2
\bibitem[{{Parker}(1993)}]{parker93}Parker E N 1993 {\it ApJ} {\bf 408} 707
\bibitem[{{Peimbert, Storey \& Torres-Peimbert}(1993)}]{peimbert93}Peimbert M, Storey P J and Torres-Peimbert S 1993 {\it ApJ} {\bf 414} 626
\bibitem[{{Peimbert et al.}(2007)}]{peimbert07}Peimbert M et al. 2007 Rev Mex. Astron. Astrof. 29 72-79
\bibitem[{{Petcov}(2009)}]{petcov09}Petcov 2009 {\it Journal Phys: conf series} {\bf 173} 012025
\bibitem[{{Piau et al.}(2010)}]{piau10}Piau L, Kervella P, Dib S and Hauschildt P 2010 {\it A\&A}  {\bf 526} 100
\bibitem[{{Pontecorvo}(1947)}]{pontecorvo47}Pontecorvo B 1947 {\it Rep. Prog. Phys.} {\bf 11} 32
\bibitem[{{Pontecorvo}(1968)}]{pontecorvo68}Pontecorvo B 1968 {\it JETP} {\bf 26} 984 
\bibitem[{{Proffitt \& Michaud}(1991)}]{proffitt91}Proffitt C R and Michaud G 1991 {\it Astrophys. J} {\bf 380} 238
\bibitem[{{Raghavan \& Pakvasa}(1988)}]{raghavan88}Raghavan R S and Pakvasa S 1988 {\it Phys. Rev. D} {\bf 37} 849; Raghavan R S {\it et al} 1988  AT\&T Bell Labs report 88-01 
\bibitem[{{Raghavan et al.}(2008)}]{raghavan08}Raghavan R S and the LENS collaboration 2008 {\it Journal of Phys} {\bf 120} 52014
\bibitem[{{Rashba et al.}(2007)}]{rashba07}Rashba T, Semikoz V B, Turck-Chi\`eze S, Valle J W F 2007 {\it MNRAS}, {\bf 377}, 453
\bibitem[{{Reines et al.}(1960)}]{reines60}Reines F et al 1960 {\it Phys. Rev.} {\bf 117} 159
\bibitem[{{Rempel}(2003)}]{rempel03}Rempel 2003 {\it A\&A} {\bf 397}  1097
\bibitem[{{Rogers, Swenson \& Iglesias}(1996)}]{rogers96}Rogers F J, Swenson F J and Iglesias C A  1996 {\it ApJ} {\bf 456} 902
\bibitem[{{Rogers \& Nayfonov}(2002)}]{rogers02}Rogers F J and Nayfonov A  2002 {\it ApJ} {\bf 576} 1064
\bibitem[{{Roxburgh}(1985)}]{roxburgh85}Roxburgh, I. W 1985, {\it Sol. Phys.}, {\bf 100} 21
\bibitem[{{Rudiger \& Kitchatinov}(1997)}]{rudiger97}Rudiger and Kitchatinov  1997 {\it Astronomische Nachrichten} {\bf 318} 273
\bibitem[{{Schatzman}(1969)}]{schatzman69}Schatzman E 1969 {\it ApL} {\bf 3}  139
\bibitem[{{Schatzman et al.}(1981)}]{schatzman81}Schatzman E, Maeder A,  Angrand, F, Glowinski R 1981{\it A\&A} {\bf 96}  1
\bibitem[{{Shaviv \& Salpeter}(1971)}]{shaviv71}Shaviv, G and  Salpeter E E 1971 {\it ApJ} {\bf 165} 171
\bibitem[{{Scherrer et al.}(1995)}]{scherrer95}Scherrer P H {\it et al} 1995 {\it Sol. Phys.} {\bf 162} 129
\bibitem[{{Scherrer et al.}(2002)}]{scherrer02}Scherrer P H {\it et al} 2002 AAS Meeting 56 {\it Bull. American Astronomical Society} {\bf 34} 735
\bibitem[{{Skumanich}(1972)}]{skumanich72}Skumanich, A. 1972, {\it ApJ}, {\bf 171} 565
\bibitem[{{Skumanich, MacGregor \& Jackson}(2004)}]{skumanich04}Skumanich A, Macgregor K B and Jackson S 2004 AAS Meeting 204 {\it Bull. American Astronomical Society} {\bf 36} 785
\bibitem[{{Solanki}(2003)}]{solanki03}Solanki S K 2003 {\it Astronomy and Astrophysics Review} {\bf 11} 153
\bibitem[{{Spiegel \& Zahn}(1992)}]{spiegel92}Spiegel and Zahn J P 1992 {\it Astronomy and Astrophysics} {\bf 265} 106
\bibitem[{{Spergel\& Press}(1985)}]{spergel85}Spergel, D. N. \& Press, W. H. 1985, ApJ, 294, 663
\bibitem[{{Spruit}(2003)}]{spruit03}Spruit H C 2003 {\it Solar Physics} {\bf 213} 1
\bibitem[{{Sturrock \& Scargle}(2001)}]{sturrock01}Sturrock P A and Scargle 2001, {\it ApJ} {\bf 550} L101
\bibitem[{{Sturrock}(2009)}]{sturrock09}Sturrock P A 2009 {\it Sol Phys} {\bf 260}  245
\bibitem[{{Thompson et al.}(2003)}]{thompson03}Thompson {\it et al} 2003 {\it ARA\&A} {\bf 41}  599
\bibitem[{{Thoul et al.}(1994)}]{thoul94}Thoul A A {\it et al} 1994 {\it Astro. Journal Part 1} {\bf 421} 828
\bibitem[{{Thuillier, Dewitte \& Schmutz}(2007)}]{thuillier07}Thuillier G Dewitte S and Schmutz W 2007 {\it Adv. Space Res. } {\bf 26} 1792
\bibitem[{{Totsuka et al.}(1991)}]{totsuka91}Totsuka {\it et al} 1991 {\it Nuclear Phys. B Proceedings Supplem.} {\bf 19} 69
\bibitem[{{Turck-Chi\`eze et al.}(1988)}]{turck88}Turck-Chi\`eze S, Cahen S, Cass\' e M and Doom C 1988 {\it ApJ} {\bf 335} 415
\bibitem[{{Turck-Chi\`eze \& Lopes}(1993)}]{turcklopes93}Turck-Chi\`eze S and Lopes I 1993 {\it ApJ} {\bf 408} 347
\bibitem[{{Turck-Chi\`eze et al.}(1993)}]{turck93}Turck-Chi\`eze S {\it et al} 1993 {\it Phys. Reports} {\bf 230} 57
\bibitem[{{Turck-Chi\`eze et al.}(2001a)}]{turck01a}Turck-Chi\`eze S, Nghiem P, Couvidat S, and Turcotte S 2001a {\it Sol. Phys} {\bf 200} 323  
\bibitem[{{Turck-Chi\`eze et al.}(2001b)}]{turck01b}Turck-Chi\`eze S {\it et al} 2001b {\it ApJ} {\bf 555} L69 
\bibitem[{{Turck-Chi\`eze et al.}(2004a)}]{turck04a}Turck-Chi\`eze S {\it et al}
2004 {\it Physical Review Letters} {\bf 93} 211102
\bibitem[{{Turck-Chi\`eze et al.}(2004b)}]{turck04b}Turck-Chi\`eze S {\it et al} 2004b {\it Astrophys. J} {\bf 604} 455; {\bf 608}, 610
\bibitem[{{Turck-Chi\`eze et al.}(2005)}]{turck05}Turck-Chi\`eze S {\it et al} 2005 {\it ESA-SP 588} {\bf 39th ESLAB symposium}  193
\bibitem[{{Turck-Chi\`eze et al.}(2006)}]{turck06}Turck-Chi\`eze S and the GOLF-NG collaboration 2006 {\it Adv Space Res} {\bf 38} 1812
\bibitem[{{Turck-Chi\`eze et al.}(2008)}]{turck08b}Turck-Chi\`eze S {\it et al}  2008 {\it Astron. Nachr} {bf 329} 521
\bibitem[{{Turck-Chi\`eze et al.}(2009a)}]{turck09a}Turck-Chi\`eze S {\it et al} 2009a
 {\it Experimental Astronomy} {\bf 23} 1017
\bibitem[{{Turck-Chi\`eze et al.}(2009b)}]{turck09b}Turck-Chi\`eze S {\it et al} 2009b {\it HEDP} {\bf 5} 132
\bibitem[{{Turck-Chi\`eze et al.}(2010a)}]{turck10a}Turck-Chi\`eze S, Palacios A, Marques J  and Nghiem P 2010 {\it ApJ} {\bf 715} 1539
\bibitem[{{Turck-Chi\`eze et al.}(2010b)}]{turck10d} Turck-Chi\`eze S  {\it et al}   2010 in Magnetic Coupling between the interior ans atmosphere of the Sun, ed S S Hasan and R J Rutten, edited by S.S. Hasan and R.J. Rutten {\it Astrophys. and Space Science} {\bf 368}
\bibitem[{{Turck-Chi\`eze, Piau \& Couvidat}(2011)}]{turck10b}Turck-Chi\`eze S, Piau L, Couvidat S 2011 {\it ApJ lett} {\bf 731} L29
\bibitem[{{Turck-Chi\`eze et al.}(2011a)}]{turck10c} Turck-Chi\`eze S {\it et al} 2011 {\it  Astrophys \& Space Science} on line, arXiv:1101.1170
\bibitem[{{Turck-Chi\`eze et al.}(2011b)}]{Turck2011} Turck-Chi\`eze S, Garcia R A, Lopes I   {\it et al}   2011, ApJ lett submitted
\bibitem[{{Valle}(2010)}]{valle10}Valle J W F 2010
{\it  J.\ Phys.\ Conf.\ Ser.\ } {\bf 203} 012009
\bibitem[{{Vitense}(1953)}]{vitense53}Vitense E 1953 {\it Pub. Of the Astro. Soc. of the Pacific}  {\bf 65} 385
\bibitem[{{Von Weizsacker}(1938)}]{vonweizsacker38}Von Weizsacker 1938 Physik {\bf 38 } 176 
\bibitem[{{Vorontsov, Baturin \& Pamiatnykh}(1991)}]{vorontsov91}Vorontsov S V, Baturin V A and  Pamiatnykh A A 1991 {\it Nature} {\bf 349} 49
\bibitem[{{Wu}(1957)}]{wu1957}Wu C S 1957  {\it Phys. Rev.} {\bf 105} 1413
\bibitem[{{Wolfenstein}(1978)}]{wolfenstein78}Wolfenstein L 1978 {\it Phys. Rev. D} {\bf 17} 2369 
\bibitem[{{Zahn}(1992)}]{zahn92}Zahn J P 1992 {\it A\&A} {\bf265} 115
\bibitem[{{Zahn, Brun \& Mathis}(2007)}]{zahn07}Zahn J P, Brun A S and Mathis S 2007 {\it A\&A} {\bf 474}
145
\bibitem[{{Zhao \& Kosovichev}(2004)}]{zhao04}Zhao J, Kosovichev A G 2004 {\it ApJ} {\bf 603} 776
\bibitem[{{Zhao, Kosovichev \& Duvall}(2001)}]{zhao01}Zhao J, Kosovichev A G, Duvall T L Jr 2001 {\it ApJ} {\bf 557} 384

\endrefs

\end{document}